\documentclass[aps,twocolumn,prc,superscriptaddress,showpacs,nofootinbib,floatfix,amssymb,amsfonts,amsmath]{revtex4-2}
\pdfoutput=1

\usepackage{graphicx}
\usepackage{dcolumn}
\usepackage{bm}
\usepackage{color}
\usepackage{amsmath}    
\usepackage{amsfonts}   
\usepackage{amssymb}
\usepackage{graphicx}   

\begin{document}
\title{Global study of separable pairing interaction in covariant density functional
theory}

\author{S.\ Teeti}
\affiliation{Department of Physics and Astronomy, Mississippi
State University, MS 39762}

\author{A.\ V.\ Afanasjev}
\affiliation{Department of Physics and Astronomy, Mississippi
State University, MS 39762}

\date{\today}

\begin{abstract}
A systematic global investigation of pairing properties  based on all 
available experimental data on pairing indicators has been performed
for the first time in the framework of covariant density functional theory.
It is based on separable pairing interaction of Ref.\ \cite{TMR.09}. The 
optimization of the scaling factors of this interaction to experimental 
data clearly reveals its isospin dependence in neutron subsystem. 
However, the situation is less certain in proton subsystem since similar 
accuracy of the description of pairing indicators can be achieved both
with isospin-dependent and mass-dependent scaling factors. The 
differences in the functional dependencies of scaling factors lead to 
the uncertainties in the prediction of proton and neutron pairing properties 
which are especially pronounced at high isospin and could have a significant
impact on some physical observables. For a given part of nuclear 
chart the scaling factors for spherical nuclei are smaller than those for 
deformed ones; this feature exists also in non-relativistic density functional
theories.  Its origin  is traced back to particle-vibration coupling in odd-$A$  
nuclei which is missing in all existing global studies of pairing.  Although the present 
investigation is based on the NL5(E) covariant energy density functional (CEDF), its 
general conclusions are expected to be valid  also for other CEDFs built at 
the Hartree level.
\end{abstract}

\maketitle

\section{Introduction}
\label{Introduc}

   Pairing correlations play an extremely important role in nuclear physics:
the inclusion of pairing (particle-particle [pp]) correlations is critical for open 
shell nuclei and this is absolute majority of the nuclei in the nuclear chart (see  
Refs. \cite{Bohr1975,Sol-book-1976,RS.80}). In principle the effective pp-interaction 
is isospin-dependent with a isoscalar (T = 0) and isovector (T = 1) parts. 
However, only isovector part is important since (i) the vast majority of pairing 
effects emerge from this part (see  Refs. \cite{Bohr1975,Sol-book-1976,RS.80})
and (ii) there is no clear indications that the pp-interaction in the T=0 channel 
is strong enough to produce a pairing condensate  (see Refs.\ \cite{A.12,FM.14}). 
That is a reason why we consider only isovector pairing interaction between like
particles in the present paper.

 It is well known that  the pairing correlations play a significant role in the 
description of the ground state properties of open-shell nuclei. However, the 
properties of   rotating nuclei and fission barriers are  especially sensitive to fine 
details of pairing interaction. For example, the experimental moments of inertia 
of low and medium spin rotational bands and their evolution with spin cannot be 
described without inclusion of pairing interaction \cite{RS.80,Szy-book}. In addition, 
the accuracy of their description sensitively depends both on the details of 
pairing interaction [such as its form [for example, quadrupole pairing \cite{SWM.94}] 
and strength (see Refs.\ \cite{GBDFH.94,CRHB,AO.13})] and on (at least, 
approximate)  particle number projection (such as Lipkin-Nogami method) 
(see Refs.\ \cite{GBDFH.94,SWM.94,AER.00,J1Rare,CRHB}). The same 
situation exists also in the description of fission barriers. It was found in Ref.\ 
\cite{KALR.10} that the pairing gap changes considerably with deformation and 
that relativistic mean field (RMF)+BCS calculations with constant gap do not 
provide an adequate description of the barriers.  Relativistic Hartree-Bogoliubov 
(RHB) calculations show that there is a substantial difference in the predicted
barrier heights between zero-range and finite range pairing forces even in the case 
when the pairing strengths of these two forces are adjusted to the same value of 
the pairing gap at the ground state (see Ref.\ \cite{KALR.10}). For zero range forces 
the barrier heights depend on the renormalization procedure. Note also that the
details of pairing are important for the description of transitional nuclei since the
modification of the strength of pairing could drive the system from transitional 
in nature to spherical one and vice versa (see the discussion of octupole deformed
nuclei in Sec. V of Ref.\  \cite{AAR.16}).

  The pairing correlations is an important building block of different density 
functional theories (DFTs).  Among these nuclear DFT's, covariant density 
functional theory (CDFT) is one of most attractive since covariant energy density
functionals (CEDFs) exploit basic properties of QCD at low energies,
such as symmetries and the separation of scales~\cite{LNP.641}. They
provide a consistent treatment of the spin degrees of freedom, they
include the complicated interplay between the large Lorentz scalar
and vector self-energies induced on the QCD level by the in-medium changes
of the scalar and vector quark condensates~\cite{CFG.92}. In addition,
these functionals include {\it nuclear magnetism} \cite{KR.89}, i.e. a
consistent description of currents and time-odd mean fields important
for odd-mass nuclei \cite{AA.10}, the excitations with unsaturated spins,
magnetic moments \cite{HR.88} and nuclear rotations \cite{AR.00,TO-rot}.
Because of Lorentz invariance no new adjustable parameters are required for
the time-odd parts of the mean fields \cite{AA.10}. Of course, at present, all attempts to 
derive these functionals, defining the particle-hole channel of the DFTs, directly 
from the bare forces~\cite{BT.92,HKL.01,SOA.05,HSR.07} do not reach the 
required accuracy. Note that the same situation exists also in the non-relativistic DFTs 
based  on zero-range Skyrme and finite-range Gogny interactions (see Ref.\ 
\cite{Drut2010_PPNP64-120, PM.14} and references quoted therein).
However, in recent years modern phenomenological CEDFs have been 
constructed ~\cite{DD-ME2,DD-PC1,DD-MEdelta,PC-PK1,AAT.19} which provide 
an excellent description of ground and excited states all over the nuclear 
chart ~\cite{VALR.05,MTZZLG.06,NVR.11,AARR.14,RDFNS-book} with 
a high  predictive power.

   The pairing correlations and pairing indicators have been subject of a 
number of studies within the framework of density functional theories (DFTs). 
For example, it was shown in Ref.\ \cite{SDN.98} using Skyrme DFT that 
odd-even staggering  (OES) of binding energies  in light atomic nuclei is 
strongly affected by both nucleonic pairing and deformed mean field. Various 
approximations in the extraction of pairing indicators as well as in the 
calculations of pairing gaps  in Skyrme DFT and their comparison with 
experimental data have been  investigated in Ref.\ \cite{BRRM.00}. 
The global analysis of pairing interaction has been performed
with the SLy4 functional and contact pairing interaction with possible 
density dependence employing different approximations such as the BCS,
Hartree-Fock-Bogoliubov (HFB) and the HFB with approximate particle
number projection by means of the Lipkin-Nogami method in Ref.\ 
\cite{BBNSS.09}.
  Other interesting results on pairing properties obtained in 
Skyrme DFT could be found in Refs.\ \cite{DBHM.01,MAB.11,BLS.12,YMSH.12,CQW.15}.
The mass dependence of the 
average pairing gap values for neutrons and protons  has been  investigated  
in large scale Gogny DFT calculations with the Gogny D1S functional in Ref.\ 
\cite{HBGSS.02}.  However, this analysis is based on the comparison 
of  experimental $\Delta^{(3)}$  indicators, extracted from binding energies,
with theoretical averaged pairing gaps calculated in even-even nuclei. This
drawback has been removed only in Refs.\ \cite{RBB.12,DABRS.15} in 
which experimental and theoretical $\Delta^{(3)}$ indicators have been 
directly compared in several isotope chains of spherical and deformed 
nuclei across nuclei chart (see Ref.\ \cite{RBB.12}) in the calculations with 
the D1S Gogny force and in deformed actinides in the calculations with the 
D1M force (see Ref.\ \cite{DABRS.15}). Note that global analysis of such 
kind is still absent in the Gogny DFT. Other publications on particular
aspects of the pairing interaction based on the Gogny forces have been recently 
reviewed in Ref.\ \cite{RRR.19}.  Note that only non-relativistic studies are mentioned 
here since the detailed discussion of the pairing studies within the CDFT is presented 
below.

In the literature on nuclear DFT
several types of effective pairing forces
$V^{pp}$ have been used. The most simple force is the seniority force of Kerman~\cite{Kerman1961_APNY12-300,RS.80}
with constant pairing matrix elements $G$.
This force is widely used, but is has many limitations, e.g. correlations in
pairs with higher angular momentum are neglected, the scattering between pairs
with different shells is not constant in realistic forces, the coupling to the
continuum is not properly taken into account and the predictive power is limited.
As an alternative a (contact)  zero-range $\delta$-force is used in many 
calculations and it is frequently made density-dependent to be more 
realistic (see, for example, Refs.\ \cite{BBNSS.09,BLS.12,YMSH.12,CQW.15}).
However, it is well known that these forces have the 
problem that, for calculations in full space, the pairing gap shows an ultraviolet 
divergence for any given strength of the interaction (see discussion in Ref.\ \cite{KALR.10} 
and  references quoted therein).

   Thus the search for more realistic pairing led to the use of finite-range
Gogny force in the pairing channel of DFTs. Gogny has derived his energy 
functional from a finite range force of the Brink-Booker type which allows to 
avoid the complicated problem of 
pairing cutoff  \cite{D1}. This is because the finite range guarantees that the 
force decreases as a function of the momentum transfer and the gap 
equation converges without any problems. This type of pairing based on the 
D1S Gogny functional \cite{D1S-a,D1S}  (further called as Gogny pairing) 
has been used in the CDFT framework  in many applications related to the 
description of the ground state properties \cite{GELR.96,VALR.05,RDFNS.16}, 
rotating nuclei \cite{CRHB,AO.13,A.14}, fission barriers  \cite{KALR.10}, the 
nuclei in the vicinity of the proton and neutron drip  lines \cite{LVR.01,VALR.05} 
etc.  Note that the pairing itself is a non-relativistic effect which affects only the 
occupation of the single-particle states in the vicinity of the Fermi level; the 
impact of the pairing field on the small components of the Dirac spinor can be 
neglected to a very good approximation \cite{SR.02}.

  However, the use of the Gogny pairing is computationally  time- and memory-consuming.
Thus, a separable pairing interaction of finite range was introduced as a simplification 
of the Gogny pairing by Tian et al in Ref.\ \cite{TMR.09}.  Its matrix elements in 
$r$-space have the form
\begin{eqnarray}
\label{Eq:TMR}
V({\bm r}_1,{\bm r}_2,{\bm r}_1',{\bm r}_2') &=& \nonumber \\
= - f\,G \delta({\bm R}-&\bm{R'}&)P(r) P(r') \frac{1}{2}(1-P^{\sigma}) 
\end{eqnarray}
with ${\bm R}=({\bm r}_1+{\bm r}_2)/2$ and ${\bm r}={\bm r}_1-{\bm r}_2$
being the center of mass and relative coordinates.
The form factor $P(r)$ is of Gaussian shape
\begin{eqnarray}
P(r)=\frac{1}{(4 \pi a^2)^{3/2}}e^{-r^2/4a^2}
\end{eqnarray}
 The factor $f$ is the scaling factor of the pairing force which in 
general is particle number dependent, i.e. $f=f(Z,N)$ (see Ref.\ \cite{AARR.14}).
The parameters $G=728$ MeV fm$^3$ and $a=0.644$ fm of this interaction,
which  are the same for protons and neutrons,
have been derived by a mapping of the $^1$S$_0$ pairing gap of infinite 
nuclear matter to that of the Gogny force D1S \cite{TMR.09} under the
condition that $f=1.0$. These parameters are also used in the present
paper. In finite nuclei, 
the pairing gaps calculated with separable pairing interaction and Gogny 
pairing are very close to each other (see Ref.\ \cite{AARR.14}).  Because
numerical calculations with separable pairing interaction are less time-consuming 
that those with Gogny pairing, its use become widespread in the CDFT 
calculations of the ground state properties \cite{TMR.09a,NRVTM.10,AARR.14,RDFNS.16}, 
fission properties \cite{AARR.17,ZLVZZ.15,TAA.20},  the properties of rotating nuclei 
\cite{Wang.17,Xiong.20} and in many beyond mean field and QRPA calculations
\cite{TMR.09b,PNLV.12,LLLYM.15,SALM.19}  based on the CDFT framework. 

\begin{figure}[htb]
\centering
\includegraphics[width=8.4cm]{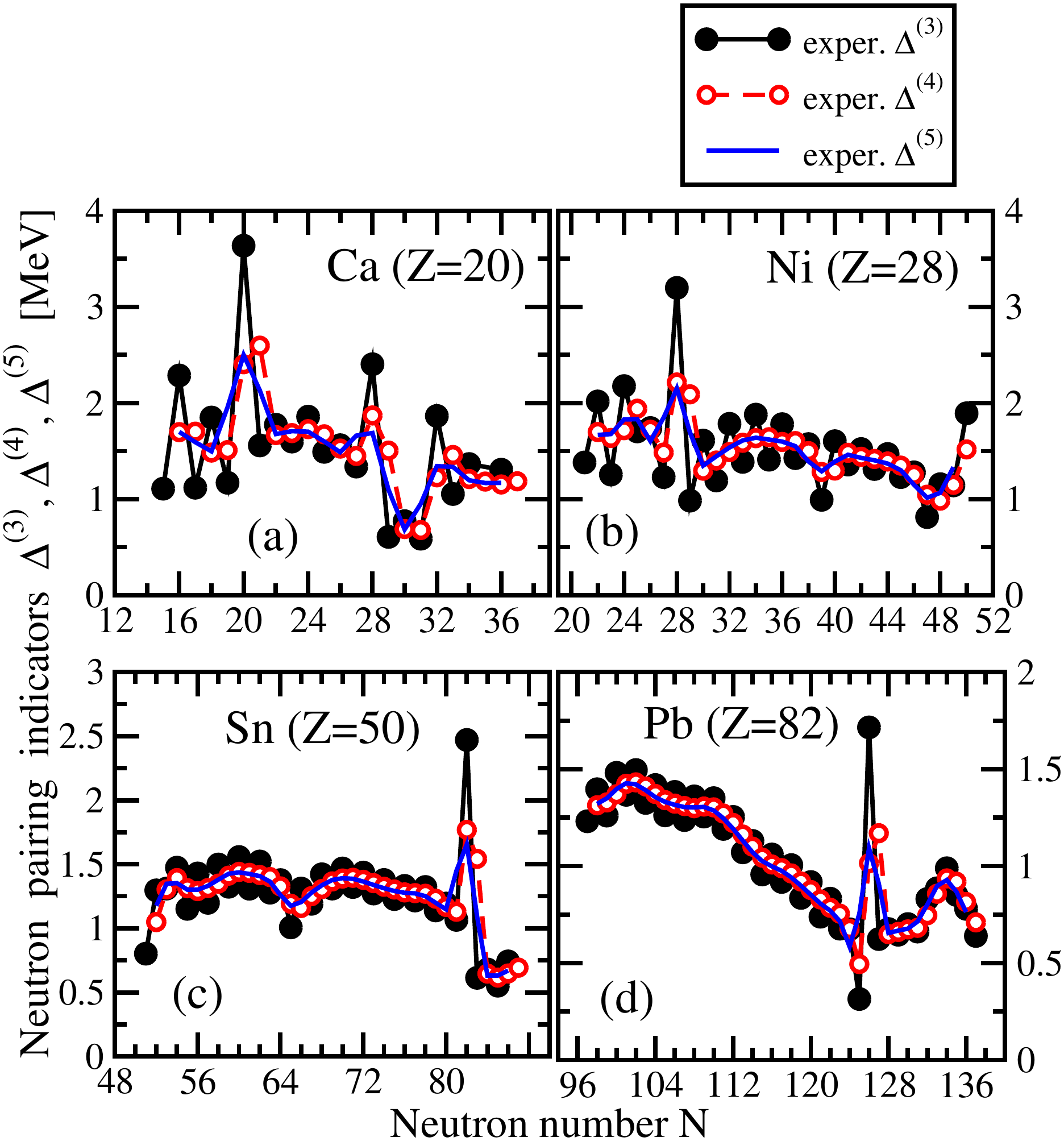}
\caption{Experimental neutron pairing indicators $\Delta^{(3)}_{\nu}(Z,N)$,
$\Delta^{(4)}_{\nu}(Z,N)$ and $\Delta^{(5)}_{\nu}(Z,N)$  as a function of the neutron number $N$
for indicated isotopic chains of spherical nuclei. 
}
\label{Delta-indic-neut}
\end{figure}
   
\begin{figure*}[ht]
\centering
\includegraphics[width=12.6cm]{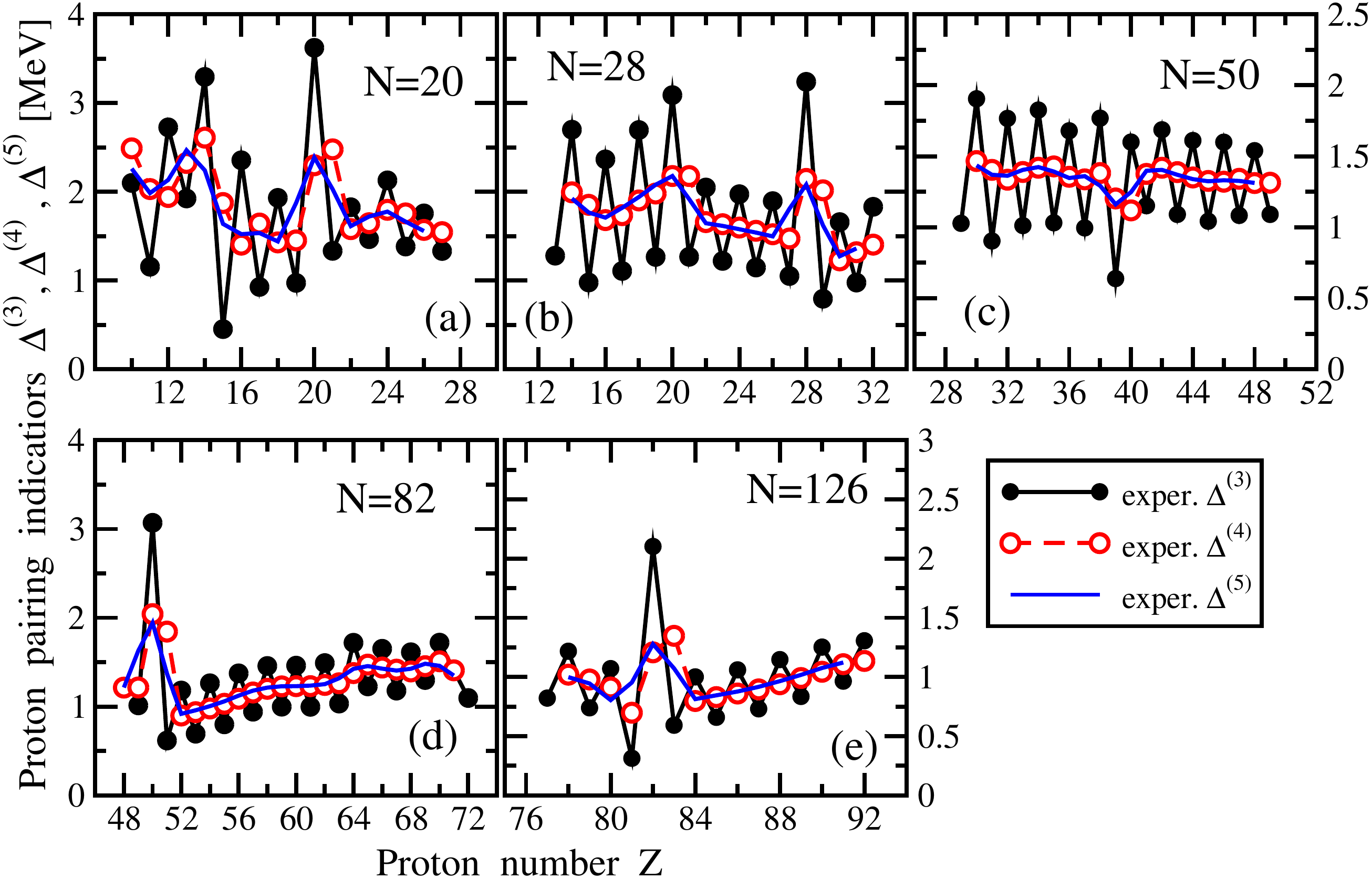}
\caption{The same as Fig.\ \ref{Delta-indic-neut} but for proton  
indicators $\Delta^{(3)}_{\pi}(Z,N)$, $\Delta^{(4)}_{\pi}(Z,N)$ and $\Delta^{(5)}_{\pi}(Z,N)$ as a
function of proton number $Z$ for indicated isotonic chains of spherical
nuclei. 
}
\label{Delta-indic-prot}
\end{figure*}

   In many CDFT applications both the Gogny and separable pairing interactions
have been used with scaling factor $f=1.0$. However, over the years it became 
clear that the pairing force based on the D1S force does not take fully into account 
the mass and particle dependencies of the experimental pairing. For the first time, this feature 
has been seen in the Cranked Relativistic Hartree-Bogoliubov (CRHB) calculations 
of rotational properties of few actinides in Ref.\ \cite{A250} and later it was confirmed 
in a more systematic CRHB calculations of Ref.\ \cite{AO.13} in the same mass
region. The analysis of
odd-even staggering in nuclear binding energies of spherical nuclei has also
revealed the need for mass or particle number dependencies of the Gogny D1S
pairing for the reproduction of experimental pairing indicators \cite{WSDL.13,AARR.14}. 
These results indicate that the pairing effects are underestimated for light nuclei 
and overestimated for heavy ones when the Gogny D1S force is used in the pairing 
channel.  Similar situation exists also for separable pairing interaction the parameters
of which are defined by the fit to the Gogny D1S force \cite{AARR.14}.

  It is necessary to recognize that the studies of particle number dependencies
of the Gogny and separable pairing in the CDFT framework have been performed 
either in localized regions of nuclear chart or for specific types of systems such as 
spherical nuclei in Refs.\ \cite{WSDL.13,AARR.14} or rotating nuclei in Refs.\ 
\cite{A250,AO.13}.  As a consequence,  even nowadays these dependencies are 
neglected in many applications of the CDFT.  Thus, the goal of the present paper is 
to perform for the first time the global study of separable pairing interaction in the 
CDFT framework and optimize its particle number dependencies.   

    The paper is organized as follows.  Experimental pairing indicators 
are  discussed in Sec.\ \ref{Exp-pair-ind}. The discussion of theoretical pairing
gaps is presented in Sec.\ \ref{Sec-th-pairing}. Section \ref{Dis-results}  is 
devoted to the analysis  of the results of the calculations and optimization of 
pairing interaction.  The extrapolation properties of optimized separable 
pairing towards very neutron-rich nuclei  and related theoretical uncertainties  are discussed 
in Sec.\ \ref{extremes}. Finally, Sec.\ \ref{Concl} summarizes the results
of our paper.

\begin{figure*}[ht]
\includegraphics[width=15.7cm,angle=0]{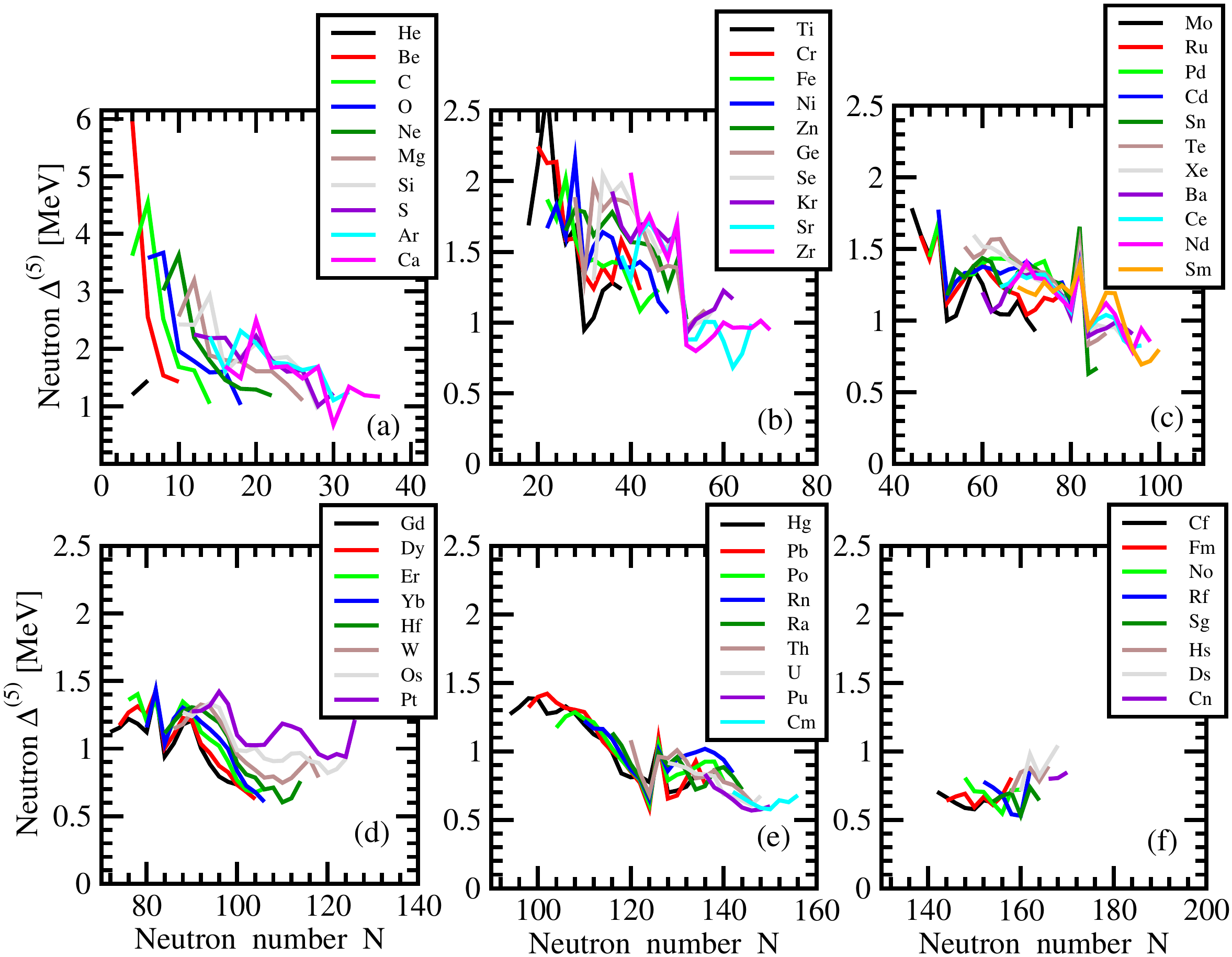}\\
\caption{Experimental neutron $\Delta^{(5)}_{\nu}(Z,N)$ indicators as a function 
of neutron number $N$.  Pairing indicators based both on measured 
and estimated binding energies (see discussion in the beginning of Sec.
\protect\ref{Dis-results}) are included. The data are grouped into six panels 
according to increasing proton number.  This is done to illustrate both the isospin
dependence and the impact of different shell closures on the $\Delta^{(5)}_{\nu}(Z,N)$ 
indicators. The ranges of the neutron number and  $\Delta^{(5)}_{\nu}(Z,N)$ 
changes are the same in the panels (b)-(f); this allows direct comparison
of the slopes of the $\Delta^{(5)}_{\nu}(Z,N)$ values as a function of neutron number
in these panels. Note that these ranges are different in panel (a).  The
experimental data on binding energies are taken from AME2016 mass
evaluation (see Ref.\ \cite{AME2016-first}). 
}
\label{global-neutron-gap}
\end{figure*}

\begin{figure*}[ht]
\includegraphics[width=15.7cm,angle=0]{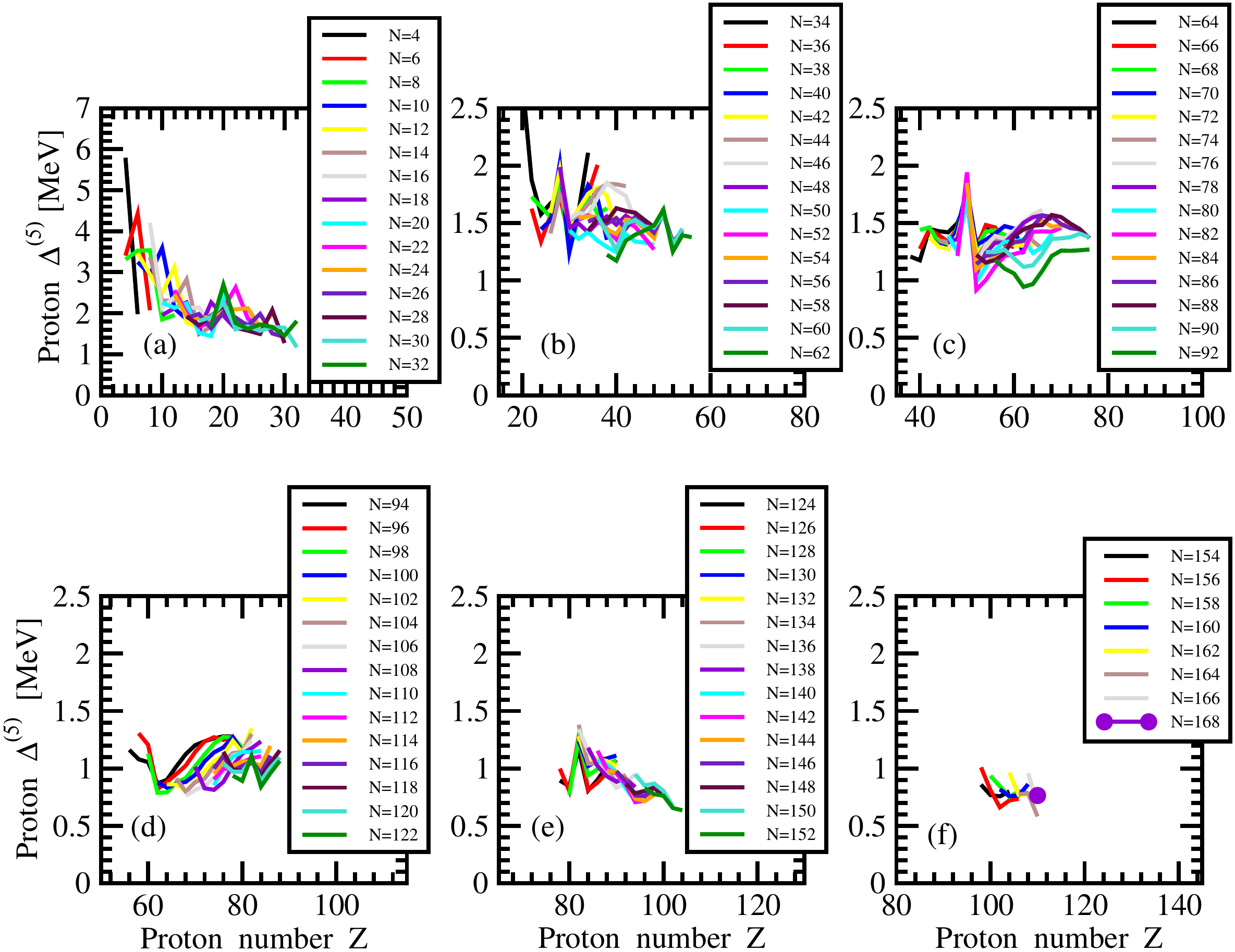}
\caption{The same as in Fig.\ \ref{global-neutron-gap} but for the proton $\Delta^{(5)}_{\pi}(Z,N)$
indicators as a function of proton number $Z$.}
\label{global-proton-gap}
\end{figure*}

\section{Experimental pairing indicators}
\label{Exp-pair-ind}

   There are several expressions for the pairing indicators in the literature. These
are 3-, 4- and 5-points indicators\footnote{Note that these pairing indicators are derived
from the Taylor expansion of the nuclear mass in nucleon-number differences \cite{BRRM.00}.
As a result, they depend on a number of assumptions some of which are strongly
violated at shell closures (see discussion in Sec. 4.2 of Ref.\ \cite{BRRM.00}).}
given by
\begin{eqnarray}
\Delta_{i}^{(3)}(Q_{0})  =   
&\frac{\pi_{Q_{0}}}{2}[B(Q_{0}-1) -& \nonumber \\
& 2B(Q_{0})+B(Q_{0}+1)], &
\label{gap-diff3}
\end{eqnarray}
\begin{eqnarray}
\Delta_{i}^{(4)}(Q_{0})=&\frac{\pi_{Q_{0}}}{4}[B(Q_{0}-2)-3B(Q_{0}-1)& \nonumber \\
&+3B(Q_{0})-B(Q_{0}+1)],&
\label{gap-diff4}
\end{eqnarray}
\begin{eqnarray}
\Delta_{i}^{(5)}(Q_{0})=&-\frac{\pi_{Q_{0}}}{8}[B(Q_{0}+2)-4B(Q_{0}+1)+6B(Q_{0})& \nonumber \\
&-4B(Q_{0}-1)+B(Q_{0}-2)],&
\label{gap-diff5}
\end{eqnarray}
which quantify the 
OES of  binding energies.
Here $Q_0$
is equal to either proton ($Z$) or neutron ($N$) number, $\pi_{Q_0}=(-1)^{Q_0}$ is the
number parity in respective subsystem and $B(Q)$ is the (negative)
binding energy of a system with $Q$ particles.  In these equations, if the 
number of protons $Z$ is fixed, then $i=\nu$ and the indicator gives the neutron 
OES. Otherwise, we obtain proton ($i=\pi$) OES  if the number of neutrons is fixed.
The number parity $\pi_{Q_0}$ is chosen in such a way that the OES is positive 
irrespective of its  centering at either even or odd particle number $Q_0$.

   These experimental indicators are illustrated for spherical nuclei in Figs.\ 
\ref{Delta-indic-neut}  and \ref{Delta-indic-prot}. The $\Delta^{(3)}$ indicator 
shows  a significant odd-even staggering. This staggering is washed out 
when the $\Delta^{(4)}$ and $\Delta^{(5)}$ indicators are used.  The 
$\Delta^{(4)}$  indicator gives an expression which is asymmetric around 
the nucleus  with $Q_{0}$ and  therefore  offers  two  choices \cite{BRRM.00}.
Thus, we use in further studies the $\Delta^{(5)}$ indicator which is symmetric 
and expected to yields better decoupling from mean-field effects \cite{BRRM.00}.

  One should note that the peaks in these indicators (which are especially 
pronounced for the $\Delta^{(3)}$ ones)  appear at $Q_{shell}$ values corresponding 
to the shell  closures or neighboring $Q$ values.  These are the peaks in neutron
pairing indicators at $N=20$ and $N=28$ in the Ca isotopes,  at $N=28$ and 
$N=50$ in the Ni isotopes, at $N=50$  in the Sn isotopes and at $N=126$ in 
the Pb isotopes (see Fig.\ \ref{Delta-indic-neut}). Similar peaks are also seen in 
proton pairing indicators at $Z=20$ in the $N=20$ and $N=28$ isotones, at $Z=50$ 
in the $N=82$ isotones and $Z=82$ in the $N=126$ isotopes (see Fig.\ 
\ref{Delta-indic-prot}).  These peaks in no way should be interpreted as an 
indication of increased pairing correlations. This is connected with the fact that 
pairing correlations either disappear or are significantly weakened at shell closures (see 
discussion in Sec. IV of Ref.\ \cite{AARR.14})  and the peaks are not produced by pairing, 
but by the large or substantially increased shell gaps for closed-shell configurations. Therefore, 
proton/neutron $\Delta^{(5)}$ quantities corresponding to such peaks at proton/neutron 
shell closures and their close vicinities  do not represent  pairing indicators and thus
are eliminated from the consideration.

  It is interesting to estimate global relative differences in the definition of 
experimental $\Delta^{(4)}$  and $\Delta^{(5)}$ indicators by means of the following 
function 
\begin{eqnarray} 
D_i = \frac{1}{n^{set}_i} \sum_{Z,N} \frac{|\Delta_i^{(5)}(Z,N) -  \Delta_i^{(4)}(Z,N)|} {\Delta_i^{(5)}(Z,N)} \times 100\% 
\end{eqnarray}
Here the sum runs over $n^{set}_i$ nuclei for which pairing indicators are available
and the subscript "i" indicates the subsystem (proton or neutron). The analysis of
all available experimental data based on AME2016 compilation of Ref.\ \cite{AME2016-first}
(see, for example, Figs.\ \ref{global-neutron-gap} and \ref{global-proton-gap} below and 
their discussion) leads to 
$D_{\pi}= 5.50$\% and $D_{\nu}=2.68\%$ if the pairing indicators at $Q=Q_{shell}$
are excluded from consideration and to improved values of
$D_{\pi}= 4.20$\% and $D_{\nu}=1.93\%$
if the pairing indicators at $Q=Q_{shell}, Q_{shell}\pm 2$
are excluded from consideration.  These numbers suggest that even 
on the level of experimental data, the pairing indicators suffer from some degree
of uncertainty and are not uniquely defined.

   It is necessary to recognize that all these definitions of the OES of the 
binding energies  do  not provide a clean measure of pairing correlations 
since they have contributions which are not directly related to them. 
These include:

\begin{itemize}

\item 
   Non-negligible contributions in time-even and time-odd mean fields
   from the response of the underlying mean field to the blocking of a
   single-particle state in odd-mass nuclei \cite{SDN.98,BRRM.00,PERN.10,AA.10}. 
   This includes also the effects from the breaking of Kramer's degeneracy 
   of the single-particle states  in odd-$A$ nuclei \cite{AA.10}.
  
  \item
 The impact of the fact that the structure (in terms of the Nilsson orbital) of the 
experimental ground states in odd-$A$ nuclei is reproduced globally only in 
approximately 40\% of the nuclei in the DFTs \cite{BQM.07,AS.11}\footnote{
The  inclusion of  particle-vibrational coupling increases the accuracy 
of the description of the single-particle configurations in odd-$A$ nuclei but such 
studies are limited to spherical nuclei (see Refs.\ \cite{LA.11,AL.15}).}. As a 
consequence, the impact of the blocked orbital  on the deformations and binding 
energies of the ground state is expected  to deviate somewhat (see, for example, 
Fig.\ 4 in Ref.\ \cite{AS.11} for deformation effects) from that expected for the case
when the ground state is based on the  single-particle orbital with correct 
structure.  These facts are ignored in absolute majority of the studies of pairing
via OES of binding energies. It is only in Ref.\ \cite{RBB.12} that this feature has been
mentioned but suggested recipe [not used by the author of Ref.\ \cite{RBB.12} themselves], namely, the 
use of binding energy of the one-quasiparticle configuration with correct Nilsson structure
in the calculations of OES, suffers from substantial uncertainties in the description of the 
energies of the single-particle states in the DFT frameworks \cite{LA.11, DABRS.15}.

\item
   The impact of (quasi)particle-vibrational coupling on the binding energies 
   of odd-mass nuclei. This impact is especially pronounced in spherical
   nuclei \cite{LA.11,AL.15}\footnote{It is interesting that such features 
   have already been mentioned in seminal article of  J. Decharg{\'e} and 
   D. Gogny \cite{DG.80} where they indicated that treating explicitly the
   residual interaction through configuration mixing in odd and even nuclei 
   is expected to lower the OES by approximately 300 keV in the Sn
   isotopes.}. According to the quasiparticle-phonon model of 
   V.~G.~Soloviev \cite{Sol-book-1976}, the admixtures of the phonons to the 
   structure of the ground states  in deformed rare-earth and actinide 
   odd-mass nuclei  is  relatively small (especially, when compared with
   spherical nuclei) \cite{GISF.73,ABNSW.88,SSMJ.15}.  
  
\item
 It was also found in Ref.\ \cite{NB.19}  that random phase 
approximation based part of the total shell correction, which accounts for the 
neutron-proton  pair-vibrational correlation energy,  contributes significantly to  
the calculated  odd-even mass differences, particularly in the light nuclei.

\end{itemize} 

   Thus, one can conclude that the comparison of experimental pairing 
indicators extracted from binding energies with theoretical results 
obtained from mean field type calculations can reveal only general trends of pairing 
evolution
as a function of particle numbers and the local differences between theory 
and experiment are expected.  One should also recognize that global 
investigations of pairing at the beyond mean field level are not possible 
nowadays. The analysis of pairing interaction at the mean field level based on 
experimental data which by definition includes beyond mean field effects 
suggest that the latter should be to a degree treated as a "noisy" data
where the noise is coming from neglected beyond mean field effects. As a 
consequence, a special attention has to be paid to the reduction of the
impact of the data which strongly deviates from general trends (for
example, by employing robust fitting procedures [see Sec.\ \ref{Dis-results} below] 
as well as on the comparative analysis of different types of the nuclei and different
regions of the nuclear chart.

\section{Theoretical pairing gaps}
\label{Sec-th-pairing}

\begin{figure*}
\centering
\includegraphics[angle=-90,width=8.5cm]{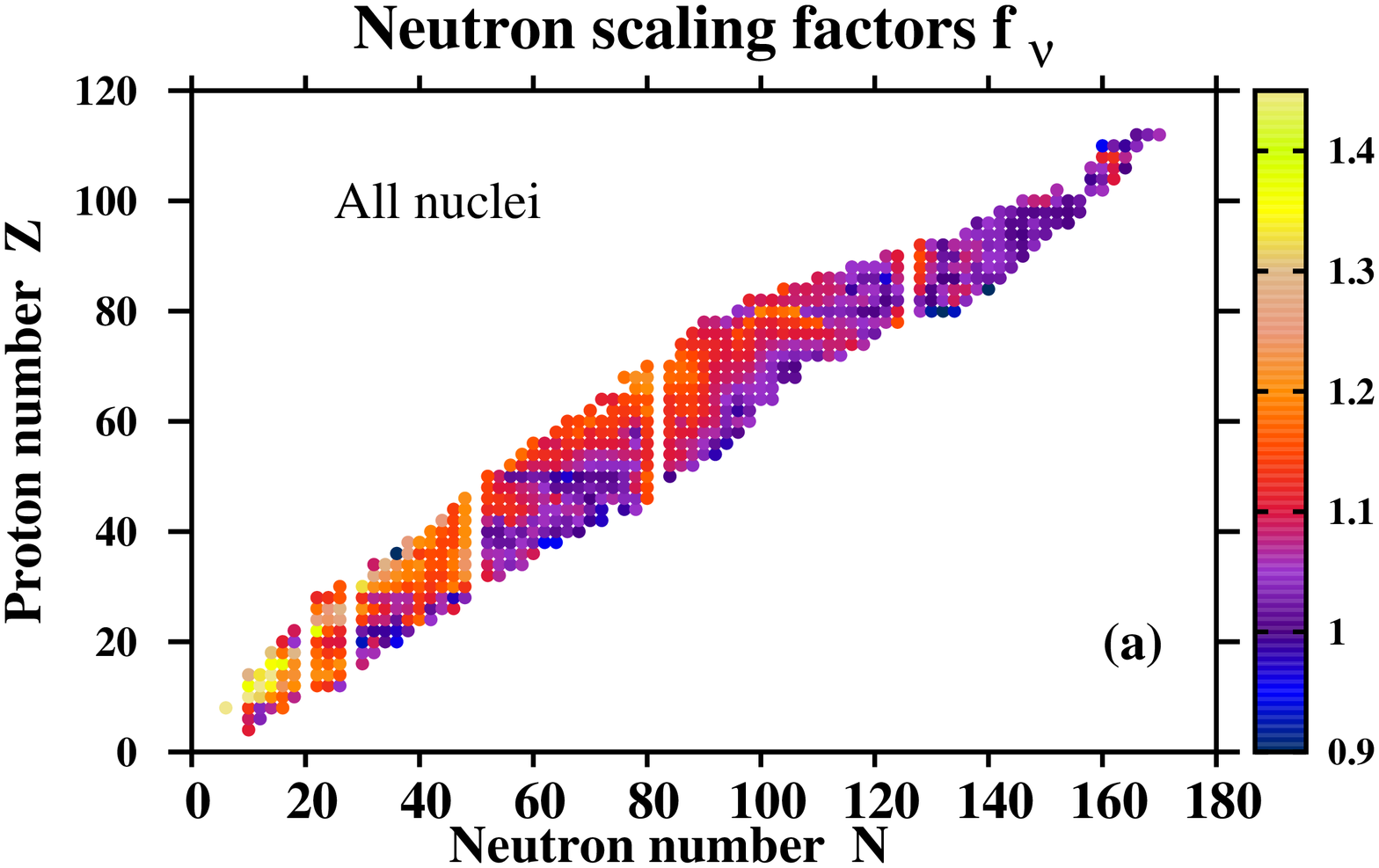}
\includegraphics[angle=-90,width=8.5cm]{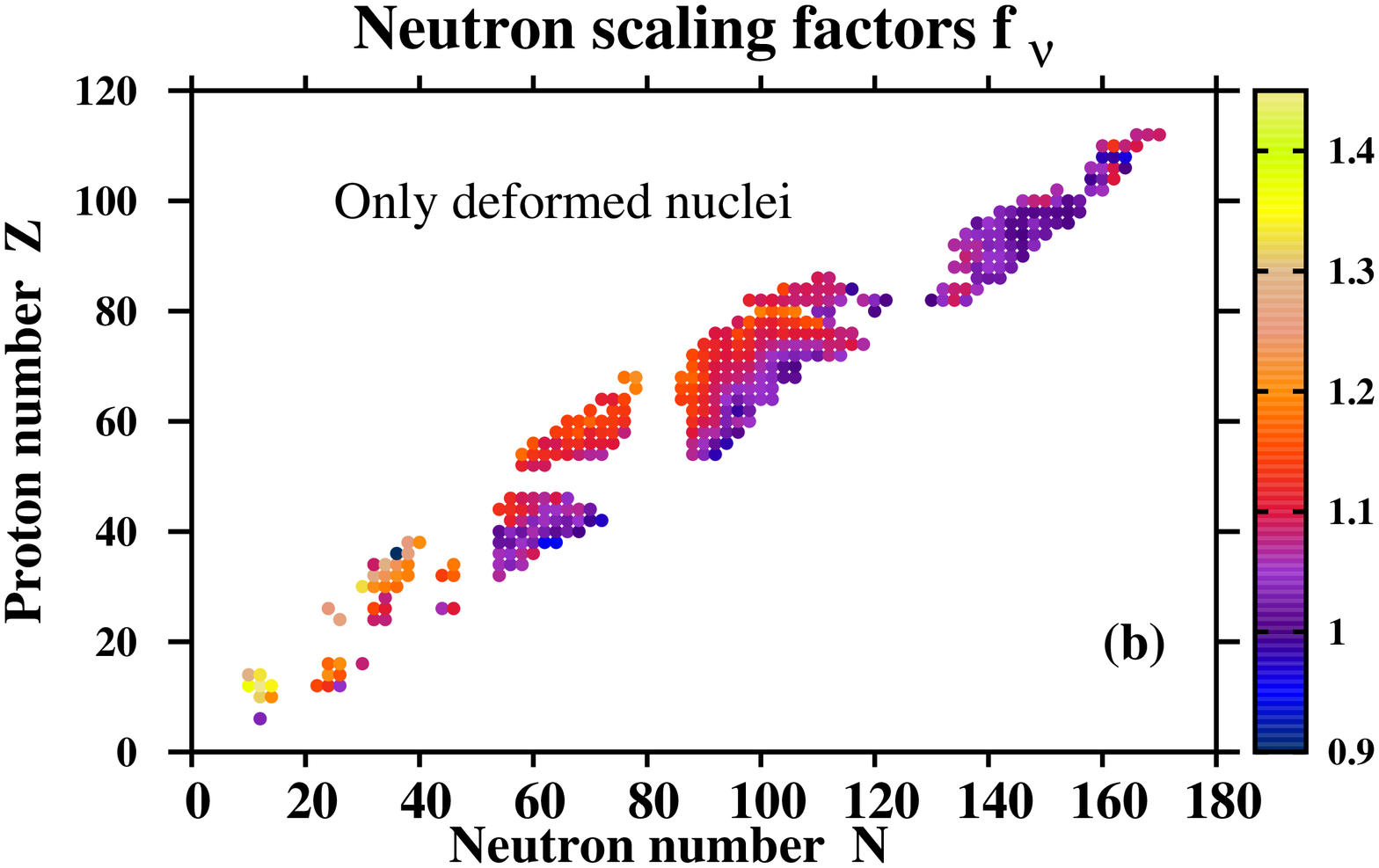}
\includegraphics[angle=-90,width=8.5cm]{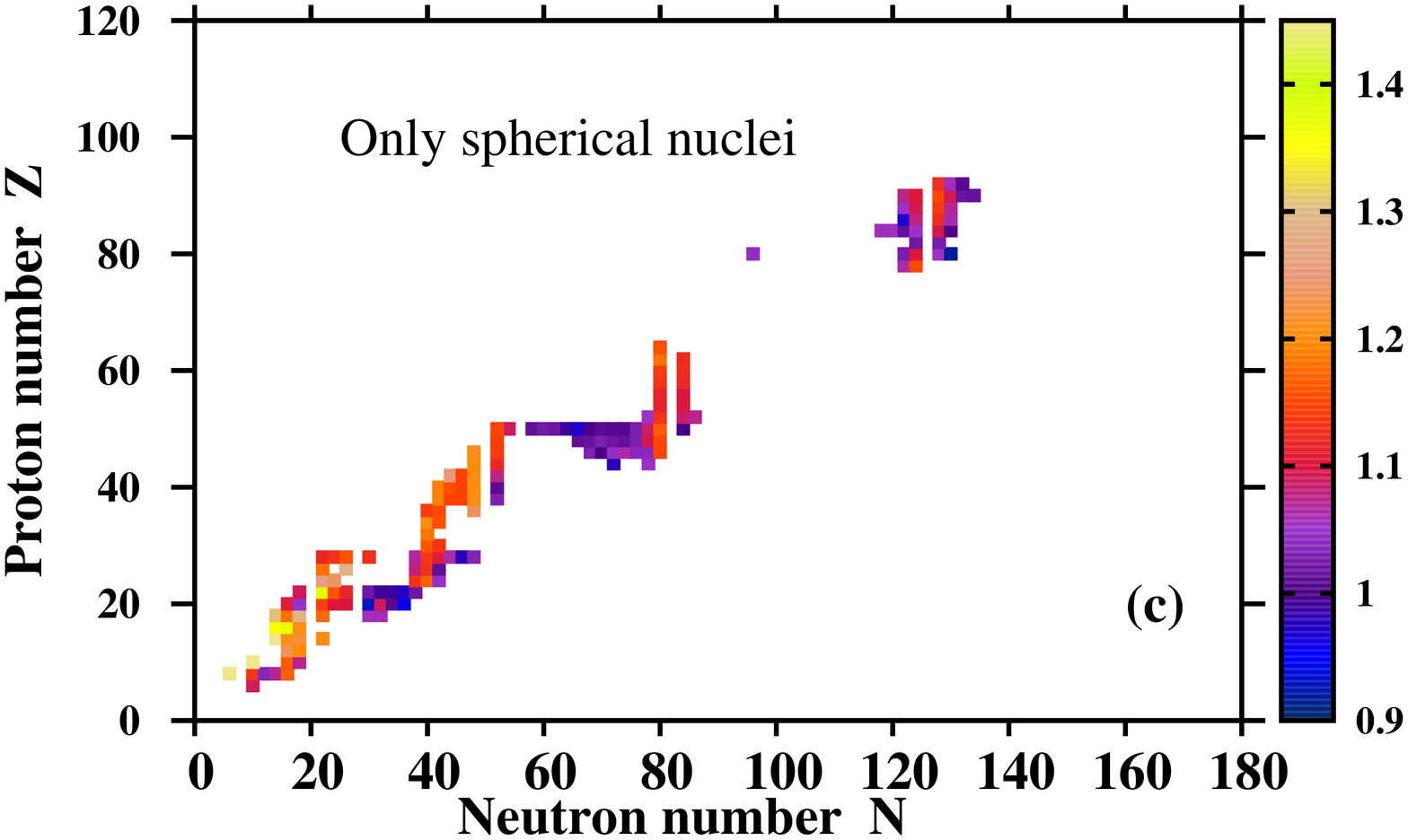}
\includegraphics[angle=-90,width=8.5cm]{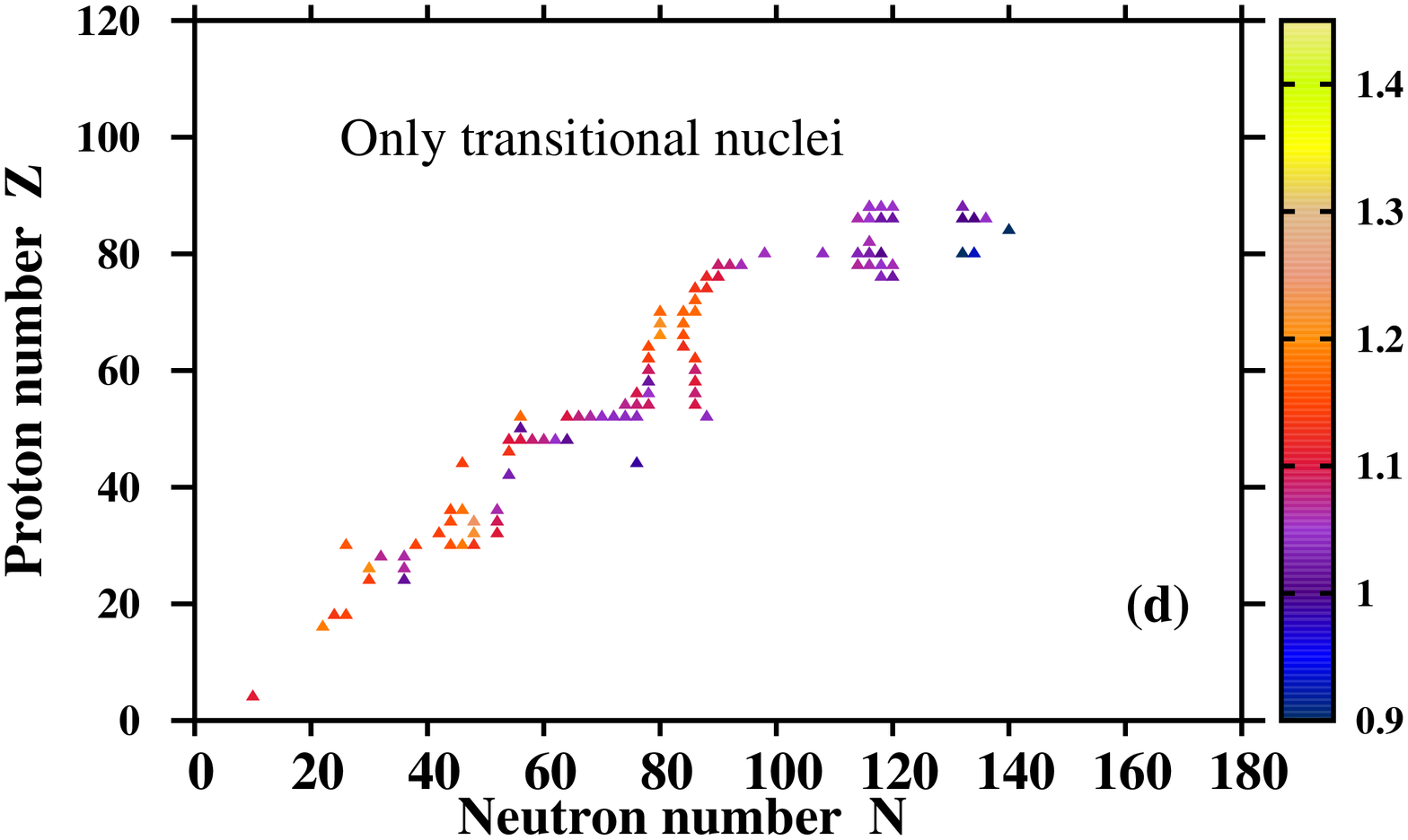}
\caption{The distribution of the scaling factors $f_{\nu}(Z,N)$ of neutron pairing in 
the nuclear chart. Panel (a) shows all nuclei for which experimental $\Delta^{(5)}_{\nu}(Z,N)$
indicators can be defined.  Note however that we exclude such indicators for 
$N=20, 50, 82$ and 126. Panels (b), (c) and (d)  show scaling factors only for the nuclei 
the ground states of which are either deformed, spherical or transitional according to the 
calculations. The nuclei with equilibrium deformations $\beta_2$ satisfying the
conditions $|\beta_2| \leq 0.02$,  $0.02< |\beta_2| < 0.15$ and $|\beta_2| \geq 0.15$
are considered as spherical, transitional and deformed, respectively.
}
\label{factors-neutron}
\end{figure*}

   In the present paper, relativistic Hartree-Bogoliubov (RHB) approach is 
used in the calculations. The formalism of this approach is discussed in detail 
in Refs.\ \cite{VALR.05,AARR.14}. Thus, we consider here only technical 
details related to pairing interaction. Note that for this study we 
employ the NL5(E) \cite{AAT.19} covariant energy density functional (CEDF) which 
globally outperforms well known NL3 \cite{NL3} and NL3* \cite{NL3*} functionals 
(see Ref.\ \cite{AAT.19}). 

  The pair field $\hat{\Delta}$ in the RHB theory is given by 
\begin{eqnarray}
\hat{\Delta} \equiv \Delta_{ n^{}_1 n^{}_2}~=~\frac{1}{2}\sum_{n_1^{\prime}n_2^\prime}
\langle n^{}_1 n^{}_2\vert V^{pp}\vert n_1^{\prime}n_2^\prime\rangle\kappa^{}_{n_1^{\prime}n_2^\prime}.%
\label{gap}
\end{eqnarray}
It contains the pairing tensor  $\kappa$
\begin{equation}
\kappa = V^{*}U^{T}
\label{kappa}
\end{equation}
and the effective interaction $V^{pp}$ in the particle-particle channel for which
separable pairing interaction of Eq.\ (\ref{Eq:TMR}) is used. In theoretical
calculations we use pairing gap\footnote{The pairing gap $\Delta_{\rm uv}$
is related to the average of the state dependent gaps over the pairing tensor.
$\Delta_{\rm uv}$ averages over $u_kv_k$, a quantity which is concentrated around
the Fermi surface.  However, because of the fact that $\kappa\sim \sum_k u_kv_k$
diverges for the seniority force and for zero range forces, $\Delta_{\rm uv}$
turns out to depend on the pairing window. This is, however, no problem for the
separable pairing of finite range used in this investigation.}
\begin{equation}
\Delta_{\rm uv}=\frac{\sum_k u_kv_k\Delta_k}{\sum_k u_kv_k}
\end{equation}
which provides a better description of experimental pairing indicators
in the CDFT framework as compared with other forms of pairing gaps [see Sec. IV 
in Ref.\ \cite{AARR.14}).
  
   For each even-even nucleus under consideration we define neutron and 
proton scaling factors $f_{\nu}(Z,N)$ and $f_{\pi}(Z,N)$ of separable pairing 
force (see Eq.\ (\ref{Eq:TMR})) from the condition
\begin{equation} 
\Delta_{\rm uv}^i (Z,N) = \Delta^{(5)}_{i-cor}(Z,N)
\label{scal-fact-cond}
\end{equation}
where $\Delta^{(5)}_{i-cor}(Z,N)$ (defined below) stands for experimental 
$\Delta^{(5)}(Z,N)$  pairing indicator corrected for the effects of time-odd 
mean fields in odd-$A$ nuclei. Here subscript "i" indicates the subsystem 
(proton or neutron).
This is done for all even-even $(Z,N)$ nuclei for which either experimental 
$\Delta^{(5)}_{\nu}(Z,N)$  indicator or experimental $\Delta^{(5)}_{\pi}(Z,N)$ 
indicator or both of them are available. 
   
   The $\Delta^{(5)}_{\nu-cor}$  and  $\Delta^{(5)}_{\pi-cor}$ indicators are 
determined by means of Eq.\ (\ref{gap-diff5}) but with experimental binding 
energies of odd nuclei  $B^{odd-i}(Z,N)$ corrected for the impact of time-odd 
mean fields ($\Delta E_i$) via
\begin{equation}
B^{odd-i}_{cor}(Z,N) =  B^{odd-i}(Z,N) + \Delta E_i
\end{equation}
This is equivalent to averaged removal of time-odd effects from odd-$A$ nuclei 
and allows to compare  the $\Delta^{(5)}_{\nu-cor}$  and  $\Delta^{(5)}_{\pi-cor}$ 
indicators with the respective calculated $\Delta_{\rm uv}$ pairing gaps which 
do not include time-odd mean fields.  Note that this leads to slight increase of the 
$\Delta^{(5)}_{\nu-cor}$  and $\Delta^{(5)}_{\pi-cor}$ values as  compared with 
the $\Delta^{(5)}_{\nu}$ and  $\Delta^{(5)}_{\pi}$ ones.

\begin{figure*}
\centering
\includegraphics[angle=-90,width=8.5cm]{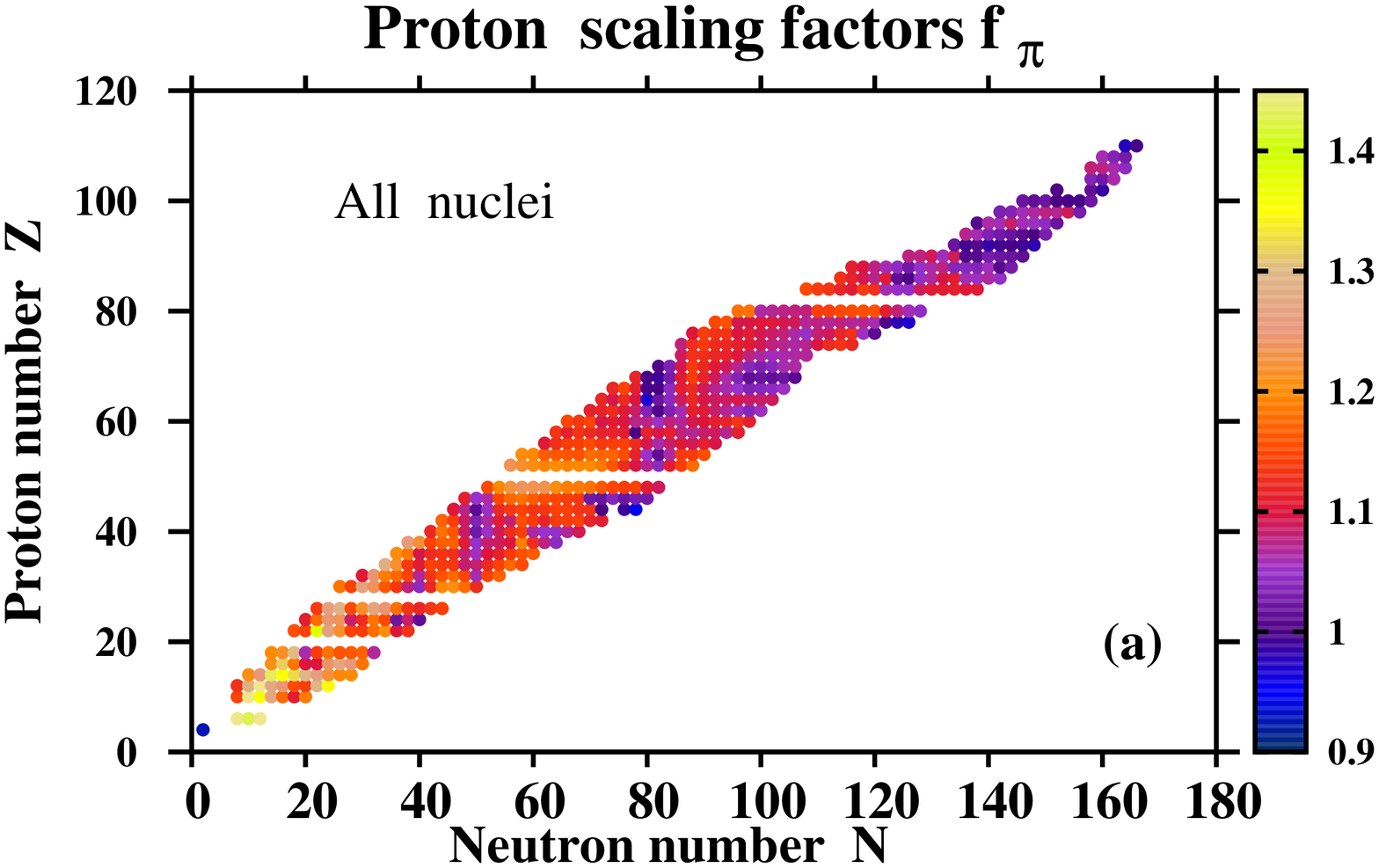}
\includegraphics[angle=-90,width=8.5cm]{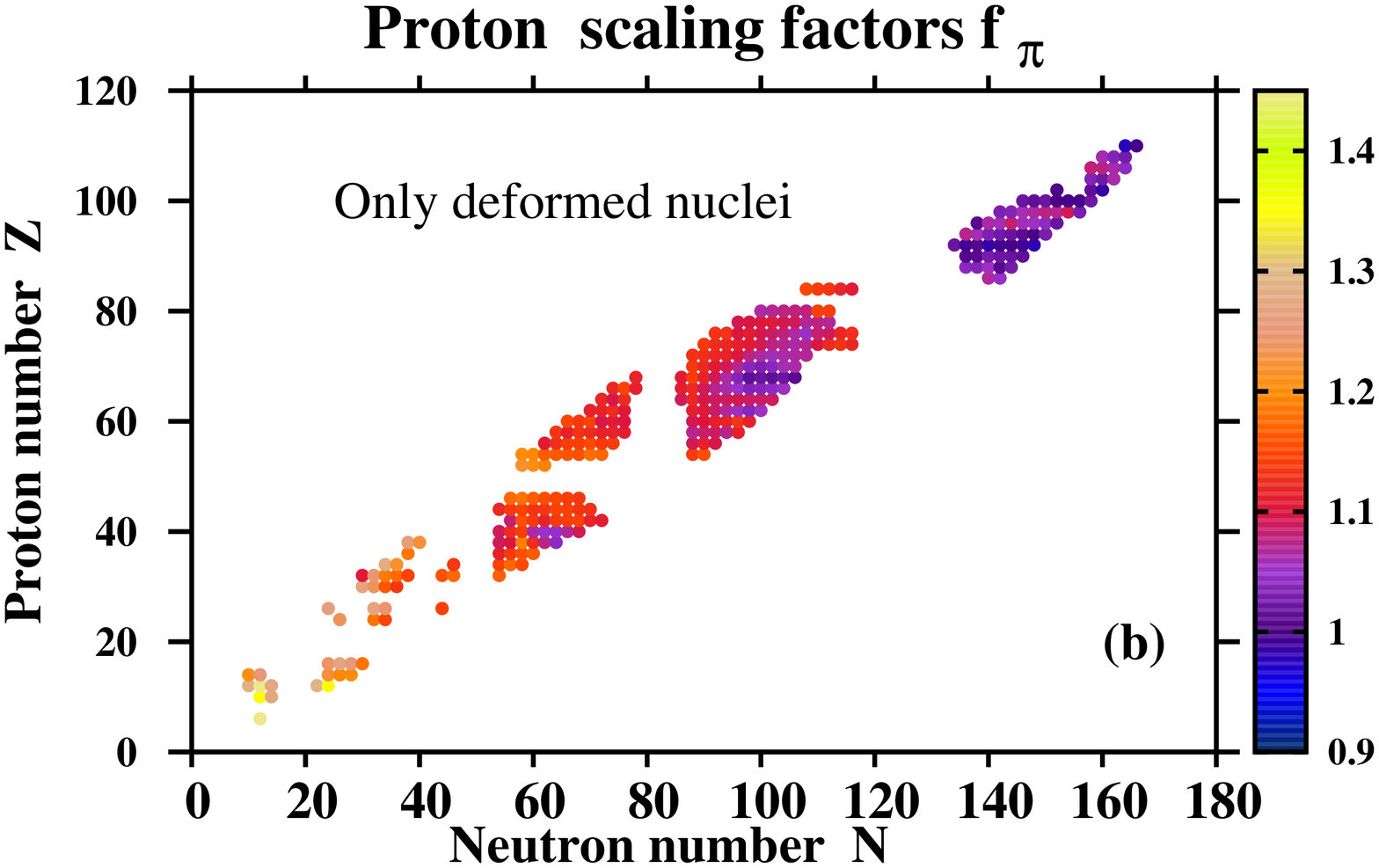}
\includegraphics[angle=-90,width=8.5cm]{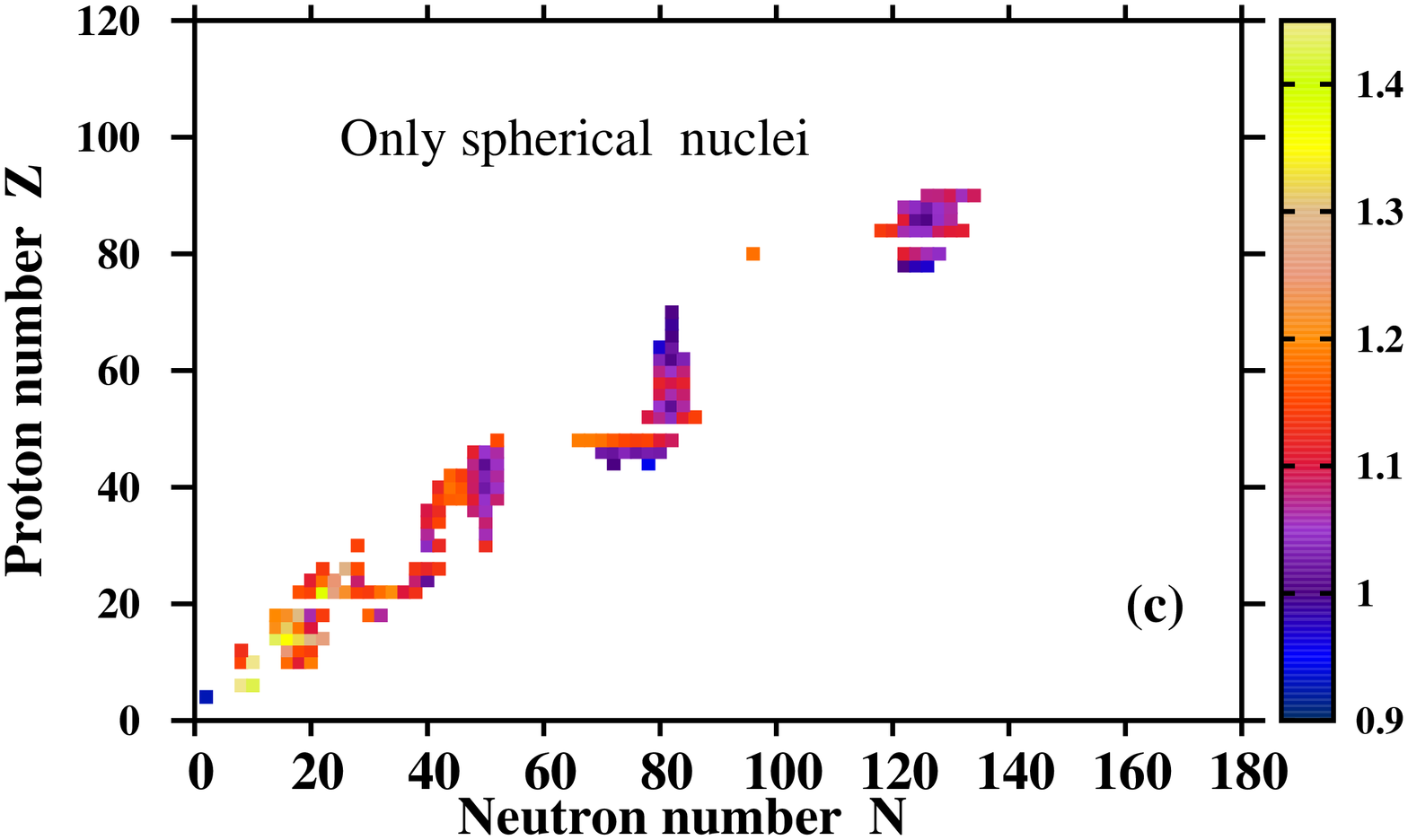}
\includegraphics[angle=-90,width=8.5cm]{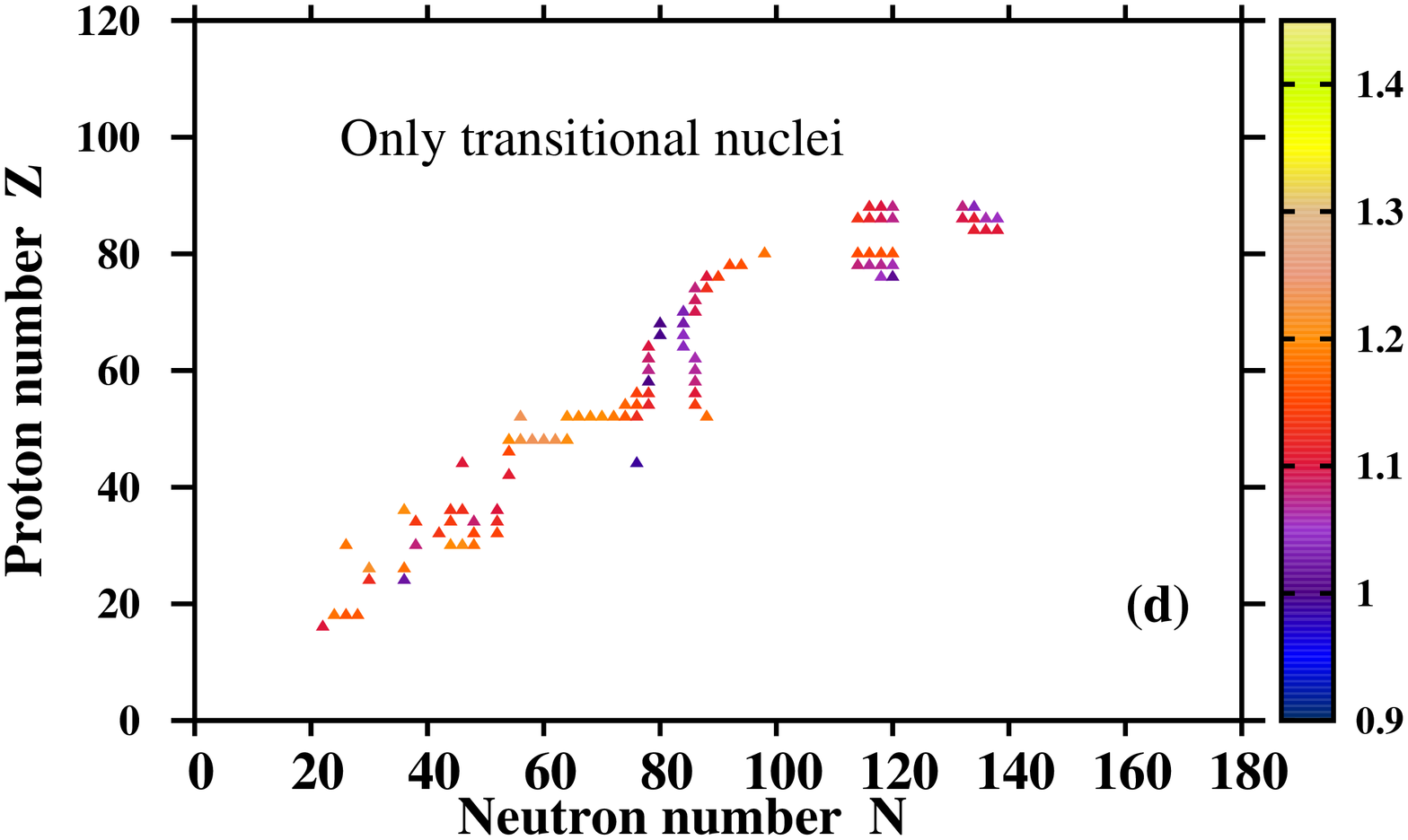}
\caption{The same as Fig.\ \protect\ref{factors-neutron} but for proton scaling 
factors $f_{\pi}(Z,N)$ of proton pairing. Note however that 
we exclude proton pairing indicators for $Z=20, 50$ and 82.} 
\label{factors-proton}
\end{figure*}

  The need for such a
correction comes from the fact that time-odd mean fields (nuclear magnetism 
in the CDFT framework [see Ref.\ \cite{KR.90,AA.10}]) provide an additional 
binding in odd-$Z$ and odd-$N$ nuclei but have no effect on binding energies 
of even-even nuclei (see Ref.\ \cite{AA.10}).  As a consequence, respective 
pairing indicators defined by Eqs.\ (\ref{gap-diff3}),  (\ref{gap-diff4}) and (\ref{gap-diff5}) 
will be modified (see also Sec. IIID in Ref.\ \cite{AA.10} and Ref.\ \cite{RBRM.99}).

   Note that additional bindings due to time-odd  mean fields only weakly depend 
on  CEDF \cite{AA.10}. Thus,  we use  averaged functions $\Delta E_N$ and  $\Delta E_Z$ for 
additional bindings  in odd-$N$ and odd-$Z$ nuclei  due  to time-odd mean fields 
as defined  for the NL3 functional in Ref.\  \cite{AA.10}, namely, 
\begin{eqnarray}
\Delta E_{\nu} &=&0.73 N^{-0.58}\,\, {\rm [MeV]}, \nonumber \\
\Delta E_{\pi} &=& 0.37 Z^{-0.54} \,\,{\rm [MeV]}.
\label{E-NM-cor} 
\end{eqnarray}  
The powers of these expressions are similar for different subsystems. On 
the other hand, the magnitudes in front of the base of power differ considerably 
between proton and neutron quantities, indicating weaker additional binding due 
to time-odd mean fields for odd-proton nuclei.

     The analysis based on the use of the $\Delta_{uv}^{i}(Z,N)$ pairing gaps
has the advantage that these quantities are calculated in even-even nuclei. This
allows to avoid the complicated problem of calculating the blocked states in 
odd-mass nuclei (see Refs.\ \cite{AS.11,AO.13} and discussion below). On the 
other hand, the validity of their use in the comparison with experimental $\Delta^{(5)}_{i}(Z,N)$ 
indicators is not obvious. It turns out, however, that away from spherical shell closures, the 
$\Delta_{\rm uv}^{i}(Z,N)$ values come close to the calculated $\Delta^{(5)}_{i}(Z,N)$ indicators
\cite{AARR.14}. To verify that for the  chains of spherical nuclei shown in Figs.
\ref{Delta-indic-neut} and \ref{Delta-indic-prot} we compared the $\Delta^{(5)}_{i}(Z,N)$ indicators,
obtained in the  calculations without time-odd mean fields, with calculated $\Delta_{uv}^{i}(Z,N)$ pairing 
gaps. In both types of the calculations the Gogny D1S force was used in the pairing channel. This analysis 
is very similar to the one presented in Sec.\ IV of Ref.\ \cite{AARR.14} but performed 
without time-odd mean fields in odd-$A$ nuclei. It turns out that the average deviation
between $\Delta^{(5)}_{i}(Z,N)$ and $\Delta_{uv}(Z,N)$ quantities for this set of nuclei
is on the level of  approximately 5\% if shell closure nuclei with $Q_{shell}$ and their 
neighbors with $Q_{shell}\pm 2$ are excluded from consideration. However, the proton
and neutron scaling factors have to be modified on average by only $\approx 2\%$ to 
compensate for this difference between  two quantities. This value defines the 
accuracy of the approach and it is sufficient for the analysis of global trends. 

\begin{table*}[htb]
\begin{center}
\caption{ 
The parameters of global functional dependencies (see text for details)
defined by the fits to the sets of "measured+estimated" and "measured" scaling factors. The standard 
non-linear least squares fitting is employed here. The bold style is used for RMSE of the best fit
in each group of functional dependencies.
}
\label{fitting-parameter-NL-fit}
\begin{tabular}{|c|c|c|c|c|c|c|c|c|c|c|c|c|} \hline \multicolumn{1}{|c|}{  } & \multicolumn{3}{c|}{Measured+estimated} & \multicolumn{3}{c|}{Measured} \\ \hline
                                                                                  &     C$_i$   &   $\alpha$$_i$   &  RMSE         &  C$_i$   &  $\alpha$$_i$     &   RMSE   \\ \hline
                                 1                                               &     2          &       3                  &   4                &     5        &    6                     &      7  \\ \hline
  $f_{ \pi}=C_{\pi}A^{\alpha_{\pi}} $                            &   1.529     &  -0.0672            & {\bf 0.0549}  & 1.528     & -0.0669              &   {\bf 0.0550} \\ \hline
  $f_{\pi}=C_{\pi}Z^{\alpha_{\pi}} $                             &   1.455     &  -0.0695            &   0.0554        & 1.455     & -0.0690              &   0.05536    \\ \hline
    $f_{ \pi}=C_{\pi}|N-Z|^{\alpha_{\pi}} $                    & 1.256        &  -0.04367          &   0.0567        & 1.256     & -0.0433              &  0.0569 \\ \hline
$f_{ \pi}=C_{\pi}e^{\alpha_{\pi} |N-Z|} $                     &  1.176       &-0.0025              &  0.0566         & 1.178     & -0.00259            &  0.0569   \\ \hline
 $f_{ \pi}=C_{\pi}e^{\alpha_{\pi} [|N-Z|/(N+Z)]} $        &   1.156      & -0.264               &   0.0685        & 1.154     & -0.244                &   0.0685     \\ \hline
                                                                                  &                  &                           &                      &               &                           &                     \\ \hline  
 $f_{\nu}=C_{\nu}A^{\alpha_{\nu}} $                          &   1.388      &  -0.0503            &   0.0614         & 1.404     &   -0.0524           &  0.0592   \\ \hline
 $f_{\nu}=C_{\nu}N^{\alpha_{\nu}} $                          &   1.369      &  -0.0536            &   0.0594         & 1.382     &   -0.0554           &   0.0572  \\ \hline
$f_{ \nu}=C_{\nu}|N-Z|^{\alpha_{\nu}} $                     &  1.249       &  -0.0471            &  {\bf 0.0529}  & 1.253     & -0.0478             &  {\bf 0.0511} \\ \hline
$f_{ \nu}=C_{\nu}e^{\alpha_{\nu} |N-Z|} $                  & 1.154        & -0.0024             &   0.0561         &  1.159    & -0.00255           &  0.0548  \\ \hline
$f_{ \nu}=C_{\nu}e^{\alpha_{\nu}  [|N-Z|/(N+Z)]} $     &   1.192      & -0.551               &  0.0577          & 1.191     & -0.535               &  0.0578      \\ \hline
\end{tabular}
\end{center}
\end{table*}

  An alternative way to above described procedure would be fully self-consistent
calculations of binding energies of even-even and odd-mass nuclei with the 
inclusion of the effects of time-odd mean fields and blocking in odd-$A$ nuclei.
It would lead to theoretical $\Delta^{(5)}_i(Z,N)$ [or lower order] indicators defined 
by Eq.\ (\ref{gap-diff5}) [Eqs.\ (\ref{gap-diff3}) and (\ref{gap-diff4})] which can directly 
be compared with experimental ones.  Such an approach has been employed for 
isotopic (with magic $Z$) and isotonic (with magic $N$) chains of spherical nuclei
and the set of deformed actinides calculated with triaxial RHB computer 
code with fixed scaling factors $f_{\nu}(Z,N)$ and  $f_{\nu}(Z,N)$ in Refs.\ 
\cite{AO.13,AARR.14}.
However, this alternative procedure has its own disadvantages.  First of 
all, systematic calculations within the CDFT \cite{AS.11} and non-relativistic Skyrme 
DFT \cite{BQM.07} showed that in more than half of odd-$A$ nuclei 
the single-particle structure of the ground states cannot be reproduced. 
This means that the polarizations effects in time-even (deformation) and time-odd 
channels affecting the binding energies are not properly defined in these 
odd-$A$ nuclei. Second, the effects of
particle-vibration coupling will modify the binding energies of odd-$A$ nuclei;
this effect is expected to be especially pronounced in spherical systems
(see Ref.\ \cite{LA.11}).  Third, the definition of the ground state in a given 
deformed odd-$A$ nucleus requires the blocking of approximately ten lowest in energy 
quasiparticle states \cite{AS.11}; somewhat smaller number of the states is needed to be considered in
spherical nuclei \cite{BRRM.00,AARR.14}. Numerical calculations of the blocked solutions in odd-$A$ nuclei
are appreciably more time consuming than the ones without blocking in 
even-even nuclei. In addition, there are problems  with the convergence of 
such blocked solutions for some configurations \cite{AS.11,AO.13,DABRS.15}.  Finally, 
global calculations along this procedure  with the optimization of scaling factors 
$f_{\nu}(Z,N)$ and $f_{\pi}(Z,N)$ are prohibitively  time-consuming.

\section{Discussion}
\label{Dis-results}

\begin{table}[htb]
\begin{center}
\caption{
The same as Table \ref{fitting-parameter-NL-fit} but for the case of robust non-linear least square 
fitting based on bisquare width regression scheme. Note that only "measured+estimated" scaling 
factors are used in the fit. The RMSEs from the column 4 of the Table \ref{fitting-parameter-NL-fit} are 
shown in the column 5 (labeled as "RMSE(A)"). 
}
\label{fitting-parameter-bisquare}
\begin{tabular}{|c|c|c|c|c|} \hline 
                                                                                 &     C$_i$   &   $\alpha$$_i$  &  RMSE   & RMSE(A)   \\ \hline
                                1                                               &         2       &        3               &      4       &       5           \\ \hline
  $f_{ \pi}=C_{\pi}A^{\alpha_{\pi}} $                           &   1.555    &  -0.0706   & 0.0515             &  {\bf 0.0549}      \\ \hline
  $f_{\pi}=C_{\pi}Z^{\alpha_{\pi}} $                             &   1.471   &  -0.0722   & 0.0529             &  0.0554      \\ \hline
  $f_{ \pi}=C_{\pi}|N-Z|^{\alpha_{\pi}} $                      & 1.248     &  -0.0421    & 0.0513             & 0.0567      \\ \hline
  $f_{ \pi}=C_{\pi}e^{\alpha_{\pi} |N-Z|} $                   &  1.169    &  -0.00235  & {\bf 0.0491}             &  0.0566      \\ \hline
  $f_{ \pi}=C_{\pi}e^{\alpha_{\pi} [|N-Z|/(N+Z)]} $       &   1.167   &  -0.358      & 0.0596             &  0.0685      \\ \hline
                                                                                  &                &                 &                          &                  \\ \hline
  $f_{\nu}=C_{\nu}A^{\alpha_{\nu}} $                         &   1.36      &  -0.0462   &  0.0598            &  0.0614      \\ \hline
  $f_{\nu}=C_{\nu}N^{\alpha_{\nu}} $                         &   1.355    &  -0.0511   &  0.0582            &  0.0594      \\ \hline
  $f_{ \nu}=C_{\nu}|N-Z|^{\alpha_{\nu}} $                   &  1.24       &  -0.0446   & 0.0501            &  {\bf 0.0529}     \\ \hline
 $f_{ \nu}=C_{\nu}e^{\alpha_{\nu} |N-Z|} $                 & 1.147      & -0.00216  & 0.0521            &   0.0561     \\ \hline
  $f_{ \nu}=C_{\nu}e^{\alpha_{\nu}  [|N-Z|/(N+Z)]} $   &   1.19      & -0.573      & {\bf 0.0485}            &   0.0577     \\ \hline
\end{tabular}
\end{center}
\end{table}

\begin{table}[htb]
\begin{center}
\caption{The same as Table \ref{fitting-parameter-bisquare} but for the case when light 
nuclei with $N<20$ and $Z<20$ are excluded from consideration. 
The RMSEs from the column 5 of the Table \ref{fitting-parameter-bisquare} are 
shown in the column 5 (labeled as "RMSE(B)"). 
}
\label{fitting-parameter-no-light}
\begin{tabular}{|c|c|c|c|c|} \hline 
                                             &  C     &   $\alpha$    & RMSE & RMSE(B)  \\ \hline
    1   & 2 & 3 & 4 & 5 \\ \hline
   $f_{ \pi}=C_{\pi}A^{\alpha_{\pi}} $     &  1.578        & -0.0739    &  0.0496    &0.0515 \\ \hline
   $f_{\pi}=C_{\pi}Z^{\alpha_{\pi}} $      & 1.458         & -0.0699    &  0.0504  & 0.0529 \\ \hline
 $f_{ \pi}=C_{\pi}|N-Z|^{\alpha_{\pi}} $    & 1.242         & -0.0406    & 0.0502  & 0.0501\\ \hline
  $f_{ \pi}=C_{\pi}e^{\alpha_{\pi} |N-Z|} $                                           & 1.165 & -0.00226   & {\bf 0.0476} &  {\bf 0.0491}  \\ \hline
 $f_{ \pi}=C_{\pi}e^{\alpha_{\pi} [|N-Z|/(N+Z)]} $  & 1.163     & -0.353    & 0.0564  &  0.0596   \\ \hline
       & & & & \\ \hline
  $f_{\nu}=C_{\nu}A^{\alpha_{\nu}} $                                                             & 1.360    &   -0.0462  & 0.0574 & 0.0598 \\ \hline
  $f_{\nu}=C_{\nu}N^{\alpha_{\nu}} $                                                           & 1.355     &   -0.0510   & 0.0557 & 0.0582      \\ \hline
  $f_{ \nu}=C_{\nu}|N-Z|^{\alpha_{\nu}} $                                                           & 1.237     & -0.0438     & 0.0479 & 0.0501   \\ \hline
  $f_{ \nu}=C_{\nu}e^{\alpha_{\nu} |N-Z|} $                                                       &  1.145     & -0.00212    & 0.0503 &  0.0521  \\ \hline
  $f_{ \nu}=C_{\nu}e^{\alpha_{\nu}  [|N-Z|/(N+Z)]} $     & 1.192     & -0.595      &  {\bf 0.0462}  &  {\bf 0.0485}  \\ \hline
\end{tabular}
\end{center}
\end{table}

   The binding energies given in the AME2016 mass evaluation \cite{AME2016-first} can 
be separated into two groups: one represents the nuclei with binding energies defined only 
from experimental data, the other contains the nuclei with binding energies depending in 
addition on either interpolation or extrapolation procedures.  As a consequence,
there are the  $\Delta^{(5)}_i(Z,N)$ indicators which are defined only by experimentally 
measured binding energies and the $\Delta^{(5)}_i(Z,N)$ indicators which in addition depend on 
estimated binding energies. For simplicity, we call the  $\Delta^{(5)}_i(Z,N)$ indicators 
(and related scaling factors) in the first 
and second groups as "measured" and "estimated".  Based on the binding energies 
available in AME2016 evaluation one can define 612 neutron  and 611 proton 
measured $\Delta^{(5)}_i(Z,N)$ indicators and  110 neutron and 53 proton estimated 
$\Delta^{(5)}_i(Z,N)$ indicators. Altogether, there are 722 and 664 proton and neutron
"measured+estimated" $\Delta^{(5)}_i(Z,N)$ indicators.  Note that
we consider only even-even and odd nuclei in this paper since the binding 
energies of odd-odd nuclei are affected by the residual neutron-proton interaction
of unpaired proton and neutron.

    Experimental neutron and proton $\Delta^{(5)}_i(Z,N)$ pairing indicators are shown 
in Figs.\ \ref{global-neutron-gap} and \ref{global-proton-gap} for different isotopic 
and isotonic chains, respectively. The analysis of these figures reveals the following 
general features. First, there is a substantial staggering of these indicators as a function 
of respective particle numbers which is especially visible in light nuclei and at spherical 
shell closures. Large peaks appear in the experimental  
$\Delta^{(5)}_i(Z,N)$ indicators at spherical shell closures. In no way they should be considered as
indication of increased pairing: this is connected with the fact that pairing correlations 
either disappear or extremely weak in closed shell nuclei in theoretical calculations. 
These peaks are not produced by pairing, but by increased shell gap for closed-shell 
configurations. Second, there is a general trend of the reduction neutron/proton $\Delta^{(5)}_i(Z,N)$ 
pairing indicators with increasing proton/neutron numbers.  Third, on average neutron 
$\Delta_{\nu}^{(5)}(Z,N)$ indicators for a given isotope chain decrease with increasing 
neutron number (or equivalently isospin) (see Fig.\ \ref{global-neutron-gap}). This
trend is disturbed at spherical shell closures, which is especially visible at $N=50$ in  
Fig.\ \ref{global-neutron-gap}(b), at $N=50$ and $N=82$ in Fig.\  \ref{global-neutron-gap}(c), 
and at $N=126$ in Figs.\ \ref{global-neutron-gap}(d) and (e). Only actinides and light superheavy
nuclei do not show this trend (see Fig.\ \ref{global-neutron-gap}(f)) but this feature is most likely
due to the presence of large deformed $N=162$ shell gap (see Ref.\ \cite{AANR.15}) which leads 
to an increase of the $\Delta_{\nu}^{(5)}(Z,N)$ values in its vicinity.  This fact may also indicate that
some fluctuations in pairing indicators of lighter nuclei are also due to deformed shell gaps.
Fourth, if to exclude light nuclei proton pairing indicators show more constant (on average) proton  
$\Delta_{\pi}^{(5)}(Z,N)$ indicators for a given isotonic chain as a function of proton number as 
compared with neutron $\Delta_{\pi}^{(5)}(Z,N)$  indicators for a given isotopic chain as a 
function of neutron number (compare Figs.\ \ref{global-proton-gap} and \ref{global-neutron-gap}.)
Again this trend is disturbed at spherical shell closures which is visible as the peaks in  
$\Delta_{\pi}^{(5)}(Z,N)$ especially at $Z=28$ in Fig.\ \ref{global-proton-gap}(b), at $Z=50$ in 
Fig.\ \ref{global-proton-gap}(c) and at $Z=82$ in Fig.\ \ref{global-proton-gap}(e).

 Neutron and proton scaling factors $f_{\nu}(Z,N)$ and $f_{\pi}(Z,N)$ defined from 
the condition  of Eq.\ (\ref{scal-fact-cond}) are presented in Figs.\ \ref{factors-neutron} 
and \ref{factors-proton}. These figures reveal several general trends. First, the scaling
factors in both subsystems decrease with increasing mass number. Second,
neutron scaling  factors $f_{\nu}(Z,N)$ also decrease with increasing isospin (see Fig.\ 
\ref{factors-neutron}(a)).  However, this trend is less pronounced in 
proton subsystem  (see Fig.\ \ref{factors-proton}(a)). 
  Third, in a given part of nuclear chart the scaling factors for spherical nuclei are smaller than 
those for deformed ones\footnote{Note that the analysis within the Skyrme DFT with 
zero-range density-dependent pairing interaction also reveals that pairing interaction extracted 
from OES of binding energies is smaller in spherical nuclei as compared with that in
deformed ones \cite{BBNSS.09}. The results of the Gogny DFT calculations for a few isotope
chains of spherical and deformed nuclei presented in Ref.\ \cite{RBB.12} reveal a similar
situation:  the strength  of pairing interaction as defined by the D1M Gogny forces has to
be decreased/increased in spherical/deformed nuclei in order to reproduce experimental
$\Delta^{(3)}$ indicators.}. This is clearly seen for neutron scaling factors $f_{\nu}(Z,N)$ near 
proton shell closures at $Z=20, 28$ and 50 [see Figs.\  \ref{factors-neutron}(a) and (c)] and 
for proton scaling factors near neutron shell closures  at $N=50, 82$ and 126 [see Figs.\ 
\ref{factors-neutron}(a) and (c)].  The comparisons of Figs.\ \ref{factors-neutron}(b) and (c) 
for neutron subsystem and  Figs.\ \ref{factors-proton}(b) and (c) for proton subsystem also 
reveal this feature. The origin of this feature could be related  to the impact of particle-vibration 
coupling (PVC)  impact on the binding energies of odd-mass nuclei.  The PVC increases the 
binding energies of odd-A nuclei (see Ref.\ \cite{DG.80,LA.11}) and  thus reduces the pairing indicators. 
Its effect is more pronounced in  spherical nuclei as compared with deformed ones and thus in 
a given region of nuclear chart it leads to a suppression of scaling factors in spherical nuclei as 
compared with the ones in deformed nuclei. An alternative possible explanation of the 
difference in pairing strength in spherical and deformed nuclei is related to the differences in  the 
single-particle densities in the vicinity of the Fermi levels of such nuclei (see Sec. IV of Ref.\ 
\cite{BBNSS.09}). However, in general this effect should be taken into account already at the
mean field level with proper treatment of pairing correlations. 
 
    It is interesting to see whether the approximate  global functional dependences 
of scaling  factors can be defined and which form of such dependences provides
the best description of extracted scaling factors on a global scale.  For that we have 
investigated the following functionals  dependencies\footnote{We 
also investigated  functional dependencies of the $f_{i}(Z,N) = C_{i} e^{\alpha_{i} N}  
e^{\beta_{i} Z}$ and $f_{i}(Z,N) = C_{i} e^{\alpha_{i} [N/(N+Z)]}   e^{\beta_{i} [Z/(N+Z)]}$
($i= \pi, \nu)$ types which depend on three parameters $C_i$, $\alpha_i$ and $\beta_i$. 
These expressions represent the generalization of  Eqs.\ 
(\ref{f-exp-N-Z}) and (\ref{f-exp-N-Z-vs-N+Z}) which depend on two parameters. 
However, the addition of third parameter does improve the goodness-of-fit only 
marginally as compared with that obtained with two parameters in Table 
\ref{fitting-parameter-NL-fit} and thus does not offer substantial benefits
in the light of existing
limitations of the method. As a consequence, they are not considered in further discussion.
}
\begin{widetext}
\begin{eqnarray} 
f_{\pi}(Z,N) = C_{\pi}A^{\alpha_{\pi}},              \qquad   &   f_{\nu}(Z,N) = C_{\nu}A^{\alpha_{\nu}} \label{f-mass},\\
f_{\pi}(Z,N) = C_{\pi}Z^{\alpha_{\pi}},              \qquad   &   f_{\nu}(Z,N) = C_{\nu}N^{\alpha_{\nu}} \label{f-part-num}, \\
f_{\pi}(Z,N) = C_{\pi}|N-Z|^{\alpha_{\pi}},        \qquad   &   f_{\nu}(Z,N) = C_{\nu}|N-Z|^{\alpha_{\nu}}, \label{f-N-Z} \\
f_{\pi}(Z,N) = C_{\pi}e^{\alpha_{\pi}|N-Z|},     \qquad   &   f_{\nu}(Z,N) = C_{\nu}e^{\alpha_{\nu}|N-Z|},     \label{f-exp-N-Z} \\
f_{\pi}(Z,N) = C_{\pi} e^{\alpha_{\pi}\frac{|N-Z|}{N+Z}}  \qquad & f_{\nu}(Z,N) = C_{\nu}e^{\alpha_{\nu} \frac{|N-Z|}{N+Z}} \label{f-exp-N-Z-vs-N+Z} 
\end{eqnarray}
\end{widetext}
in order to see whether mass, particle number and isospin dependencies can be disentangled. Here the
indices $\pi$ and $\nu$ stands for proton and neutron quantities, respectively. Separate scaling factors
for proton and neutron subsystems reflect explicit  breaking of the isospin symmetry in the pairing energy 
functional which is standard in nearly all pairing forces (see, for example, Refs.\ 
\cite{MN.92,BRRM.00,BBNSS.09}).

\begin{figure*}[ht]
\centering
\includegraphics[angle=0,width=8.5cm]{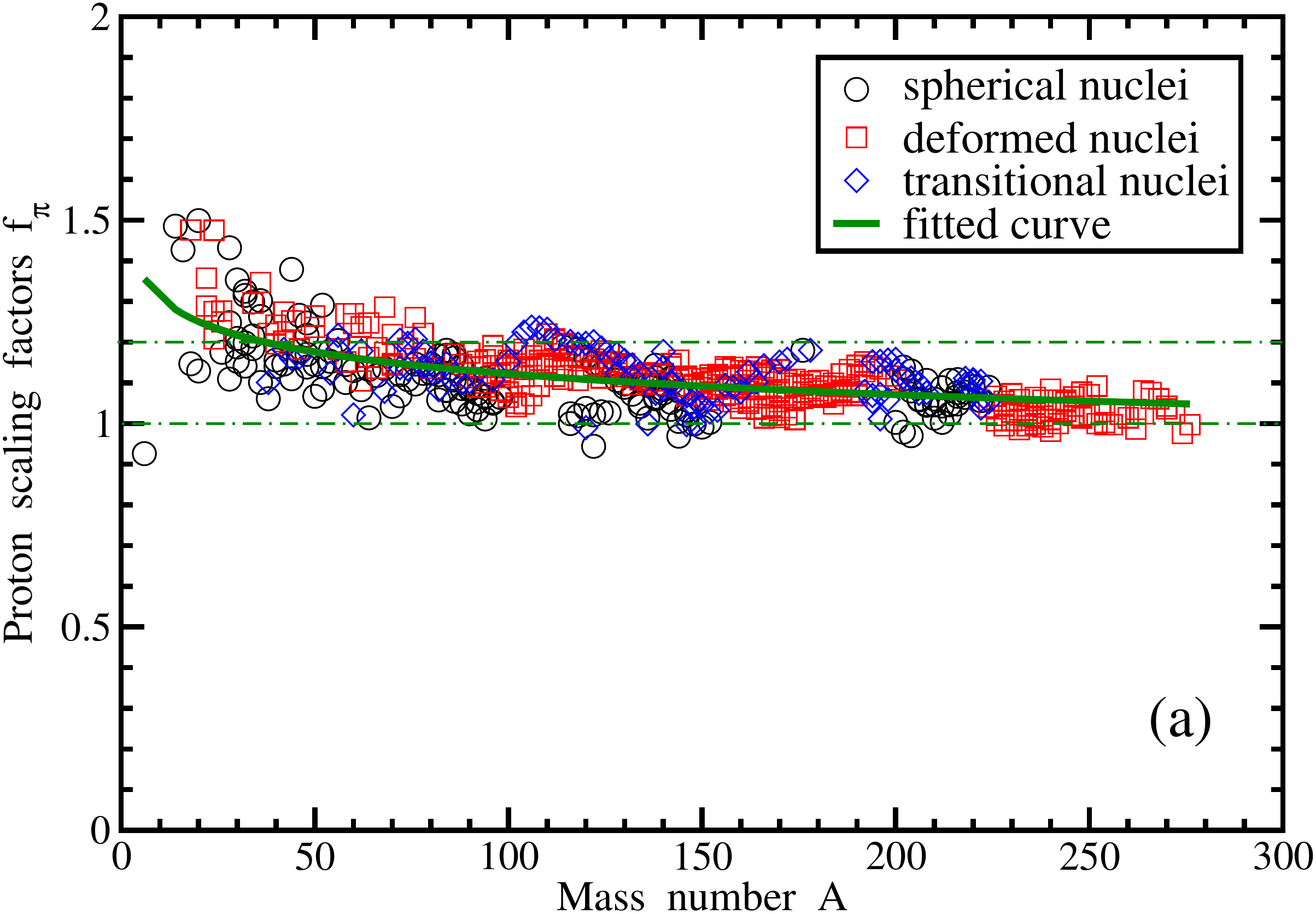}
\includegraphics[angle=0,width=8.5cm]{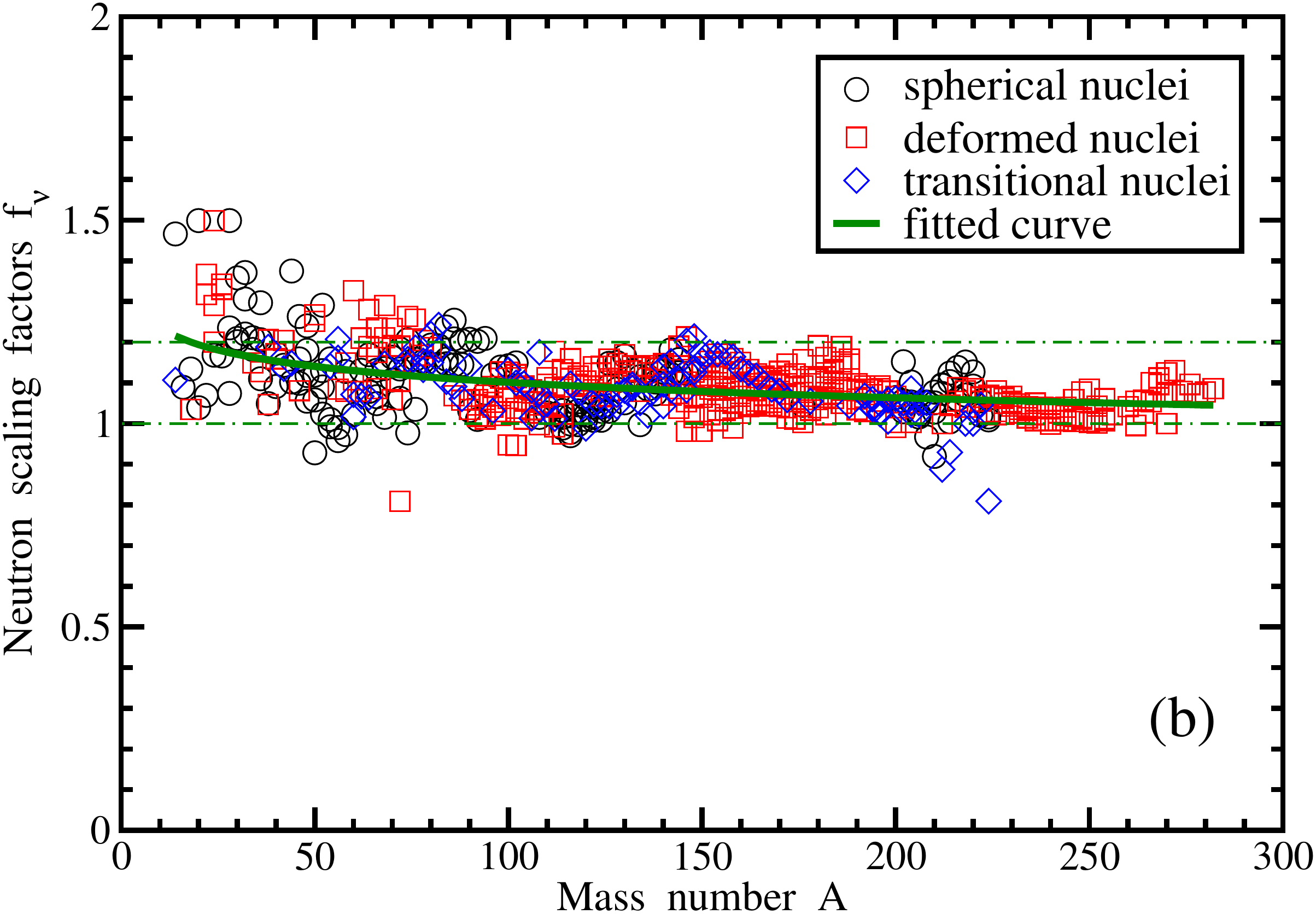}
\includegraphics[angle=0,width=8.5cm]{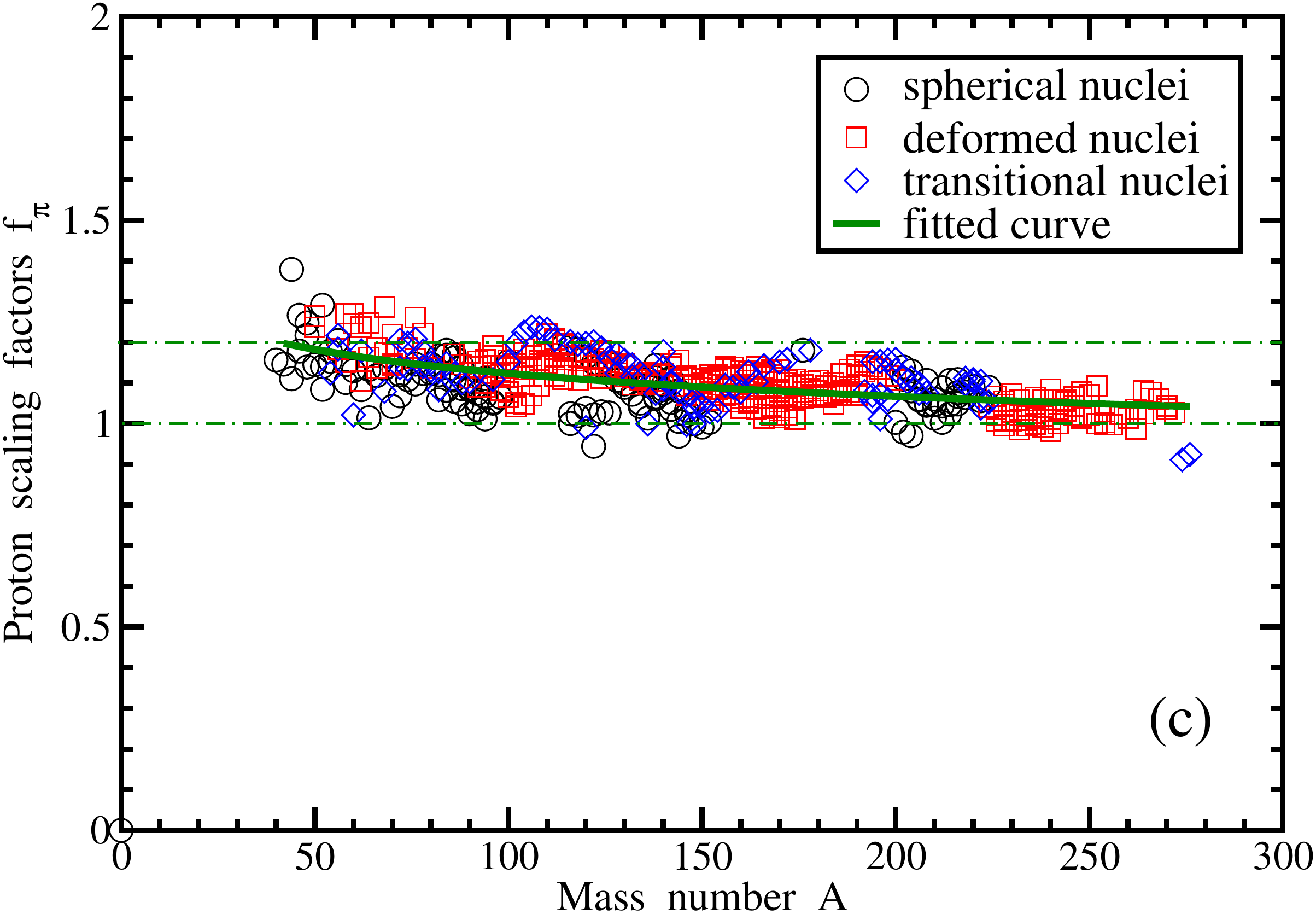}
\includegraphics[angle=0,width=8.5cm]{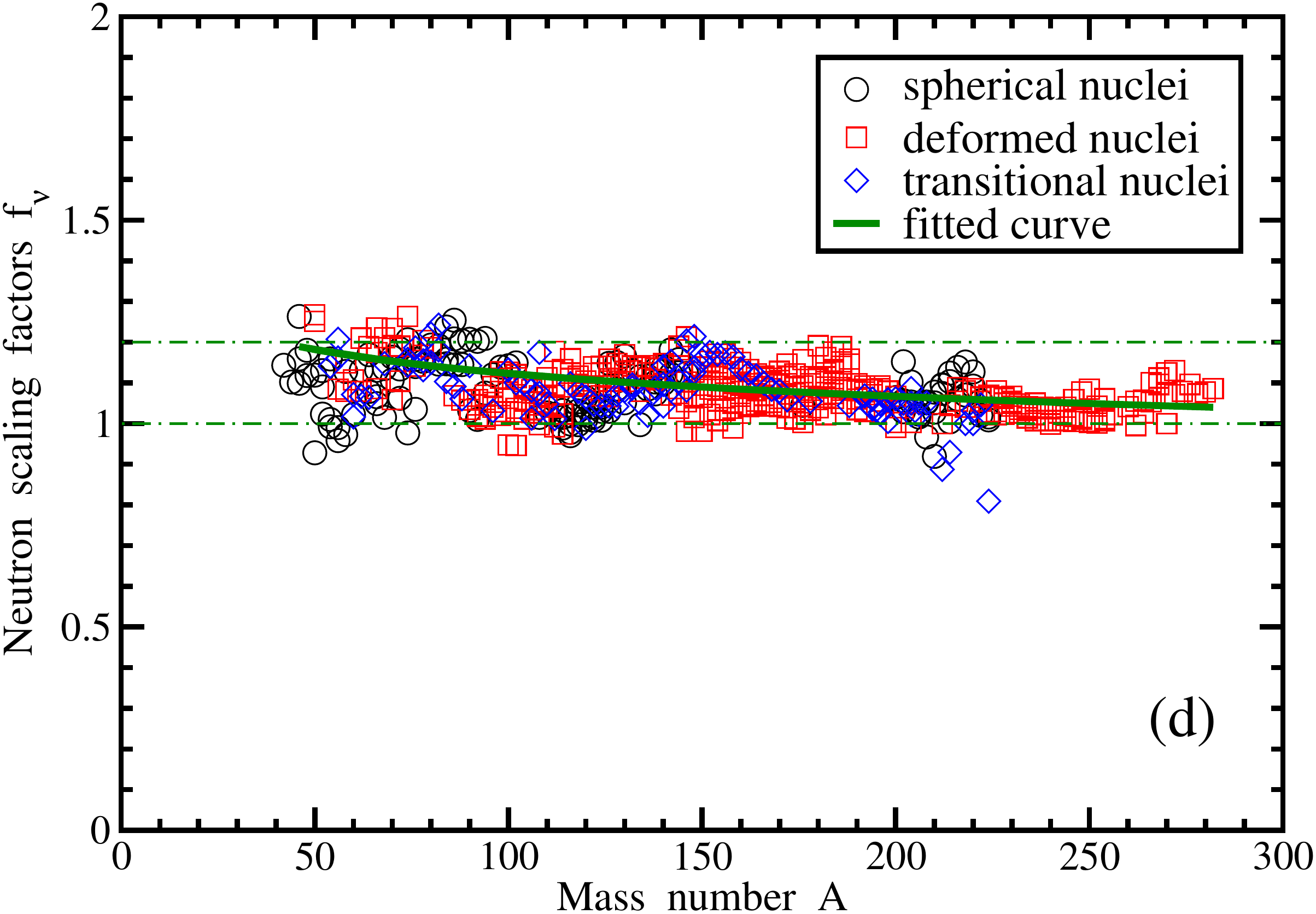}
\caption{Proton ($f_{\pi}(Z,N)$) and neutron ($f_{\nu}(Z,N)$) scaling factors as a 
function of mass numbers $A$. Open circles, squares and diamonds are used for spherical, transitional 
and deformed nuclei, respectively. Top panels include all available scaling factors, while bottom 
panels show  only those for the  nuclei with $Z> 20, N> 20$. Solid green lines 
correspond to global functional dependencies given by Eqs.\ (\protect\ref{f-mass})  with the 
parameters from the Table \ref{fitting-parameter-NL-fit} for top panels and from the Table 
\ref{fitting-parameter-no-light} for bottom panels.
\label{scaling-fact-vs-mass}
}
\end{figure*}

 The functional dependencies given by Eqs. (\ref{f-mass}) - (\ref{f-exp-N-Z-vs-N+Z}) 
are determined by means of non-linear least-squares fitting to the set of proton and neutron scaling
factors presented in Figs.\ \ref{factors-neutron} and \ref{factors-proton}. Note that we 
eliminate from this data set the scaling factors for $N=Z$ since they show enhanced
values as compared with neighboring nuclei [see Figs.\ \ref{factors-neutron}(a) and 
\ref{factors-proton}(a)].  This enhancement is due to Wigner energy which affects both 
binding  energies of the $N=Z$ nuclei and related pairing indicators. These  features
also a reason why in functional dependencies including isospin we use $|N-Z|$ instead
of $(N-Z)$. However, only few scaling factors are available for light nuclei on the 
proton-rich side of the $N=Z$ line (see Figs.\ \ref{factors-neutron}(a) and \ref{factors-proton}(a)) 
for which $(N-Z)<0$.  Thus, in general, the impact of modulus in $|N-Z|$ on the quality 
of the fit is expected to be small.

   The fitting is performed  by Curve Fitting Toolbox$^{\rm TM}$ software of Matlab.
The goodness-of-fit is defined  by root mean squared error (RMSE). The main disadvantage 
of least squares fitting is its sensitivity to outliers. Outliers can have a large influence on the  
fit because squaring the residuals magnifies the effects of these extreme data points.  Thus,
in addition to standard non-linear least squares fitting we also employ robust non-linear
least squares fitting with bisquare width regression scheme.  This scheme minimizes 
a weighted sum of squares, where the weight given to each data point depends on how far 
the point is from the fitted line. The points located near the line get full weight and the points 
farther from the line obtain reduced weight. The points that are farther from the line than would 
be expected by random chance get zero weight.

   The results of these fits are presented in Tables  \ref{fitting-parameter-NL-fit}, 
\ref{fitting-parameter-bisquare} and \ref{fitting-parameter-no-light}. One can see that the 
parameters and the goodness-of-fits depend very little on whether "measured+estimated" 
or only "measured" scaling factors  are used (see Table  \ref{fitting-parameter-NL-fit}).
The RMSEs are almost the same in proton subsystem for the fits to both sets of data
but in neutron subsystem they slightly increase when going on from the fit to only 
"measured" scaling factors to the fit which includes "measured+estimated" ones.
Thus, in the following discussion we will consider only the fits to "measured+estimated" 
scaling factors in order to have an access to the larger set of data with high isospin.

\begin{figure*}[ht]
\centering
\includegraphics[angle=0,width=8.5cm]{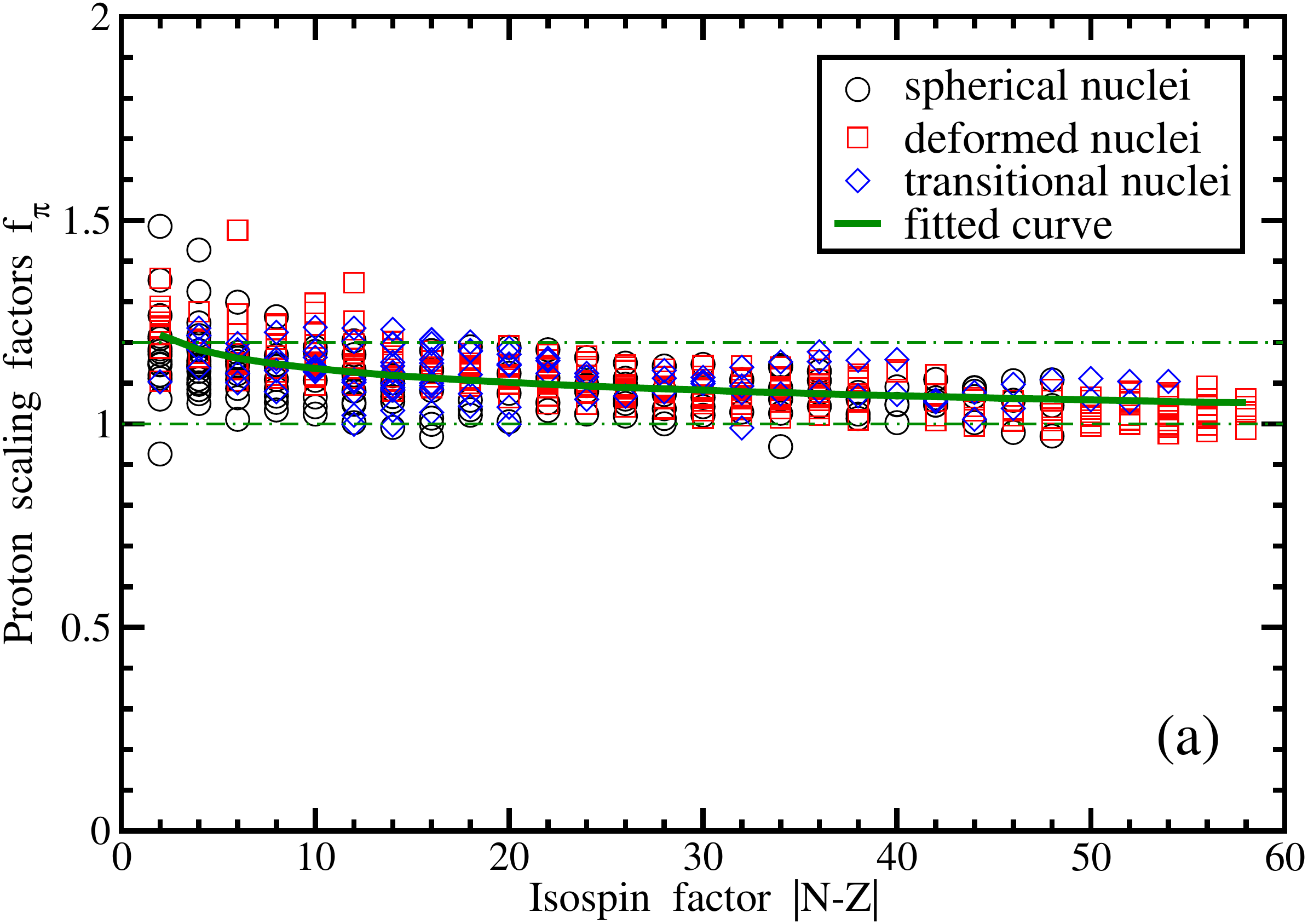}
\includegraphics[angle=0,width=8.5cm]{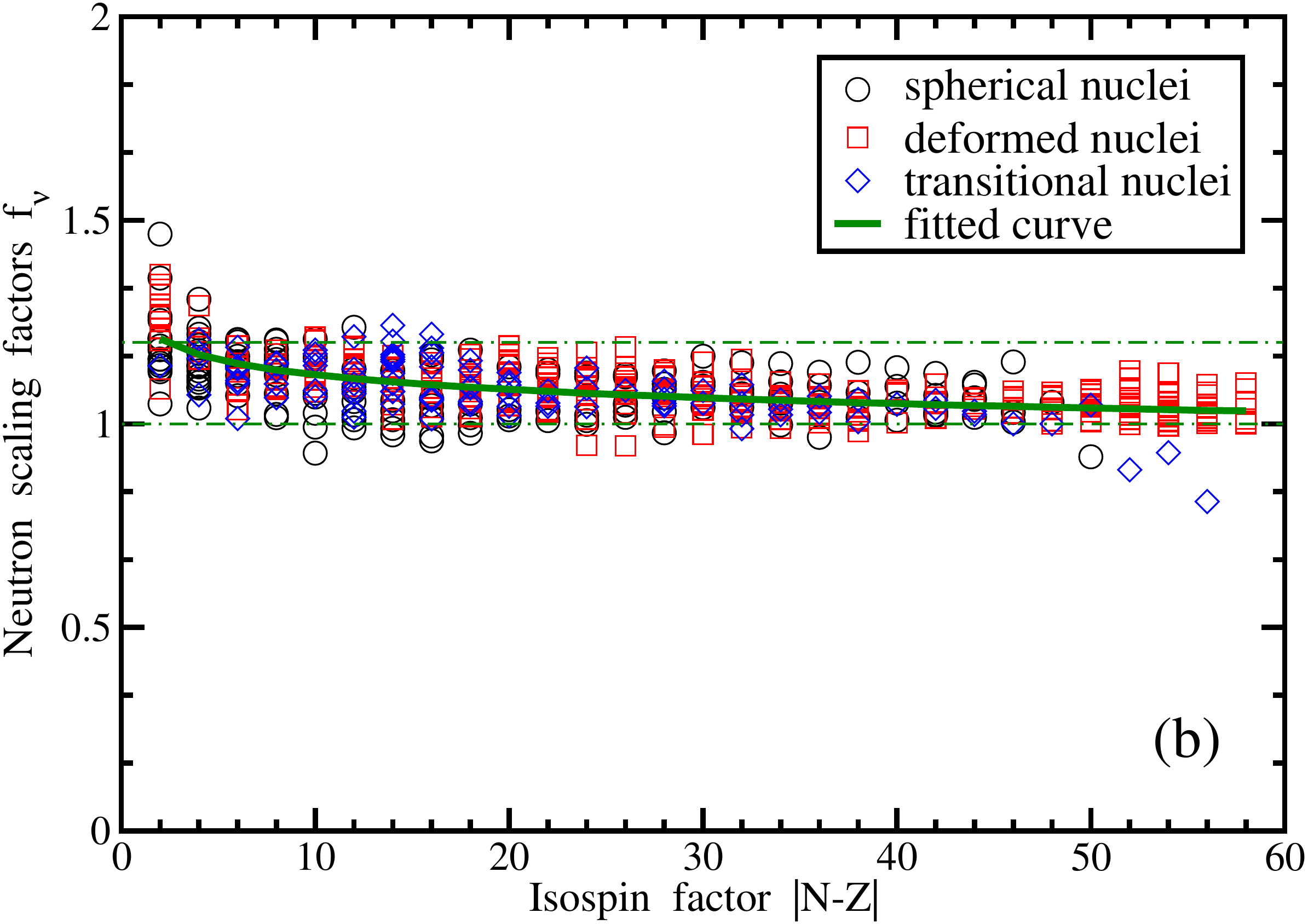}
\includegraphics[angle=0,width=8.5cm]{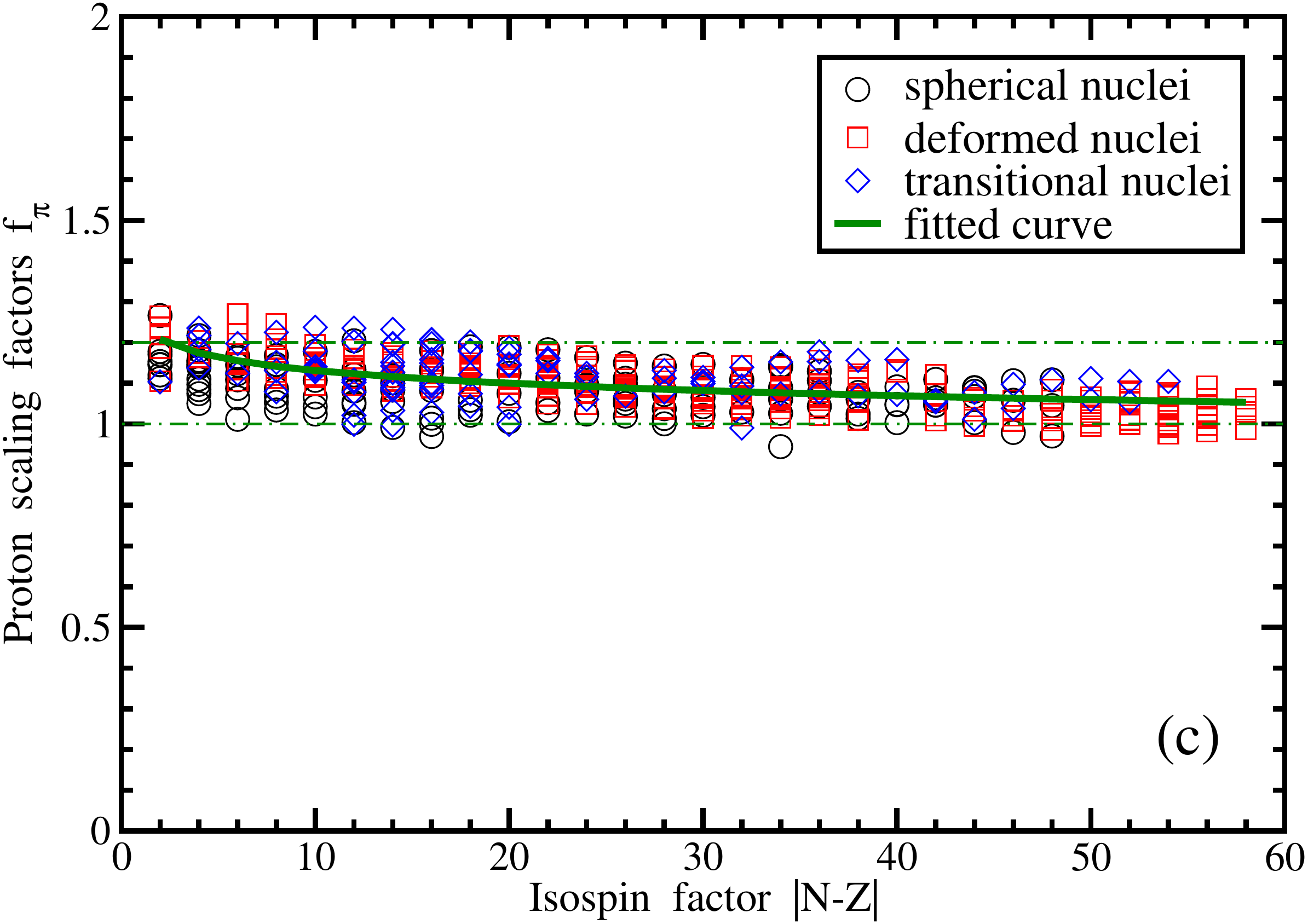}
\includegraphics[angle=0,width=8.5cm]{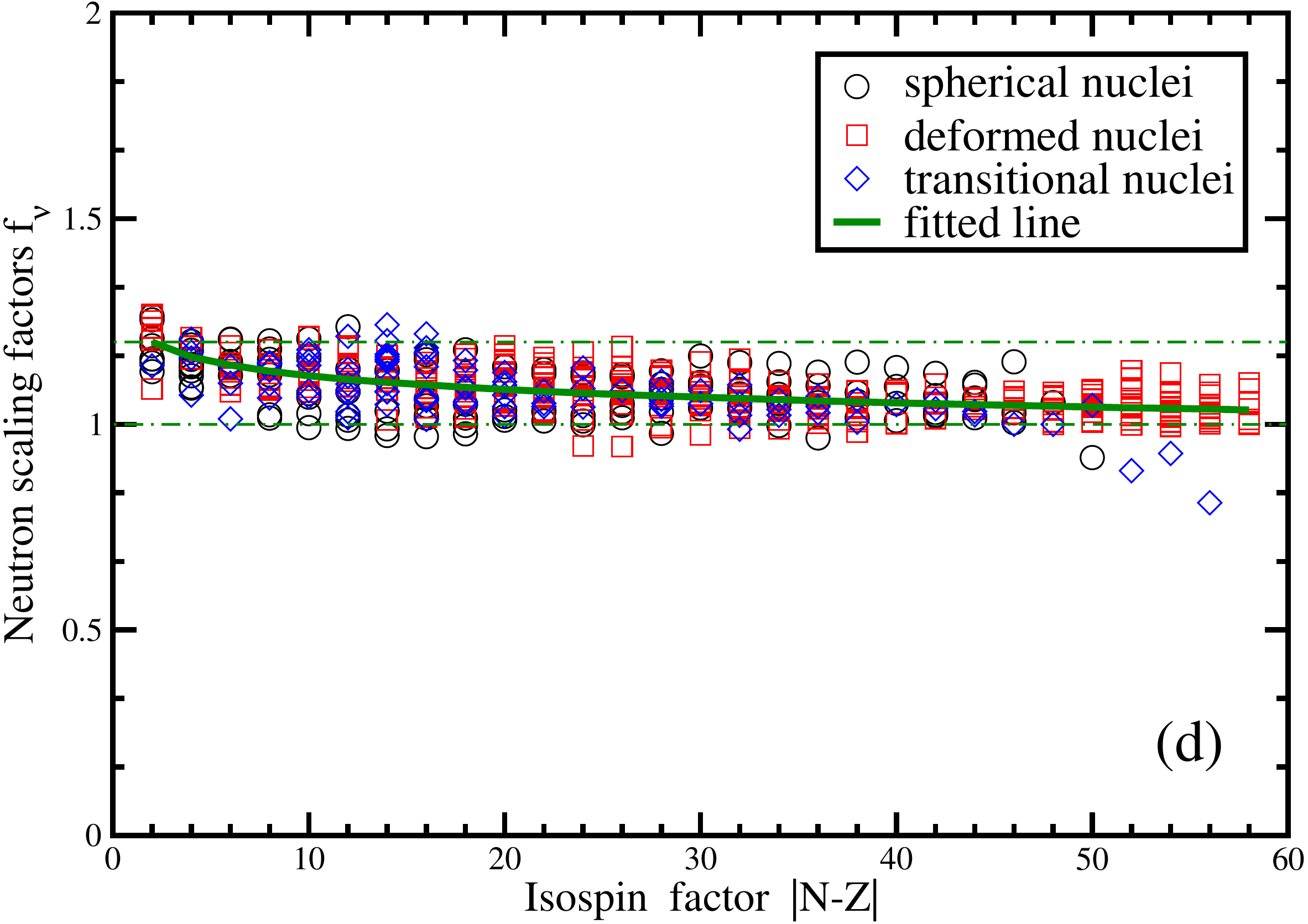}
\caption{The same as Fig.\ \ref{scaling-fact-vs-mass} but as a function of isospin factor $|N-Z|$. Solid green 
lines correspond to global functional dependencies given by Eqs.\ (\protect\ref{f-N-Z})  with the 
parameters from the Table \ref{fitting-parameter-NL-fit} for top panels and from the Table 
\ref{fitting-parameter-no-light} for bottom panels.
}
\label{Scaling-N-Z}
\end{figure*}

\begin{figure*}[ht]
\centering
\includegraphics[angle=0,width=8.5cm]{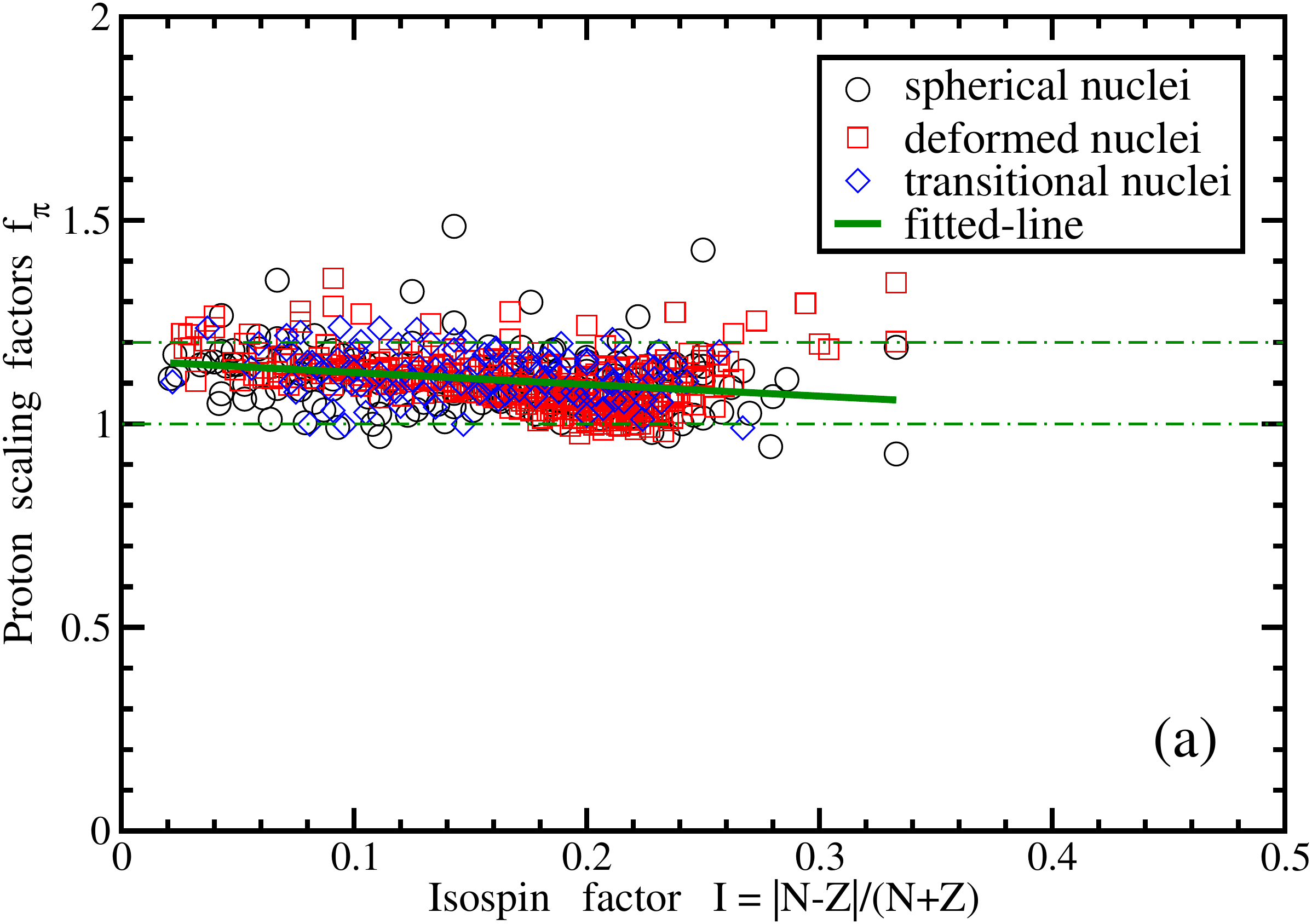}
\includegraphics[angle=0,width=8.5cm]{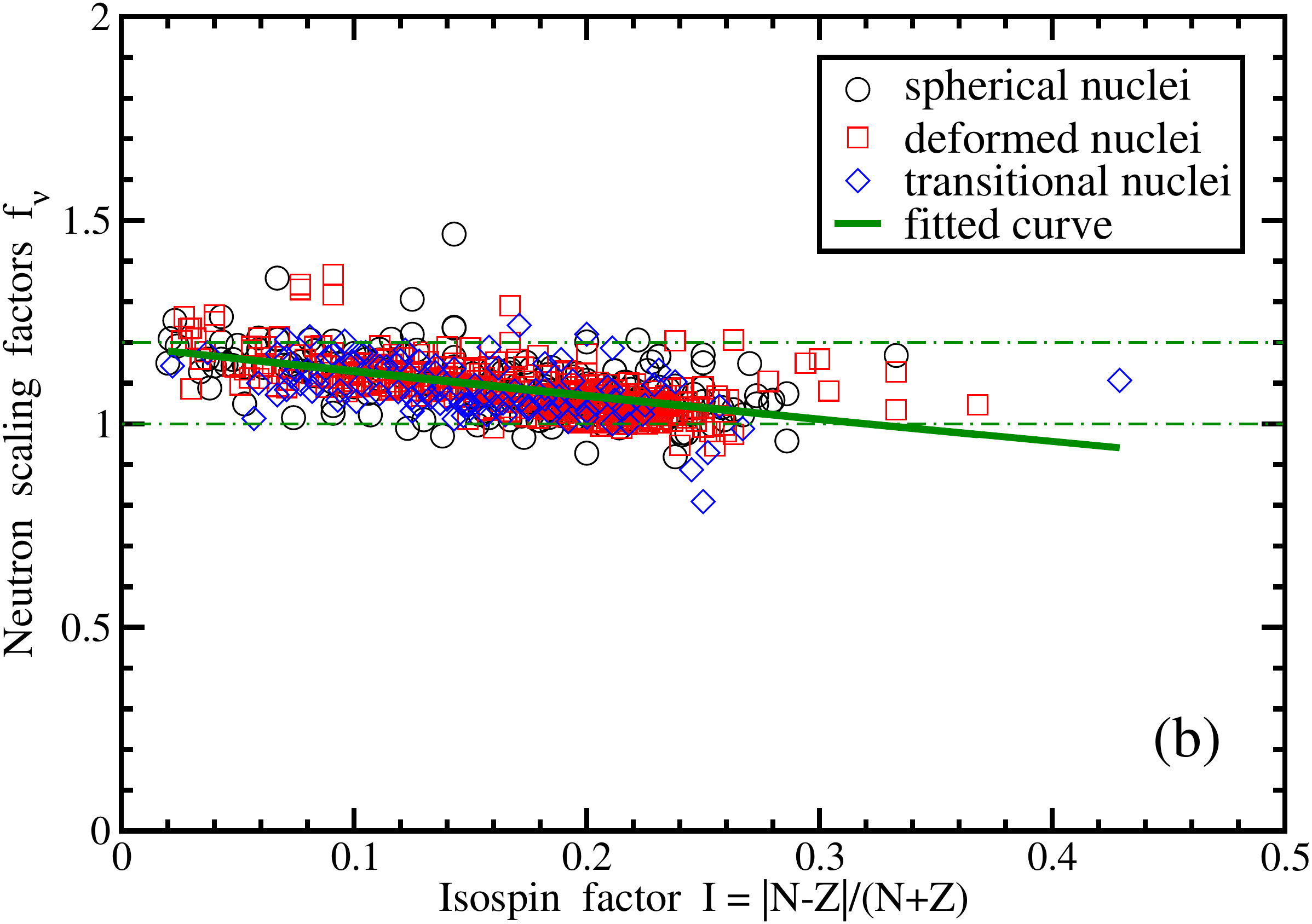}
\includegraphics[angle=0,width=8.5cm]{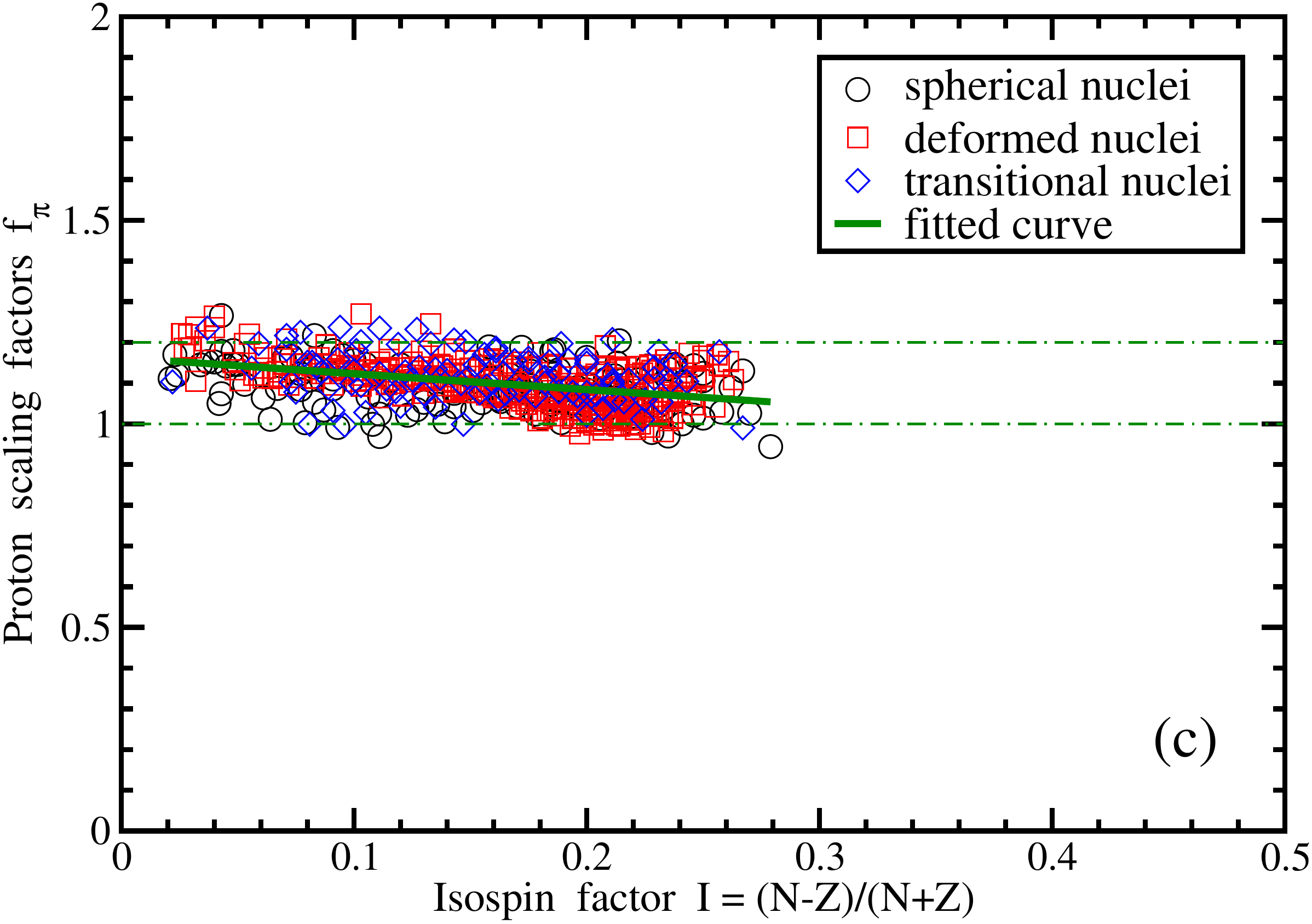}
\includegraphics[angle=0,width=8.5cm]{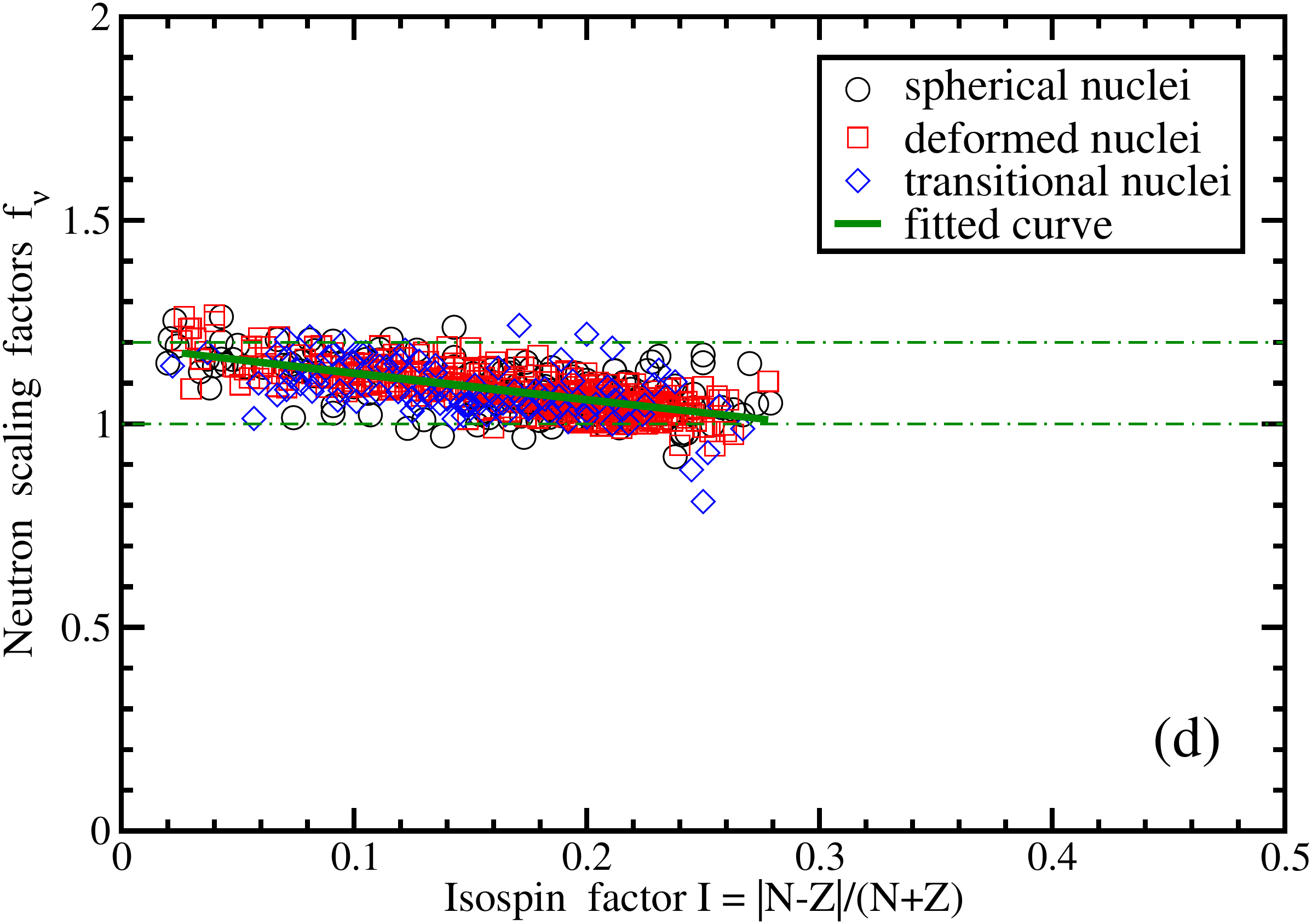}
\caption{The same as Fig.\ \ref{scaling-fact-vs-mass} but as a function of isospin factor 
$I=\frac{|N-Z|}{N+Z}$. Solid green lines correspond to global functional dependencies given 
by Eqs.\ (\protect\ref{f-exp-N-Z-vs-N+Z})  with the parameters from the Table \ref{fitting-parameter-NL-fit} 
for top panels and from the Table \ref{fitting-parameter-no-light} for bottom panels.
}
\label{Scaling-isospin}
\end{figure*}

   For proton subsystem, the best fit is obtained for functional dependence of Eq.\ (\ref{f-mass})
(see Table \ref{fitting-parameter-NL-fit}). Fig.\ \ref{scaling-fact-vs-mass}(a) compares the distribution of scaling 
factors as a function  of mass number $A$ with fitted curve. They both show a general trend of the decrease of 
scaling factors with increasing mass number. In addition, the oscillations of the scaling factors, averaged for a 
given range of $A$, with respect of fitted curve as a function of mass number  are clearly visible, especially  for 
spherical and transitional nuclei. The largest spread of scaling factors  from fitted curve are seen for the light nuclei. 
This is not surprising considering the fact that these nuclei are soft in deformation degrees of freedom so that 
the correlations beyond mean field are expected to play enhanced role in their structure. Note also that because 
of this reason the quality of mass description in light nuclei in the CDFT is also lower than in medium and heavy 
mass  regions (see Ref.\ \cite{AARR.14}).    Comparable to Eq.\ (\ref{f-mass}) goodness-of-fit is obtained also 
with functional dependencies given by Eqs.\ (\ref{f-part-num}), (\ref{f-N-Z}) and (\ref{f-exp-N-Z}) (see Table 
\ref{fitting-parameter-NL-fit}). The quality of the fit and the spreads of scaling factors with respect of the fitted curve
are illustrated on the example of the functional dependence of the $f_{\pi}(Z,N) = C_{\pi}e^{\alpha_{\pi}|N-Z|}$ 
type in Fig.\ \ref{Scaling-N-Z}.  There is a general trend of the decrease of the fitted $f_{\pi}$ curve and individual 
scaling factors with the increase of isospin factor $|N-Z|$.  The largest spread of the scaling factors with respect 
of the fitted curve is observed at low  isospin. The worst fit of proton scaling factors is provided by the functional
dependence of Eq.\ (\ref{f-exp-N-Z-vs-N+Z}) (see Table \ref{fitting-parameter-NL-fit}) and Fig.\ \ref{Scaling-isospin}(a)).
  
   It is interesting to see how these conclusions will be modified if the least reliable data points are excluded 
from consideration.  These are typically the points (outliers) which are located far away  from the fitted 
curve and these substantial deviations can be caused, for example, by the limitations of the mean field 
approximation and the neglect of beyond mean field effects. These effects are expected to be
especially important in light nuclei or for the pairing indicators the determination of  which involves two types of 
the nuclei (for example, transitional and deformed ones).  Table \ref{fitting-parameter-bisquare} shows how 
the RMSEs are modified when robust non-linear least square fitting based on bisquare width regression
scheme is used instead of standard one.  One can see that in all cases  it improves RMSEs. The fitting protocol 
based on the $f_{\pi}(Z,N) = C_{\pi}e^{\alpha_{\pi}|N-Z|}$ functional  dependence becomes the best which is 
closely followed by the functional dependences of Eqs.\ (\ref{f-mass}), (\ref{f-part-num}) and (\ref{f-exp-N-Z}).   
The exclusion of light nuclei leads to further improvement in the RMSEs (see Table \ref{fitting-parameter-no-light}) 
but does not modify the conclusions obtained based on the results of Table \ref{fitting-parameter-bisquare}. 
 Figures  
\ref{scaling-fact-vs-mass}(c),  \ref{Scaling-N-Z}(c) and \ref{Scaling-isospin}(c) compare fitted curves based on the 
parameters of Table \ref{fitting-parameter-no-light}  with the distribution of respective individual scaling factors. 
One can see that in all cases the number of outliers is decreasing drastically (as compared with Figs.\ 
\ref{scaling-fact-vs-mass}(a),  \ref{Scaling-N-Z}(a) and \ref{Scaling-isospin}(a)) and the distribution of individual 
scaling factors becomes more condensed around fitted curve.

   Similar features have also  been observed for the neutron subsystem.
The majority of outliers  are produced again by the light nuclei  (compare Figs.\  
\ref{scaling-fact-vs-mass}(d),  \ref{Scaling-N-Z}(d) and \ref{Scaling-isospin}(d) with Figs.\ 
\ref{scaling-fact-vs-mass}(b),  \ref{Scaling-N-Z}(b) and \ref{Scaling-isospin}(b)) and their removal 
leads to the improvement of goodness-of-fit (compare Tables \ref{fitting-parameter-bisquare} 
and \ref{fitting-parameter-no-light}).  Robust non-linear least square fitting based on bisquare 
width regression scheme improves RMSEs in all fits as compared with standard one (compare 
Tables \ref{fitting-parameter-NL-fit} and \ref{fitting-parameter-bisquare}). However, the principal
difference as compared with proton subsystem is the fact that isospin dependence (in one or
another form) of neutron scaling factors is substantially favored as compared with the mass
or neutron number dependences given by Eqs.\ (\ref{f-mass}) and (\ref{f-part-num}) in all 
fits.  The best RMSEs are provided either by the $f_{\nu}(Z,N) = C_{\nu}|N-Z|^{\alpha_{\nu}}$ 
(Table \ref{fitting-parameter-NL-fit}) or by the $f_{\nu}(Z,N) = C_{\nu}e^{\alpha_{\nu} \frac{|N-Z|}{N+Z}}$
(Tables  \ref{fitting-parameter-bisquare} and \ref{fitting-parameter-no-light}) functional dependencies.
However, in the light of their closeness in terms of RMSEs and the approximations used in this
study it is impossible to give a clear preference to one or another.

   The present study clearly favors the functional dependencies of scaling factors (and, thus 
of pairing interaction) which depend on isospin in neutron subsystem. The situation is more mixed in the 
case of proton subsystem since functional dependencies of scaling factors on mass/particle numbers 
produce RMSEs which are only slightly above those produced by functional dependencies which
include isospin.    The origin of this feature  is not completely clear.  However, 
the possible origin  could be related to the fact that the RPA neutron-proton pair-vibrational 
correlation  energy is expected to decrease numerically with increasing neutron excess 
due to an increasing mismatch of the occupations of  single-neutron and 
single-proton levels (see Ref.\ \cite{NB.19}). This could lead to more 
pronounced  isospin dependence of neutron pairing as compared with
proton one.

 To our knowledge, the isospin dependence of the Gogny pairing interaction in the 
Gogny DFT has not been studied so far.  The isospin dependence of pairing interaction has been 
revealed by several global studies in the framework of Skyrme DFT (see Refs.\ \cite{BLS.12,YMSH.12}). 
However, it is necessary to point out that employed isoscalar-density dependence used in these works
is phenomenological and it is not motivated by any arguments based on the microscopic theory
of effective interaction \cite{YMSH.12}.

\begin{table*}[htb]
\begin{center}
\caption{ 
The parameters of global functional dependencies defined by the fits to the set of 
"measured+estimated" scaling factors. Note that in this fit only scaling factors defined from 
deformed nuclei are used. Both the standard (labeled as "standard fit") and robust (labeled 
as "robust fit") non-linear least squares fits are used here.  The results of the fits are presented
in the "A/B" format. Here "A" corresponds to the results of the fit to the set of data including all 
scaling factors of deformed nuclei, while "B" to the ones which exclude all light nuclei with
$Z<20$ and $N<20$.  The bold style is used for RMSE of the best fit in each group of functional 
dependencies.
}
\label{fitting-parameter-deformed}
\begin{tabular}{|c|c|c|c|c|c|c|} \hline \multicolumn{1}{|c|}{  } & \multicolumn{3}{c|}{standard fit} & \multicolumn{3}{c|}{ robust fit} \\ \hline
                                                                                &     C$_i$           &   $\alpha$$_i$          &  RMSE                            &  C$_i$              &  $\alpha$$_i$        &   RMSE                    \\ \hline
                                 1                                             &     2                  &       3                         &   4                                    &     5                  &    6                         &      7                         \\ \hline
  $f_{ \pi}=C_{\pi}A^{\alpha_{\pi}} $                          &   1.801/1.840    &  -0.0992/-0.1030     & {\bf 0.0364}/{\bf 0.0349}  &  1.792/1.877    &  -0.0980/-0.1072    & {\bf 0.0373}/0.0361         \\ \hline
    $f_{\pi}=C_{\pi}Z^{\alpha_{\pi}} $                         &   1.646/1.671   &  -0.0983/-0.1019      &   0.0392/0.0386               &  1.642/1.684    &  -0.0978/-0.1039    &  0.0409/0.0400        \\ \hline
        $f_{ \pi}=C_{\pi}|N-Z|^{\alpha_{\pi}} $              &   1.323/1.289   &  -0.0589/-0.0518      &   0.0434/0.0355               &  1.315/1.296    &  -0.0576/-0.0535    & 0.0387/0.0356          \\ \hline
        $f_{ \pi}=C_{\pi}e^{\alpha_{\pi} |N-Z|} $           &   1.201/1.180     &-0.0031/-0.0026      &  0.0434/0.0335                &  1.187/1.178    &  -0.00279/-0.0026  & 0.0379/{\bf 0.0348}   \\ \hline   
$f_{ \pi}=C_{\pi}e^{\alpha_{\pi} [|N-Z|/(N+Z)]} $       &   1.163/1.188   & -0.3050/-0.4870           &   0.0651/0.0459               &  1.190/1.189    &  -0.4970/-0.5020    & 0.0511/0.0473           \\ \hline                                 
                                                                                &                         &                                  &                                         &                         &                               &                                   \\ \hline   
 $f_{\nu}=C_{\nu}A^{\alpha_{\nu}} $                        &   1.465/1.414   &  -0.0603/-0.0533      &   0.0531/0.0501               &  1.482/1.414    &  -0.0626/-0.0534    &  0.0531/0.0513          \\ \hline  
 $f_{\nu}=C_{\nu}N^{\alpha_{\nu}} $                        &   1.444/1.410     &  -0.0644/-0.0592    &   0.0506/0.0481              &   1.474/1.421    &  -0.0688/-0.0608    &  0.0503/0.0490          \\ \hline
 $f_{ \nu}=C_{\nu}|N-Z|^{\alpha_{\nu}} $                  &   1.283/1.256   &  -0.0538/-0.0474      &  {\bf 0.0420}/0.0393   &  1.282/1.264    &  -0.0535/-0.0495    & 0.0406/0.0388           \\ \hline
 $f_{ \nu}=C_{\nu}e^{\alpha_{\nu} |N-Z|} $               &   1.084/1.147   & -0.0376/-0.0021       &   0.0481/0.0425              &  1.082/1.145     & -0.0330/-0.0020    & 0.0433/0.0417            \\ \hline
$f_{ \nu}=C_{\nu}e^{\alpha_{\nu}  [|N-Z|/(N+Z)]} $   &   1.084/1.209   & -0.0378/-0.673         &  0.0474/{\bf 0.0347}               &  1.077/1.208     & -0.0395/-0.6740    & {\bf 0.0370}/{\bf 0.0352}     \\ \hline
\end{tabular}
\end{center}
\end{table*}

    In the light of substantial differences in the scaling factors of spherical and deformed 
nuclei within the given region of nuclear chart discussed above it is interesting to 
repeat the analysis but with the restriction to only scaling factors defined in deformed nuclei.
The results of such analysis are summarized in Table \ref{fitting-parameter-deformed}. 
One can see that the exclusion of spherical and transitional nuclei from  the fitting leads to a substantial 
improvement in RMSEs (compare Table \ref{fitting-parameter-deformed} with Table
\ref{fitting-parameter-NL-fit}).
This is especially the case for proton functional dependencies which on average 
improve by approximately 25\% with the best ($\approx 30$\%) and worst ($\approx 10$\%)  improvements 
provided by the 
$f_{\pi}(Z,N) = C_{\pi}A^{\alpha}$ and $f_{\pi}(Z,N) = C_{\pi}e^{\alpha_{\pi} \frac{|N-Z|}{N+Z}}$
functional dependencies. The improvements in RMSEs are smaller in the neutron subsystem with an average
improvement over the set of considered functional dependencies being on the level of $\approx 14$\%.
The best improvements (around 20\%) is achieved for the 
$f_{\nu}(Z,N) = C_{\nu}|N-Z|^{\alpha_{\nu}}$  and $f_{\nu}(Z,N) = C_{\nu}e^{\alpha_{\nu} \frac{|N-Z|}{N+Z}}$
functional dependencies.
The exclusion of light  nuclei with $Z<20$ and $N<20$, which are expected to be 
significantly affected by beyond mean fields effects, leads to further improvement in the RMSEs (see Table 
\ref{fitting-parameter-deformed}) and this effect is especially pronounced in the standard non-linear least
squares fit. One should also note that the restriction to deformed nuclei reduces the impact of outliers. 
This is seen in the fact that the transition from standard to robust fit not always reduces RMSEs; this
feature is especially pronounced for the set which excludes light nuclei.

   Despite all these changes, the general conclusions obtained for full set of data on scaling factors
including different types of nuclei are not affected by the transition to the data which is based only
on deformed nuclei.  In the neutron subsystem,  the $f_{\nu}(Z,N) = C_{\nu}|N-Z|^{\alpha_{\nu}}$ 
and $f_{\nu}(Z,N) = C_{\nu}e^{\alpha_{\nu} \frac{|N-Z|}{N+Z}}$ functional dependencies are still
favored over other types of functional dependencies and there is still no way to give a clear 
preference to one or another. In the proton subsystem, the best description of the data is still 
provided by two competing functional dependencies of the 
$f_{\pi}(Z,N) = C_{\pi}A^{\alpha}$ and $f_{\pi}(Z,N) = C_{\pi}e^{\alpha_{\pi} \frac{|N-Z|}{N+Z}}$ types
which have different dependence on the isospin.

\section{Extrapolation to very neutron-rich nuclei.}
\label{extremes}
  
   It is important to evaluate the evolution of expected uncertainties 
in the predictions of pairing properties with increasing of neutron number $N$ 
and approaching neutron drip line. For this purpose we selected the Yb isotopes 
chain in which such evolution has been studied for different classes of CEDFs 
for constant $f_{\pi}=f_{\nu}=1.075$ in Ref.\ \cite{AARR.15}.

\begin{figure*}[ht]
\centering
\includegraphics[angle=0,width=12.5cm]{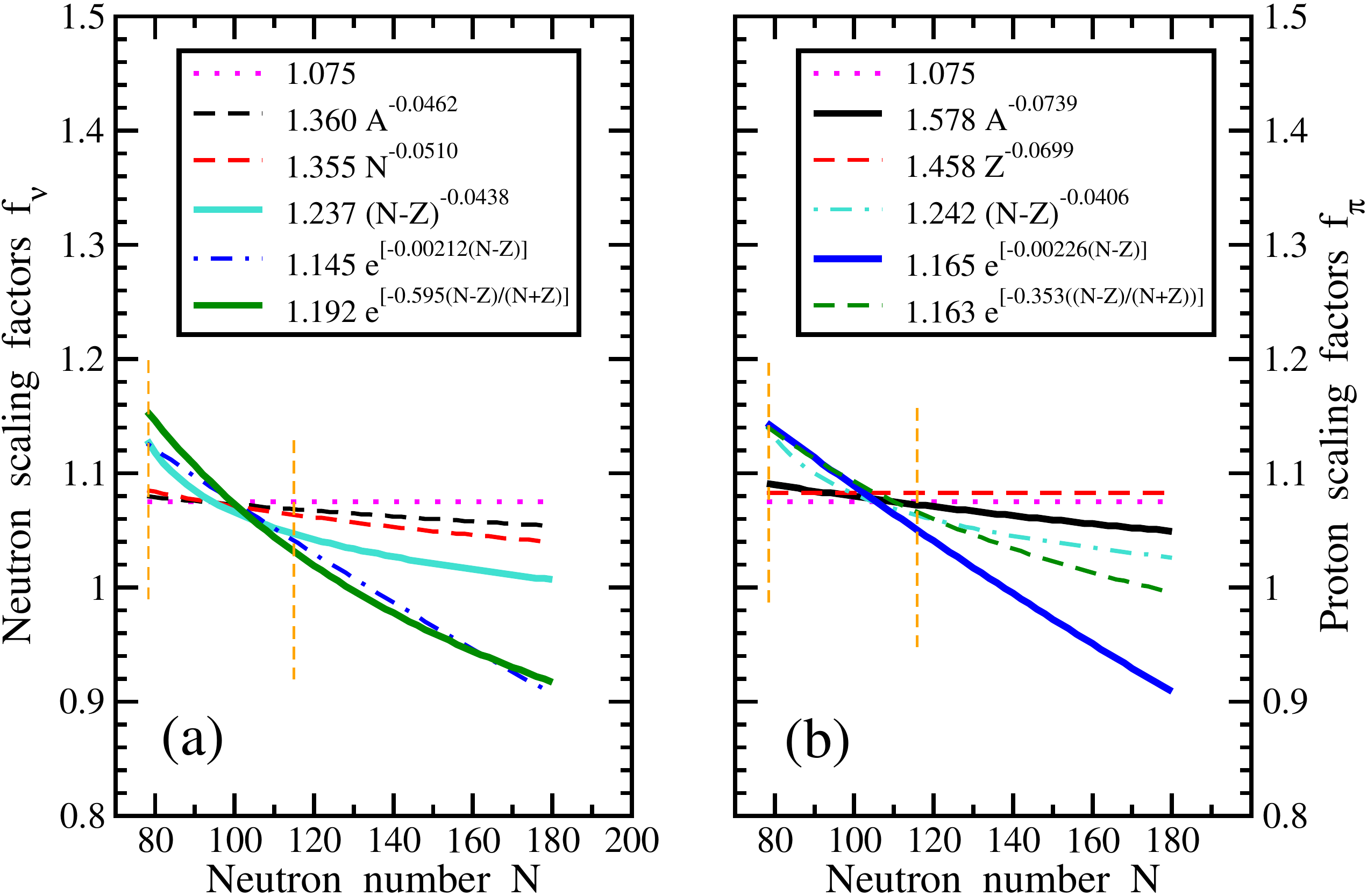}
\caption{The variations of neutron and proton scaling factors as a function 
of neutron number $N$ in the Yb ($Z=70$) isotopes. Functional dependencies 
of Eqs. (\ref{f-mass})-(\ref{f-exp-N-Z-vs-N+Z}) are given here in insert with the parameters 
defined in Table \ref{fitting-parameter-no-light}. The pairs of  functional dependencies with 
the best and worst RMSEs are  shown by thick solid and thin dashed lines. Dot-dashed thin 
line is used for the functional dependence located (in terms of RMSE) in between of these 
pairs. Vertical orange dashed lines are used to indicate the span of  neutron  numbers for 
which experimental information on binding energies  are available in the Yb isotopes.
\label{Yb-functional-dep}
}
\end{figure*}

   Fig.\ \ref{Yb-functional-dep} compares the variations of neutron and proton scaling 
factors as a function of neutron number $N$ for different functional dependencies
studied in the present paper. For neutron subsystem, the best RMSEs are provided
by  $f_{\nu}(Z,N) = C_{\nu}e^{\alpha_{\nu} \frac{|N-Z|}{N+Z}}$ and 
$f_{\nu}(Z,N) = C_{\nu}|N-Z|^{\alpha_{\nu}}$  functional dependencies (see Tables 
\ref{fitting-parameter-NL-fit}, \ref{fitting-parameter-bisquare} and  \ref{fitting-parameter-no-light}).
However, the preference of one over another depends on whether the
standard or robust non-linear least square fitting is used and whether light nuclei are
excluded from consideration. While neutron scaling factors obtained with these two 
dependencies are similar [within the accuracy of the present method] for experimentally 
known nuclei, they differ substantially for very neutron-rich nuclei. 
The scaling factors defined by $f_{\nu}(Z,N) = C_{\nu}e^{\alpha_{\nu} \frac{|N-Z|}{N+Z}}$
functional dependence show significant reduction on approaching neutron drip line, while 
those based on $f_{\nu}(Z,N) = C_{\nu}|N-Z|^{\alpha_{\nu}}$ reveal smaller decrease.
The total reduction of the scaling factors on going on from $N=78$ to $N=180$
is 20.2\% and 10.7\% for these two types of functional dependencies. The functional 
dependencies of the $f_{\nu}(Z,N) = C_{\nu}e^{\alpha_{\nu}|N-Z|}$,  
$f_{\nu}(Z,N)=C_{\nu} A^{\alpha_{\nu}}$ and $f_{\nu}(Z,N)=C_{\nu} N^{\alpha_{\nu}}$ 
types provide less accurate fits (see Tables \ref{fitting-parameter-NL-fit}, 
\ref{fitting-parameter-bisquare} and  \ref{fitting-parameter-no-light}).
In the Yb isotopes, the former one provides scaling factors which are very similar
to those obtained by $f_{\nu}(Z,N) = C_{\nu}e^{\alpha_{\nu} \frac{|N-Z|}{N+Z}}$,
while the last two deliver neutron scaling factors which only weakly decrease
with increasing neutron number and which are close to constant $f_{\nu}=1.075$
used in Ref.\ \cite{AARR.15}.
 
   The situation is more complex in the proton subsystem because contrary
to the neutron one the functional dependencies which include isospin
do not get a clear preference over those which depend
on mass or proton numbers (see Tables \ref{fitting-parameter-NL-fit}, 
\ref{fitting-parameter-bisquare} and  \ref{fitting-parameter-no-light}). As a 
consequence, two best (in terms of RMSEs) functional dependencies,
namely, $f_{\pi}(Z,N) = C_{\pi}e^{\alpha_{\pi}|N-Z|}$ and 
$f_{\pi}(Z,N) = C_{\pi}A^{\alpha_{\pi}}$ provide drastically different 
predictions for evolution of proton scaling factors as a function of
neutron number (see Fig.\ \ref{Yb-functional-dep}(b)). The 
former one shows drastic decrease of $f_{\pi}$ with increasing neutron number,
while the latter one is characterized by only modest decrease with an average
value close to constant $f_{\pi}=1.075$ used in Ref.\ \cite{AARR.15}.
The least accurate functional dependencies given by Eqs.\ (\ref{f-N-Z}) 
and (\ref{f-exp-N-Z-vs-N+Z}) provide predictions located between these two 
cases (see Fig.\ \ref{Yb-functional-dep}(b)). The  $f_{\pi}(Z,N) = C_{\pi}Z^{\alpha_{\pi}}$ 
functional dependence provides $N$-independent $f_{\pi}$ value which is close 
to the $f_{\pi}=1.075$ one used in Ref.\ \cite{AARR.15}.

    Table \ref{Yb170-sc-factor} shows proton and neutron scaling factors and 
pairing gaps in very neutron-rich $^{240}$Yb nucleus, located in the 
vicinity of two-neutron drip line, calculated with different functional dependencies 
for  scaling factors. In neutron subsystem,  two best functional dependencies given 
by  $f_{\nu}(Z,N) = C_{\nu}|N-Z|^{\alpha_{\nu}}$ and 
$f_{\nu}(Z,N) = C_{\nu}e^{\alpha_{\nu} \frac{|N-Z|}{N+Z}}$ provide the predictions 
for neutron scaling factors which differ by $\approx 8\%$. This results into neutron 
pairing gaps which differ by more than 30\%.   Note, however, that these two functional 
dependencies predict neutron scaling factors and pairing gaps which are substantially 
lower than those obtained by functional dependencies with scaling factors  
dependent either on mass or particle numbers.
 As discussed above, the situation is more 
complex in the proton subsystem where our analysis cannot reveal a clear preference 
for functional dependence of proton scaling factors. This is also seen in Table 
\ref{Yb170-sc-factor} which shows that two best dependencies given by
$f_{\pi}(Z,N) = C_{\pi}e^{\alpha_{\pi}|N-Z|}$ and $f_{\pi}(Z,N) = C_{\pi}A^{\alpha_{\pi}}$
provide quite drastic differences in the predicted values of  proton scaling factors and pairing
gaps in $^{170}$Yb.

   These uncertainties in proton and neutron pairing
at high isospin could have some impact on the predicted location of two-neutron drip line 
(see Ref.\ \cite{AARR.15}) and substantial impact on the heights of fission barriers of neutron-rich nuclei.
The latter sensitively depend on the strength of pairing interaction (see Ref.\ \cite{KALR.10})
and the reduction of only neutron pairing at high isospin (similar to the one seen in 
$^{170}$Yb) could increase the static fission  barriers by approximately 1 MeV. This
could have a significant effect on the r-process in actinides and superheavy nuclei (see 
discussion in Ref.\ \cite{TAA.20}).

\begin{table}[htb]
\begin{center}
\caption{Proton and neutron scaling factors and pairing gaps ($\Delta_{uv}^{\pi}$
and $\Delta_{uv}^{\nu}$) in the $^{240}$Yb nucleus calculated with indicated functional 
dependencies.  The $f_i=const(Z)=1.075$ are scaling factors used for the Yb isotopes 
in Ref.\ \cite{AARR.15}. The results obtained with two best/two worst (in terms of RMSEs of 
Table \ref{fitting-parameter-no-light}) functional dependencies for scaling factors are show 
by bold/italic text.
}
\label{Yb170-sc-factor}
\begin{tabular}{|c|c|c|c|c|} \hline 
                                                                                      &  $f_{\pi} $ & $\Delta_{uv}^{\pi}$ &  $f_{\nu}$ &  $\Delta_{uv}^{\nu}$    \\ \hline
                       1                                                              & 2 & 3 & 4 & 5 \\ \hline
   $f_{i}=const(Z)$                                                          &1.075     &1.519    & 1.075    & 1.373 \\ \hline                                                          
   $f_{i}=C_{i}A^{\alpha_{i}} $                                         & {\bf 1.052}   &   {\bf 1.400}   & {\it 1.056}    &  {\it 1.279}  \\ \hline
   $f_{\pi}=C_{\pi}Z^{\alpha_{\pi}}$ or  $f_{\nu}=C_{\nu}(N)^{\alpha_{\nu}} $         
                                                                                       & {\it 1.083}    &{\it 1.567}   & {\it 1.042}   & {\it 1.215} \\ \hline
 $f_{i}=C_{i}|N-Z|^{\alpha_{i}} $                                      &1.030     & 1.272   & {\bf 1.011}   & {\bf 1.042} \\ \hline
  $f_{i}=C_{i}e^{\alpha_{i} |N-Z|} $                                  & {\bf 0.929}     & {\bf 0.764 }  &0.926   & 0.704 \\ \hline
 $f_{i}=C_{i}e^{\alpha_{i} [|N-Z|/(N+Z)]} $                       &{\it 1.004}    & {\it 1.155}   & {\bf 0.930}  & {\bf 0.720}  \\ \hline
\end{tabular}
\end{center}
\end{table}

\section{Conclusions}
\label{Concl}

  A systematic global investigation of pairing properties based on all 
available experimental data on pairing indicators has been performed
for the first time in the framework of covariant density functional theory.
It is based on separable pairing interaction of Ref.\ \cite{TMR.09} and 
covariant energy density functional NL5(E) \cite{AAT.19}. The main results 
can be summarized as follows:

\begin{itemize}
\item
  The optimization of functional dependencies of proton and neutron 
scaling factors has been performed across experimentally known nuclear chart.  
It clearly reveals isospin dependence of neutron pairing with the
following forms of neutron scaling factors 
$f_{\nu}(Z,N) = C_{\nu}e^{\alpha_{\nu} \frac{|N-Z|}{N+Z}}$ 
and 
$f_{\nu}(Z,N) = C_{\nu}|N-Z|^{\alpha_{\nu}}$  
providing the best and comparable description of experimental data.
The situation is less clear in the proton subsystem since two best 
(and comparable in terms of the description of experimental data) 
functional dependencies, namely, $f_{\pi}(Z,N) = C_{\pi}e^{\alpha_{\pi}|N-Z|}$ 
and $f_{\pi}(Z,N) = C_{\pi}A^{\alpha_{\pi}}$ differ drastically with respect of the 
impact of isospin. The direct comparison of theoretical
and experimental pairing indicators extracted from binding energies
may help to resolve these uncertainties. However, there is no guarantee 
that this can be achieved at the mean field level because of potential 
impact of beyond mean field effects on the binding energies.

\item
 The differences in the functional dependencies of scaling factors lead to 
the uncertainties in the prediction of proton and neutron pairing properties 
which are expected to become especially pronounced at high isospin. 
They are expected to have some impact on the predicted 
location of two-neutron drip line (see Ref.\ \cite{AARR.15}). However,
even higher impact is expected on physical observables which extremely
sensitively depend on the pairing. These are the moments of inertia of 
rotational bands (see Refs.\ \cite{CRHB,AO.13})  and the heights of fission 
barriers (see Ref.\ \cite{KALR.10}).  For example, the reduction of only neutron 
pairing at high isospin (similar to the one seen in $^{170}$Yb) could increase 
the static fission  barriers by approximately 1 MeV and have a significant effect 
on fission cycling in the r-process in actinides and superheavy nuclei (see
discussion in Ref.\ \cite{TAA.20}).

\item
  Our analysis reveals that in a given part of nuclear chart the scaling factors for spherical 
nuclei are smaller than those for deformed ones. This result is similar to the ones obtained
earlier in non-relativistic Skyrme and Gogny DFTs (see Refs.\ \cite{BBNSS.09,RBB.12}. Thus, its
origin has to be traced back to a missing physics which is neglected in all three approaches. 
The particle-vibration coupling in odd-mass nuclei is the most likely source of such physics. It is well known
that it increases the binding energies of odd-$A$ nuclei (see Refs.\ \cite{DG.80,LA.11}) but does
not have an impact on binding energies of even-even nuclei. As a consequence, when extracted
from experimental binding energies [which by default include PVC effects], the pairing indicators
are reduced as compared with those which do not include PVC effects.  This effect is especially 
pronounced in spherical nuclei. Since scaling factors $f_{\nu}$ and $f_{\pi}$ are fitted to experimental  
pairing indicators, this leads to a suppression of scaling factors in spherical nuclei as compared 
with the ones in deformed nuclei.
 
\item 
The present analysis is based on the NL5(E) CEDF which is similar in structure
to well known NL3 and NL3* functionals but provides better description of the ground state 
properties on a global scale \cite{AAT.19}. However, it is reasonable to expect that the results 
about functional dependencies of scaling factors for separable pairing interaction will also be 
valid for the majority of CEDFs built at the Hartree level, such as those used in global studies 
of Refs.\ \cite{GTM.05,AARR.14,LLLYM.15}. This is due to the fact that all of them have 
low Lorentz effective masses and thus similar densities  of the single-particle states
in the vicinity of the Fermi levels as well as similar evolution of these states as a function 
of particle numbers across the nuclear chart (see Refs.\ 
\cite{AS.11,AARR.15,AANR.15,DABRS.15}).
 
\end{itemize}

   The need for optimization of separable pairing interaction of Ref.\ \cite{TMR.09}
across the nuclear chart should not be that surprising. First of all, there is no microscopic 
theory of pairing which would provide required form and strength of pairing for the CDFT 
calculations.  The discussion on the use of the same interaction in the particle-particle
and particle-hole channel is still on-going but this issue is still dependent on the point
of view of the DFT practitioners (see, for example, the discussion in Sec. 2.1 of Ref.
\cite{RRR.19} and in Sec. 2.1 of Ref.\ \cite{CRHB}).  The common stance in the CDFT and 
in non-relativistic Skyrme DFT is that there is no fundamental reason to have the same 
interaction both in the particle–hole and  particle–particle channel because of the use of 
effective forces.   Thus, the Gogny D1S force has been adopted in the 90ties
of the last century for the CDFT applications \cite{GELR.96,CRHB} and later replaced by less
numerically demanding separable pairing interaction of Ref.\ \cite{TMR.09}.  Some differences 
between the CDFT calculations with Gogny D1S force in the pairing channel  and Gogny 
DFT calculations are expected, but interestingly enough they are not that significant
(see discussion in Refs.\ \cite{A250,J1Rare,DABRS.15}).
 
   Although the Gogny force is considered as a benchmark for effective pairing forces 
\cite{RRR.19}, it is unclear how well this force performs globally even in the Gogny DFT
framework when theoretical and experimental pairing indicators extracted from binding 
energies are confronted.  This is because of the absence of such global studies. One should 
also remember that the calibration of the matrix elements  of the Gogny D1S force in the  
pairing channel has been based on OES in tin isotopes \cite{D1} and the description of pairing 
indicators  deviate appreciably from experimental data in some isotopic chains calculated
in Ref.\ \cite{RBB.12}.  Thus, in no way the D1S Gogny to which the separable pairing interaction
 of Ref.\ \cite{TMR.09} has been fitted should be considered as a force optimized in the pairing
 channel especially in the context of the CDFT studies. For example, non-uniqueness of the definition
 of the pairing channel by different Gogny forces is seen in the calculations of rotational
 properties of superdeformed bands of the $A\sim 190$ mass region in the CDFT framework 
 \cite{CRHB}.  The present approach improves the description of the pairing by separable
 interaction of Ref.\ \cite{TMR.09} based on the D1S Gogny force by introducing phenomenological  
 isospin and particle/mass dependencies of the strength of this interaction.

\section{ACKNOWLEDGMENTS}

This material is based upon work supported by the U.S. Department of Energy,  Office of Science, 
Office of Nuclear Physics under Award No. DE-SC0013037.

\bibliography{references-29-PRC-global-pairing}

\begin{thebibliography}{91}%
\makeatletter
\providecommand \@ifxundefined [1]{%
 \@ifx{#1\undefined}
}%
\providecommand \@ifnum [1]{%
 \ifnum #1\expandafter \@firstoftwo
 \else \expandafter \@secondoftwo
 \fi
}%
\providecommand \@ifx [1]{%
 \ifx #1\expandafter \@firstoftwo
 \else \expandafter \@secondoftwo
 \fi
}%
\providecommand \natexlab [1]{#1}%
\providecommand \enquote  [1]{``#1''}%
\providecommand \bibnamefont  [1]{#1}%
\providecommand \bibfnamefont [1]{#1}%
\providecommand \citenamefont [1]{#1}%
\providecommand \href@noop [0]{\@secondoftwo}%
\providecommand \href [0]{\begingroup \@sanitize@url \@href}%
\providecommand \@href[1]{\@@startlink{#1}\@@href}%
\providecommand \@@href[1]{\endgroup#1\@@endlink}%
\providecommand \@sanitize@url [0]{\catcode `\\12\catcode `\$12\catcode
  `\&12\catcode `\#12\catcode `\^12\catcode `\_12\catcode `\%12\relax}%
\providecommand \@@startlink[1]{}%
\providecommand \@@endlink[0]{}%
\providecommand \url  [0]{\begingroup\@sanitize@url \@url }%
\providecommand \@url [1]{\endgroup\@href {#1}{\urlprefix }}%
\providecommand \urlprefix  [0]{URL }%
\providecommand \Eprint [0]{\href }%
\providecommand \doibase [0]{http://dx.doi.org/}%
\providecommand \selectlanguage [0]{\@gobble}%
\providecommand \bibinfo  [0]{\@secondoftwo}%
\providecommand \bibfield  [0]{\@secondoftwo}%
\providecommand \translation [1]{[#1]}%
\providecommand \BibitemOpen [0]{}%
\providecommand \bibitemStop [0]{}%
\providecommand \bibitemNoStop [0]{.\EOS\space}%
\providecommand \EOS [0]{\spacefactor3000\relax}%
\providecommand \BibitemShut  [1]{\csname bibitem#1\endcsname}%
\let\auto@bib@innerbib\@empty
\bibitem [{\citenamefont {Tian}\ \emph
  {et~al.}(2009{\natexlab{a}})\citenamefont {Tian}, \citenamefont {Ma},\ and\
  \citenamefont {Ring}}]{TMR.09}%
  \BibitemOpen
  \bibfield  {author} {\bibinfo {author} {\bibfnamefont {Y.}~\bibnamefont
  {Tian}}, \bibinfo {author} {\bibfnamefont {Z.~Y.}\ \bibnamefont {Ma}}, \ and\
  \bibinfo {author} {\bibfnamefont {P.}~\bibnamefont {Ring}},\ }\href@noop {}
  {\bibfield  {journal} {\bibinfo  {journal} {Phys.\ Lett. B}\ }\textbf
  {\bibinfo {volume} {676}},\ \bibinfo {pages} {44} (\bibinfo {year}
  {2009}{\natexlab{a}})}\BibitemShut {NoStop}%
\bibitem [{\citenamefont {Bohr}\ and\ \citenamefont
  {Mottelson}(1975)}]{Bohr1975}%
  \BibitemOpen
  \bibfield  {author} {\bibinfo {author} {\bibfnamefont {A.}~\bibnamefont
  {Bohr}}\ and\ \bibinfo {author} {\bibfnamefont {B.~R.}\ \bibnamefont
  {Mottelson}},\ }\href@noop {} {\emph {\bibinfo {title} {NUCLEAR STRUCTURE
  Volume II: Nuclear Deformation}}}\ (\bibinfo  {publisher} {W. A. Benjamin,
  Inc.},\ \bibinfo {year} {1975})\BibitemShut {NoStop}%
\bibitem [{\citenamefont {Soloviev}(1976)}]{Sol-book-1976}%
  \BibitemOpen
  \bibfield  {author} {\bibinfo {author} {\bibfnamefont {V.~G.}\ \bibnamefont
  {Soloviev}},\ }\href@noop {} {\emph {\bibinfo {title} {Theory of complex
  nuclei}}}\ (\bibinfo  {publisher} {Pergamon},\ \bibinfo {year}
  {1976})\BibitemShut {NoStop}%
\bibitem [{\citenamefont {Ring}\ and\ \citenamefont {Schuck}(1980)}]{RS.80}%
  \BibitemOpen
  \bibfield  {author} {\bibinfo {author} {\bibfnamefont {P.}~\bibnamefont
  {Ring}}\ and\ \bibinfo {author} {\bibfnamefont {P.}~\bibnamefont {Schuck}},\
  }\href@noop {} {\bibfield  {journal} {\bibinfo  {journal} {{\em The Nuclear
  Many-Body Problem} (Springer-Verlag, Berlin)}\ } (\bibinfo {year}
  {1980})}\BibitemShut {NoStop}%
\bibitem [{\citenamefont {Afanasjev}(2013)}]{A.12}%
  \BibitemOpen
  \bibfield  {author} {\bibinfo {author} {\bibfnamefont {A.~V.}\ \bibnamefont
  {Afanasjev}},\ }\href@noop {} {\bibfield  {journal} {\bibinfo  {journal}
  {chapter 11 in the book "50 Years of Nuclear BCS'', (World Scientific
  Publishing Co, Singapore, 2013), p. 138, see also nuclear theory arkhive
  arXiv:1205.2134.}\ } (\bibinfo {year} {2013})}\BibitemShut {NoStop}%
\bibitem [{\citenamefont {Frauendorf}\ and\ \citenamefont
  {Macchiavelli}(2014)}]{FM.14}%
  \BibitemOpen
  \bibfield  {author} {\bibinfo {author} {\bibfnamefont {S.}~\bibnamefont
  {Frauendorf}}\ and\ \bibinfo {author} {\bibfnamefont {A.}~\bibnamefont
  {Macchiavelli}},\ }\href@noop {} {\bibfield  {journal} {\bibinfo  {journal}
  {Prog.\ Part.\ Nucl.\ Phys.}\ }\textbf {\bibinfo {volume} {78}},\ \bibinfo
  {pages} {24} (\bibinfo {year} {2014})}\BibitemShut {NoStop}%
\bibitem [{\citenamefont {Szyma{\'n}ski}(1983)}]{Szy-book}%
  \BibitemOpen
  \bibfield  {author} {\bibinfo {author} {\bibfnamefont {Z.}~\bibnamefont
  {Szyma{\'n}ski}},\ }\href@noop {} {\emph {\bibinfo {title} {Fast nuclear
  rotation}}}\ (\bibinfo  {publisher} {Clarendon Press, Oxford},\ \bibinfo
  {year} {1983})\BibitemShut {NoStop}%
\bibitem [{\citenamefont {Satuła}\ \emph {et~al.}(1994)\citenamefont
  {Satuła}, \citenamefont {Wyss},\ and\ \citenamefont {Magierski}}]{SWM.94}%
  \BibitemOpen
  \bibfield  {author} {\bibinfo {author} {\bibfnamefont {W.}~\bibnamefont
  {Satuła}}, \bibinfo {author} {\bibfnamefont {R.}~\bibnamefont {Wyss}}, \
  and\ \bibinfo {author} {\bibfnamefont {P.}~\bibnamefont {Magierski}},\ }\href
  {\doibase https://doi.org/10.1016/0375-9474(94)90968-7} {\bibfield  {journal}
  {\bibinfo  {journal} {Nucl. Phys. A}\ }\textbf {\bibinfo {volume} {578}},\
  \bibinfo {pages} {45 } (\bibinfo {year} {1994})}\BibitemShut {NoStop}%
\bibitem [{\citenamefont {Gall}\ \emph {et~al.}(1994)\citenamefont {Gall},
  \citenamefont {Bonche}, \citenamefont {Dobaczewski}, \citenamefont
  {Flocard},\ and\ \citenamefont {Heenen}}]{GBDFH.94}%
  \BibitemOpen
  \bibfield  {author} {\bibinfo {author} {\bibfnamefont {B.}~\bibnamefont
  {Gall}}, \bibinfo {author} {\bibfnamefont {P.}~\bibnamefont {Bonche}},
  \bibinfo {author} {\bibfnamefont {J.}~\bibnamefont {Dobaczewski}}, \bibinfo
  {author} {\bibfnamefont {H.}~\bibnamefont {Flocard}}, \ and\ \bibinfo
  {author} {\bibfnamefont {P.-H.}\ \bibnamefont {Heenen}},\ }\href@noop {}
  {\bibfield  {journal} {\bibinfo  {journal} {Zeit.~ Phys. A}\ }\textbf
  {\bibinfo {volume} {348}},\ \bibinfo {pages} {183} (\bibinfo {year}
  {1994})}\BibitemShut {NoStop}%
\bibitem [{\citenamefont {Afanasjev}\ \emph
  {et~al.}(2000{\natexlab{a}})\citenamefont {Afanasjev}, \citenamefont {Ring},\
  and\ \citenamefont {K{\"o}nig}}]{CRHB}%
  \BibitemOpen
  \bibfield  {author} {\bibinfo {author} {\bibfnamefont {A.~V.}\ \bibnamefont
  {Afanasjev}}, \bibinfo {author} {\bibfnamefont {P.}~\bibnamefont {Ring}}, \
  and\ \bibinfo {author} {\bibfnamefont {J.}~\bibnamefont {K{\"o}nig}},\
  }\href@noop {} {\bibfield  {journal} {\bibinfo  {journal} {Nucl.\ Phys.}\
  }\textbf {\bibinfo {volume} {A676}},\ \bibinfo {pages} {196} (\bibinfo {year}
  {2000}{\natexlab{a}})}\BibitemShut {NoStop}%
\bibitem [{\citenamefont {Afanasjev}\ and\ \citenamefont
  {Abdurazakov}(2013)}]{AO.13}%
  \BibitemOpen
  \bibfield  {author} {\bibinfo {author} {\bibfnamefont {A.~V.}\ \bibnamefont
  {Afanasjev}}\ and\ \bibinfo {author} {\bibfnamefont {O.}~\bibnamefont
  {Abdurazakov}},\ }\href@noop {} {\bibfield  {journal} {\bibinfo  {journal}
  {Phys.\ Rev. C}\ }\textbf {\bibinfo {volume} {88}},\ \bibinfo {pages}
  {014320} (\bibinfo {year} {2013})}\BibitemShut {NoStop}%
\bibitem [{\citenamefont {Valor}\ \emph {et~al.}(2000)\citenamefont {Valor},
  \citenamefont {Egido},\ and\ \citenamefont {Robledo}}]{AER.00}%
  \BibitemOpen
  \bibfield  {author} {\bibinfo {author} {\bibfnamefont {A.}~\bibnamefont
  {Valor}}, \bibinfo {author} {\bibfnamefont {J.~L.}\ \bibnamefont {Egido}}, \
  and\ \bibinfo {author} {\bibfnamefont {L.~M.}\ \bibnamefont {Robledo}},\
  }\href@noop {} {\bibfield  {journal} {\bibinfo  {journal} {Nucl.\ Phys. A}\
  }\textbf {\bibinfo {volume} {665}},\ \bibinfo {pages} {46} (\bibinfo {year}
  {2000})}\BibitemShut {NoStop}%
\bibitem [{\citenamefont {Afanasjev}\ \emph
  {et~al.}(2000{\natexlab{b}})\citenamefont {Afanasjev}, \citenamefont
  {K{\"o}nig}, \citenamefont {Ring}, \citenamefont {Robledo},\ and\
  \citenamefont {Egido}}]{J1Rare}%
  \BibitemOpen
  \bibfield  {author} {\bibinfo {author} {\bibfnamefont {A.~V.}\ \bibnamefont
  {Afanasjev}}, \bibinfo {author} {\bibfnamefont {J.}~\bibnamefont
  {K{\"o}nig}}, \bibinfo {author} {\bibfnamefont {P.}~\bibnamefont {Ring}},
  \bibinfo {author} {\bibfnamefont {L.~M.}\ \bibnamefont {Robledo}}, \ and\
  \bibinfo {author} {\bibfnamefont {J.~L.}\ \bibnamefont {Egido}},\ }\href@noop
  {} {\bibfield  {journal} {\bibinfo  {journal} {Phys.\ Rev. C}\ }\textbf
  {\bibinfo {volume} {62}},\ \bibinfo {pages} {054306} (\bibinfo {year}
  {2000}{\natexlab{b}})}\BibitemShut {NoStop}%
\bibitem [{\citenamefont {Karatzikos}\ \emph {et~al.}(2010)\citenamefont
  {Karatzikos}, \citenamefont {Afanasjev}, \citenamefont {Lalazissis},\ and\
  \citenamefont {Ring}}]{KALR.10}%
  \BibitemOpen
  \bibfield  {author} {\bibinfo {author} {\bibfnamefont {S.}~\bibnamefont
  {Karatzikos}}, \bibinfo {author} {\bibfnamefont {A.~V.}\ \bibnamefont
  {Afanasjev}}, \bibinfo {author} {\bibfnamefont {G.~A.}\ \bibnamefont
  {Lalazissis}}, \ and\ \bibinfo {author} {\bibfnamefont {P.}~\bibnamefont
  {Ring}},\ }\href@noop {} {\bibfield  {journal} {\bibinfo  {journal} {Phys.\
  Lett. B}\ }\textbf {\bibinfo {volume} {689}},\ \bibinfo {pages} {72}
  (\bibinfo {year} {2010})}\BibitemShut {NoStop}%
\bibitem [{\citenamefont {Agbemava}\ \emph {et~al.}(2016)\citenamefont
  {Agbemava}, \citenamefont {Afanasjev},\ and\ \citenamefont {Ring}}]{AAR.16}%
  \BibitemOpen
  \bibfield  {author} {\bibinfo {author} {\bibfnamefont {S.~E.}\ \bibnamefont
  {Agbemava}}, \bibinfo {author} {\bibfnamefont {A.~V.}\ \bibnamefont
  {Afanasjev}}, \ and\ \bibinfo {author} {\bibfnamefont {P.}~\bibnamefont
  {Ring}},\ }\href@noop {} {\bibfield  {journal} {\bibinfo  {journal} {Phys.
  Rev. C}\ }\textbf {\bibinfo {volume} {93}},\ \bibinfo {pages} {044304}
  (\bibinfo {year} {2016})}\BibitemShut {NoStop}%
\bibitem [{LNP(2004)}]{LNP.641}%
  \BibitemOpen
  \href@noop {} {\bibfield  {journal} {\bibinfo  {journal} {Extended Density
  Functionals in Nuclear Structure Physics, {\em Lecture Notes in Physics,
  edited by G.\ A.\ Lalazissis, P. Ring, and D. Vretenar (Springer-Verlag,
  Heidelberg, 2004)}}\ }\textbf {\bibinfo {volume} {Vol. 641}} (\bibinfo {year}
  {2004})}\BibitemShut {NoStop}%
\bibitem [{\citenamefont {Cohen}\ \emph {et~al.}(1992)\citenamefont {Cohen},
  \citenamefont {Furnstahl},\ and\ \citenamefont {Griegel}}]{CFG.92}%
  \BibitemOpen
  \bibfield  {author} {\bibinfo {author} {\bibfnamefont {T.~D.}\ \bibnamefont
  {Cohen}}, \bibinfo {author} {\bibfnamefont {R.~J.}\ \bibnamefont
  {Furnstahl}}, \ and\ \bibinfo {author} {\bibfnamefont {K.}~\bibnamefont
  {Griegel}},\ }\href@noop {} {\bibfield  {journal} {\bibinfo  {journal}
  {Phys.\ Rev. C}\ }\textbf {\bibinfo {volume} {45}},\ \bibinfo {pages} {1881}
  (\bibinfo {year} {1992})}\BibitemShut {NoStop}%
\bibitem [{\citenamefont {Koepf}\ and\ \citenamefont {Ring}(1989)}]{KR.89}%
  \BibitemOpen
  \bibfield  {author} {\bibinfo {author} {\bibfnamefont {W.}~\bibnamefont
  {Koepf}}\ and\ \bibinfo {author} {\bibfnamefont {P.}~\bibnamefont {Ring}},\
  }\href@noop {} {\bibfield  {journal} {\bibinfo  {journal} {Nucl.\ Phys. A}\
  }\textbf {\bibinfo {volume} {493}},\ \bibinfo {pages} {61} (\bibinfo {year}
  {1989})}\BibitemShut {NoStop}%
\bibitem [{\citenamefont {Afanasjev}\ and\ \citenamefont
  {Abusara}(2010{\natexlab{a}})}]{AA.10}%
  \BibitemOpen
  \bibfield  {author} {\bibinfo {author} {\bibfnamefont {A.~V.}\ \bibnamefont
  {Afanasjev}}\ and\ \bibinfo {author} {\bibfnamefont {H.}~\bibnamefont
  {Abusara}},\ }\href@noop {} {\bibfield  {journal} {\bibinfo  {journal}
  {Phys.\ Rev. C}\ }\textbf {\bibinfo {volume} {81}},\ \bibinfo {pages}
  {014309} (\bibinfo {year} {2010}{\natexlab{a}})}\BibitemShut {NoStop}%
\bibitem [{\citenamefont {Hofmann}\ and\ \citenamefont {Ring}(1988)}]{HR.88}%
  \BibitemOpen
  \bibfield  {author} {\bibinfo {author} {\bibfnamefont {U.}~\bibnamefont
  {Hofmann}}\ and\ \bibinfo {author} {\bibfnamefont {P.}~\bibnamefont {Ring}},\
  }\href@noop {} {\bibfield  {journal} {\bibinfo  {journal} {Phys.\ Lett. B}\
  }\textbf {\bibinfo {volume} {214}},\ \bibinfo {pages} {307} (\bibinfo {year}
  {1988})}\BibitemShut {NoStop}%
\bibitem [{\citenamefont {Afanasjev}\ and\ \citenamefont {Ring}(2000)}]{AR.00}%
  \BibitemOpen
  \bibfield  {author} {\bibinfo {author} {\bibfnamefont {A.~V.}\ \bibnamefont
  {Afanasjev}}\ and\ \bibinfo {author} {\bibfnamefont {P.}~\bibnamefont
  {Ring}},\ }\href@noop {} {\bibfield  {journal} {\bibinfo  {journal} {Phys.\
  Rev. C}\ }\textbf {\bibinfo {volume} {62}},\ \bibinfo {pages} {031302(R)}
  (\bibinfo {year} {2000})}\BibitemShut {NoStop}%
\bibitem [{\citenamefont {Afanasjev}\ and\ \citenamefont
  {Abusara}(2010{\natexlab{b}})}]{TO-rot}%
  \BibitemOpen
  \bibfield  {author} {\bibinfo {author} {\bibfnamefont {A.~V.}\ \bibnamefont
  {Afanasjev}}\ and\ \bibinfo {author} {\bibfnamefont {H.}~\bibnamefont
  {Abusara}},\ }\href@noop {} {\bibfield  {journal} {\bibinfo  {journal}
  {Phys.\ Rev. C}\ }\textbf {\bibinfo {volume} {82}},\ \bibinfo {pages}
  {034329} (\bibinfo {year} {2010}{\natexlab{b}})}\BibitemShut {NoStop}%
\bibitem [{\citenamefont {Brockmann}\ and\ \citenamefont {Toki}(1992)}]{BT.92}%
  \BibitemOpen
  \bibfield  {author} {\bibinfo {author} {\bibfnamefont {R.}~\bibnamefont
  {Brockmann}}\ and\ \bibinfo {author} {\bibfnamefont {H.}~\bibnamefont
  {Toki}},\ }\href@noop {} {\bibfield  {journal} {\bibinfo  {journal} {Phys.\
  Rev.\ Lett.}\ }\textbf {\bibinfo {volume} {68}},\ \bibinfo {pages} {3408}
  (\bibinfo {year} {1992})}\BibitemShut {NoStop}%
\bibitem [{\citenamefont {Hofmann}\ \emph {et~al.}(2001)\citenamefont
  {Hofmann}, \citenamefont {Keil},\ and\ \citenamefont {Lenske}}]{HKL.01}%
  \BibitemOpen
  \bibfield  {author} {\bibinfo {author} {\bibfnamefont {F.}~\bibnamefont
  {Hofmann}}, \bibinfo {author} {\bibfnamefont {C.~M.}\ \bibnamefont {Keil}}, \
  and\ \bibinfo {author} {\bibfnamefont {H.}~\bibnamefont {Lenske}},\
  }\href@noop {} {\bibfield  {journal} {\bibinfo  {journal} {Phys.\ Rev. C}\
  }\textbf {\bibinfo {volume} {64}},\ \bibinfo {pages} {034314} (\bibinfo
  {year} {2001})}\BibitemShut {NoStop}%
\bibitem [{\citenamefont {Serra}\ \emph {et~al.}(2005)\citenamefont {Serra},
  \citenamefont {Otsuka}, \citenamefont {Akaishi}, \citenamefont {Ring},\ and\
  \citenamefont {Hirose}}]{SOA.05}%
  \BibitemOpen
  \bibfield  {author} {\bibinfo {author} {\bibfnamefont {M.}~\bibnamefont
  {Serra}}, \bibinfo {author} {\bibfnamefont {T.}~\bibnamefont {Otsuka}},
  \bibinfo {author} {\bibfnamefont {Y.}~\bibnamefont {Akaishi}}, \bibinfo
  {author} {\bibfnamefont {P.}~\bibnamefont {Ring}}, \ and\ \bibinfo {author}
  {\bibfnamefont {S.}~\bibnamefont {Hirose}},\ }\href@noop {} {\bibfield
  {journal} {\bibinfo  {journal} {Prog.\ Theor.\ Phys.}\ }\textbf {\bibinfo
  {volume} {113}},\ \bibinfo {pages} {1009} (\bibinfo {year}
  {2005})}\BibitemShut {NoStop}%
\bibitem [{\citenamefont {Hirose}\ \emph {et~al.}(2007)\citenamefont {Hirose},
  \citenamefont {Serra}, \citenamefont {Ring}, \citenamefont {Otsuka},\ and\
  \citenamefont {Akaishi}}]{HSR.07}%
  \BibitemOpen
  \bibfield  {author} {\bibinfo {author} {\bibfnamefont {S.}~\bibnamefont
  {Hirose}}, \bibinfo {author} {\bibfnamefont {M.}~\bibnamefont {Serra}},
  \bibinfo {author} {\bibfnamefont {P.}~\bibnamefont {Ring}}, \bibinfo {author}
  {\bibfnamefont {T.}~\bibnamefont {Otsuka}}, \ and\ \bibinfo {author}
  {\bibfnamefont {Y.}~\bibnamefont {Akaishi}},\ }\href@noop {} {\bibfield
  {journal} {\bibinfo  {journal} {Phys.\ Rev. C}\ }\textbf {\bibinfo {volume}
  {75}},\ \bibinfo {pages} {024301} (\bibinfo {year} {2007})}\BibitemShut
  {NoStop}%
\bibitem [{\citenamefont {Drut}\ \emph {et~al.}(2010)\citenamefont {Drut},
  \citenamefont {Furnstahl},\ and\ \citenamefont
  {Platter}}]{Drut2010_PPNP64-120}%
  \BibitemOpen
  \bibfield  {author} {\bibinfo {author} {\bibfnamefont {J.~E.}\ \bibnamefont
  {Drut}}, \bibinfo {author} {\bibfnamefont {R.~J.}\ \bibnamefont {Furnstahl}},
  \ and\ \bibinfo {author} {\bibfnamefont {L.}~\bibnamefont {Platter}},\
  }\href@noop {} {\bibfield  {journal} {\bibinfo  {journal} {Prog. Part. Nucl.
  Phys.}\ }\textbf {\bibinfo {volume} {64}},\ \bibinfo {pages} {120} (\bibinfo
  {year} {2010})}\BibitemShut {NoStop}%
\bibitem [{\citenamefont {Peru}\ and\ \citenamefont {Martini}(2014)}]{PM.14}%
  \BibitemOpen
  \bibfield  {author} {\bibinfo {author} {\bibfnamefont {S.}~\bibnamefont
  {Peru}}\ and\ \bibinfo {author} {\bibfnamefont {M.}~\bibnamefont {Martini}},\
  }\href@noop {} {\bibfield  {journal} {\bibinfo  {journal} {Eur.\ Phys. J}\
  }\textbf {\bibinfo {volume} {50}},\ \bibinfo {pages} {88} (\bibinfo {year}
  {2014})}\BibitemShut {NoStop}%
\bibitem [{\citenamefont {Lalazissis}\ \emph {et~al.}(2005)\citenamefont
  {Lalazissis}, \citenamefont {Nik{\v{s}}i{\'{c}}}, \citenamefont {Vretenar},\
  and\ \citenamefont {Ring}}]{DD-ME2}%
  \BibitemOpen
  \bibfield  {author} {\bibinfo {author} {\bibfnamefont {G.~A.}\ \bibnamefont
  {Lalazissis}}, \bibinfo {author} {\bibfnamefont {T.}~\bibnamefont
  {Nik{\v{s}}i{\'{c}}}}, \bibinfo {author} {\bibfnamefont {D.}~\bibnamefont
  {Vretenar}}, \ and\ \bibinfo {author} {\bibfnamefont {P.}~\bibnamefont
  {Ring}},\ }\href@noop {} {\bibfield  {journal} {\bibinfo  {journal} {Phys.\
  Rev. C}\ }\textbf {\bibinfo {volume} {71}},\ \bibinfo {pages} {024312}
  (\bibinfo {year} {2005})}\BibitemShut {NoStop}%
\bibitem [{\citenamefont {Nik\v{s}i\'{c}}\ \emph {et~al.}(2008)\citenamefont
  {Nik\v{s}i\'{c}}, \citenamefont {Vretenar},\ and\ \citenamefont
  {Ring}}]{DD-PC1}%
  \BibitemOpen
  \bibfield  {author} {\bibinfo {author} {\bibfnamefont {T.}~\bibnamefont
  {Nik\v{s}i\'{c}}}, \bibinfo {author} {\bibfnamefont {D.}~\bibnamefont
  {Vretenar}}, \ and\ \bibinfo {author} {\bibfnamefont {P.}~\bibnamefont
  {Ring}},\ }\href@noop {} {\bibfield  {journal} {\bibinfo  {journal} {Phys.\
  Rev. C}\ }\textbf {\bibinfo {volume} {78}},\ \bibinfo {pages} {034318}
  (\bibinfo {year} {2008})}\BibitemShut {NoStop}%
\bibitem [{\citenamefont {Roca-Maza}\ \emph {et~al.}(2011)\citenamefont
  {Roca-Maza}, \citenamefont {Vi{\~n}as}, \citenamefont {Centelles},
  \citenamefont {Ring},\ and\ \citenamefont {Schuck}}]{DD-MEdelta}%
  \BibitemOpen
  \bibfield  {author} {\bibinfo {author} {\bibfnamefont {X.}~\bibnamefont
  {Roca-Maza}}, \bibinfo {author} {\bibfnamefont {X.}~\bibnamefont
  {Vi{\~n}as}}, \bibinfo {author} {\bibfnamefont {M.}~\bibnamefont
  {Centelles}}, \bibinfo {author} {\bibfnamefont {P.}~\bibnamefont {Ring}}, \
  and\ \bibinfo {author} {\bibfnamefont {P.}~\bibnamefont {Schuck}},\
  }\href@noop {} {\bibfield  {journal} {\bibinfo  {journal} {Phys.\ Rev. C}\
  }\textbf {\bibinfo {volume} {84}},\ \bibinfo {pages} {054309} (\bibinfo
  {year} {2011})}\BibitemShut {NoStop}%
\bibitem [{\citenamefont {Zhao}\ \emph {et~al.}(2010)\citenamefont {Zhao},
  \citenamefont {Li}, \citenamefont {Yao},\ and\ \citenamefont
  {Meng}}]{PC-PK1}%
  \BibitemOpen
  \bibfield  {author} {\bibinfo {author} {\bibfnamefont {P.~W.}\ \bibnamefont
  {Zhao}}, \bibinfo {author} {\bibfnamefont {Z.~P.}\ \bibnamefont {Li}},
  \bibinfo {author} {\bibfnamefont {J.~M.}\ \bibnamefont {Yao}}, \ and\
  \bibinfo {author} {\bibfnamefont {J.}~\bibnamefont {Meng}},\ }\href@noop {}
  {\bibfield  {journal} {\bibinfo  {journal} {Phys.\ Rev. C}\ }\textbf
  {\bibinfo {volume} {82}},\ \bibinfo {pages} {054319} (\bibinfo {year}
  {2010})}\BibitemShut {NoStop}%
\bibitem [{\citenamefont {Agbemava}\ \emph {et~al.}(2019)\citenamefont
  {Agbemava}, \citenamefont {Afanasjev},\ and\ \citenamefont
  {Taninah}}]{AAT.19}%
  \BibitemOpen
  \bibfield  {author} {\bibinfo {author} {\bibfnamefont {S.~E.}\ \bibnamefont
  {Agbemava}}, \bibinfo {author} {\bibfnamefont {A.~V.}\ \bibnamefont
  {Afanasjev}}, \ and\ \bibinfo {author} {\bibfnamefont {A.}~\bibnamefont
  {Taninah}},\ }\href@noop {} {\bibfield  {journal} {\bibinfo  {journal} {Phys.
  Rev. C}\ }\textbf {\bibinfo {volume} {99}},\ \bibinfo {pages} {014318}
  (\bibinfo {year} {2019})}\BibitemShut {NoStop}%
\bibitem [{\citenamefont {Vretenar}\ \emph {et~al.}(2005)\citenamefont
  {Vretenar}, \citenamefont {Afanasjev}, \citenamefont {Lalazissis},\ and\
  \citenamefont {Ring}}]{VALR.05}%
  \BibitemOpen
  \bibfield  {author} {\bibinfo {author} {\bibfnamefont {D.}~\bibnamefont
  {Vretenar}}, \bibinfo {author} {\bibfnamefont {A.~V.}\ \bibnamefont
  {Afanasjev}}, \bibinfo {author} {\bibfnamefont {G.~A.}\ \bibnamefont
  {Lalazissis}}, \ and\ \bibinfo {author} {\bibfnamefont {P.}~\bibnamefont
  {Ring}},\ }\href@noop {} {\bibfield  {journal} {\bibinfo  {journal} {Phys.\
  Rep.}\ }\textbf {\bibinfo {volume} {409}},\ \bibinfo {pages} {101} (\bibinfo
  {year} {2005})}\BibitemShut {NoStop}%
\bibitem [{\citenamefont {Meng}\ \emph {et~al.}(2006)\citenamefont {Meng},
  \citenamefont {Toki}, \citenamefont {Zhou}, \citenamefont {Zhang},
  \citenamefont {Long},\ and\ \citenamefont {Geng}}]{MTZZLG.06}%
  \BibitemOpen
  \bibfield  {author} {\bibinfo {author} {\bibfnamefont {J.}~\bibnamefont
  {Meng}}, \bibinfo {author} {\bibfnamefont {H.}~\bibnamefont {Toki}}, \bibinfo
  {author} {\bibfnamefont {S.~G.}\ \bibnamefont {Zhou}}, \bibinfo {author}
  {\bibfnamefont {S.~Q.}\ \bibnamefont {Zhang}}, \bibinfo {author}
  {\bibfnamefont {W.~H.}\ \bibnamefont {Long}}, \ and\ \bibinfo {author}
  {\bibfnamefont {L.~S.}\ \bibnamefont {Geng}},\ }\href@noop {} {\bibfield
  {journal} {\bibinfo  {journal} {Prog. Part. Nucl. Phys.}\ }\textbf {\bibinfo
  {volume} {57}},\ \bibinfo {pages} {470} (\bibinfo {year} {2006})}\BibitemShut
  {NoStop}%
\bibitem [{\citenamefont {Nik\v{s}i\'{c}}\ \emph {et~al.}(2011)\citenamefont
  {Nik\v{s}i\'{c}}, \citenamefont {Vretenar},\ and\ \citenamefont
  {Ring}}]{NVR.11}%
  \BibitemOpen
  \bibfield  {author} {\bibinfo {author} {\bibfnamefont {T.}~\bibnamefont
  {Nik\v{s}i\'{c}}}, \bibinfo {author} {\bibfnamefont {D.}~\bibnamefont
  {Vretenar}}, \ and\ \bibinfo {author} {\bibfnamefont {P.}~\bibnamefont
  {Ring}},\ }\href@noop {} {\bibfield  {journal} {\bibinfo  {journal} {Prog.\
  Part.\ Nucl.\ Phys.}\ }\textbf {\bibinfo {volume} {66}},\ \bibinfo {pages}
  {519} (\bibinfo {year} {2011})}\BibitemShut {NoStop}%
\bibitem [{\citenamefont {Agbemava}\ \emph {et~al.}(2014)\citenamefont
  {Agbemava}, \citenamefont {Afanasjev}, \citenamefont {Ray},\ and\
  \citenamefont {Ring}}]{AARR.14}%
  \BibitemOpen
  \bibfield  {author} {\bibinfo {author} {\bibfnamefont {S.~E.}\ \bibnamefont
  {Agbemava}}, \bibinfo {author} {\bibfnamefont {A.~V.}\ \bibnamefont
  {Afanasjev}}, \bibinfo {author} {\bibfnamefont {D.}~\bibnamefont {Ray}}, \
  and\ \bibinfo {author} {\bibfnamefont {P.}~\bibnamefont {Ring}},\ }\href@noop
  {} {\bibfield  {journal} {\bibinfo  {journal} {Phys.\ Rev. C}\ }\textbf
  {\bibinfo {volume} {89}},\ \bibinfo {pages} {054320} (\bibinfo {year}
  {2014})}\BibitemShut {NoStop}%
\bibitem [{RDF(2016{\natexlab{a}})}]{RDFNS-book}%
  \BibitemOpen
  \href@noop {} {\emph {\bibinfo {title} {Relativistic Density Functional for
  Nuclear Structure, \rm Int. Rev. Nucl. Phys. vol. 10}}}\ (\bibinfo
  {publisher} {World Scientific Publishing Co, Edited by Jie Meng},\ \bibinfo
  {year} {2016})\BibitemShut {NoStop}%
\bibitem [{\citenamefont {Satula}\ \emph {et~al.}(1998)\citenamefont {Satula},
  \citenamefont {Dobaczewski},\ and\ \citenamefont {Nazarewicz}}]{SDN.98}%
  \BibitemOpen
  \bibfield  {author} {\bibinfo {author} {\bibfnamefont {W.}~\bibnamefont
  {Satula}}, \bibinfo {author} {\bibfnamefont {J.}~\bibnamefont {Dobaczewski}},
  \ and\ \bibinfo {author} {\bibfnamefont {W.}~\bibnamefont {Nazarewicz}},\
  }\href@noop {} {\bibfield  {journal} {\bibinfo  {journal} {Phys. Rev. Lett.}\
  }\textbf {\bibinfo {volume} {81}},\ \bibinfo {pages} {3599} (\bibinfo {year}
  {1998})}\BibitemShut {NoStop}%
\bibitem [{\citenamefont {Bender}\ \emph {et~al.}(2000)\citenamefont {Bender},
  \citenamefont {Rutz}, \citenamefont {Reinhard},\ and\ \citenamefont
  {Maruhn}}]{BRRM.00}%
  \BibitemOpen
  \bibfield  {author} {\bibinfo {author} {\bibfnamefont {M.}~\bibnamefont
  {Bender}}, \bibinfo {author} {\bibfnamefont {K.}~\bibnamefont {Rutz}},
  \bibinfo {author} {\bibfnamefont {P.-G.}\ \bibnamefont {Reinhard}}, \ and\
  \bibinfo {author} {\bibfnamefont {J.~A.}\ \bibnamefont {Maruhn}},\
  }\href@noop {} {\bibfield  {journal} {\bibinfo  {journal} {Eur. Phys. J. A}\
  }\textbf {\bibinfo {volume} {8}},\ \bibinfo {pages} {59} (\bibinfo {year}
  {2000})}\BibitemShut {NoStop}%
\bibitem [{\citenamefont {Bertsch}\ \emph {et~al.}(2009)\citenamefont
  {Bertsch}, \citenamefont {Bertulani}, \citenamefont {W.~Nazarewicz},\ and\
  \citenamefont {Stoitsov}}]{BBNSS.09}%
  \BibitemOpen
  \bibfield  {author} {\bibinfo {author} {\bibfnamefont {G.~F.}\ \bibnamefont
  {Bertsch}}, \bibinfo {author} {\bibfnamefont {C.~A.}\ \bibnamefont
  {Bertulani}}, \bibinfo {author} {\bibfnamefont {N.~S.}\ \bibnamefont
  {W.~Nazarewicz}}, \ and\ \bibinfo {author} {\bibfnamefont {M.~V.}\
  \bibnamefont {Stoitsov}},\ }\href@noop {} {\bibfield  {journal} {\bibinfo
  {journal} {Phys.\ Rev. C}\ }\textbf {\bibinfo {volume} {79}},\ \bibinfo
  {pages} {034306} (\bibinfo {year} {2009})}\BibitemShut {NoStop}%
\bibitem [{\citenamefont {Duguet}\ \emph {et~al.}(2001)\citenamefont {Duguet},
  \citenamefont {Bonche}, \citenamefont {Heenen},\ and\ \citenamefont
  {Meyer}}]{DBHM.01}%
  \BibitemOpen
  \bibfield  {author} {\bibinfo {author} {\bibfnamefont {T.}~\bibnamefont
  {Duguet}}, \bibinfo {author} {\bibfnamefont {P.}~\bibnamefont {Bonche}},
  \bibinfo {author} {\bibfnamefont {P.-H.}\ \bibnamefont {Heenen}}, \ and\
  \bibinfo {author} {\bibfnamefont {J.}~\bibnamefont {Meyer}},\ }\href@noop {}
  {\bibfield  {journal} {\bibinfo  {journal} {Phys. Rev. C}\ }\textbf {\bibinfo
  {volume} {65}},\ \bibinfo {pages} {014310} (\bibinfo {year}
  {2001})}\BibitemShut {NoStop}%
\bibitem [{\citenamefont {Mukherjee}\ \emph {et~al.}(2011)\citenamefont
  {Mukherjee}, \citenamefont {Alhassid},\ and\ \citenamefont
  {Bertsch}}]{MAB.11}%
  \BibitemOpen
  \bibfield  {author} {\bibinfo {author} {\bibfnamefont {A.}~\bibnamefont
  {Mukherjee}}, \bibinfo {author} {\bibfnamefont {Y.}~\bibnamefont {Alhassid}},
  \ and\ \bibinfo {author} {\bibfnamefont {G.~F.}\ \bibnamefont {Bertsch}},\
  }\href@noop {} {\bibfield  {journal} {\bibinfo  {journal} {Phys.\ Rev. C}\
  }\textbf {\bibinfo {volume} {83}},\ \bibinfo {pages} {014319} (\bibinfo
  {year} {2011})}\BibitemShut {NoStop}%
\bibitem [{\citenamefont {Bertulani}\ \emph {et~al.}(2012)\citenamefont
  {Bertulani}, \citenamefont {Liu},\ and\ \citenamefont {Sagawa}}]{BLS.12}%
  \BibitemOpen
  \bibfield  {author} {\bibinfo {author} {\bibfnamefont {C.~A.}\ \bibnamefont
  {Bertulani}}, \bibinfo {author} {\bibfnamefont {H.}~\bibnamefont {Liu}}, \
  and\ \bibinfo {author} {\bibfnamefont {H.}~\bibnamefont {Sagawa}},\ }\href
  {\doibase 10.1103/PhysRevC.85.014321} {\bibfield  {journal} {\bibinfo
  {journal} {Phys. Rev. C}\ }\textbf {\bibinfo {volume} {85}},\ \bibinfo
  {pages} {014321} (\bibinfo {year} {2012})}\BibitemShut {NoStop}%
\bibitem [{\citenamefont {Yamagami}\ \emph {et~al.}(2012)\citenamefont
  {Yamagami}, \citenamefont {Margueron}, \citenamefont {Sagawa},\ and\
  \citenamefont {Hagino}}]{YMSH.12}%
  \BibitemOpen
  \bibfield  {author} {\bibinfo {author} {\bibfnamefont {M.}~\bibnamefont
  {Yamagami}}, \bibinfo {author} {\bibfnamefont {J.}~\bibnamefont {Margueron}},
  \bibinfo {author} {\bibfnamefont {H.}~\bibnamefont {Sagawa}}, \ and\ \bibinfo
  {author} {\bibfnamefont {K.}~\bibnamefont {Hagino}},\ }\href {\doibase
  10.1103/PhysRevC.86.034333} {\bibfield  {journal} {\bibinfo  {journal} {Phys.
  Rev. C}\ }\textbf {\bibinfo {volume} {86}},\ \bibinfo {pages} {034333}
  (\bibinfo {year} {2012})}\BibitemShut {NoStop}%
\bibitem [{\citenamefont {Changizi}\ \emph {et~al.}(2015)\citenamefont
  {Changizi}, \citenamefont {Qi},\ and\ \citenamefont {Wyss}}]{CQW.15}%
  \BibitemOpen
  \bibfield  {author} {\bibinfo {author} {\bibfnamefont {S.}~\bibnamefont
  {Changizi}}, \bibinfo {author} {\bibfnamefont {C.}~\bibnamefont {Qi}}, \ and\
  \bibinfo {author} {\bibfnamefont {R.}~\bibnamefont {Wyss}},\ }\href {\doibase
  http://dx.doi.org/10.1016/j.nuclphysa.2015.04.010} {\bibfield  {journal}
  {\bibinfo  {journal} {Nucl. Phys. A}\ }\textbf {\bibinfo {volume} {940}},\
  \bibinfo {pages} {210 } (\bibinfo {year} {2015})}\BibitemShut {NoStop}%
\bibitem [{\citenamefont {Hilaire}\ \emph {et~al.}(2002)\citenamefont
  {Hilaire}, \citenamefont {Berger}, \citenamefont {Girod}, \citenamefont
  {Satula},\ and\ \citenamefont {Schuck}}]{HBGSS.02}%
  \BibitemOpen
  \bibfield  {author} {\bibinfo {author} {\bibfnamefont {S.}~\bibnamefont
  {Hilaire}}, \bibinfo {author} {\bibfnamefont {J.-F.}\ \bibnamefont {Berger}},
  \bibinfo {author} {\bibfnamefont {M.}~\bibnamefont {Girod}}, \bibinfo
  {author} {\bibfnamefont {W.}~\bibnamefont {Satula}}, \ and\ \bibinfo {author}
  {\bibfnamefont {P.}~\bibnamefont {Schuck}},\ }\href@noop {} {\bibfield
  {journal} {\bibinfo  {journal} {Phys.\ Lett. B}\ }\textbf {\bibinfo {volume}
  {531}},\ \bibinfo {pages} {61} (\bibinfo {year} {2002})}\BibitemShut
  {NoStop}%
\bibitem [{\citenamefont {Robledo}\ \emph {et~al.}(2012)\citenamefont
  {Robledo}, \citenamefont {Bernard},\ and\ \citenamefont {Bertsch}}]{RBB.12}%
  \BibitemOpen
  \bibfield  {author} {\bibinfo {author} {\bibfnamefont {L.~M.}\ \bibnamefont
  {Robledo}}, \bibinfo {author} {\bibfnamefont {R.}~\bibnamefont {Bernard}}, \
  and\ \bibinfo {author} {\bibfnamefont {G.~F.}\ \bibnamefont {Bertsch}},\
  }\href {\doibase 10.1103/PhysRevC.86.064313} {\bibfield  {journal} {\bibinfo
  {journal} {Phys. Rev. C}\ }\textbf {\bibinfo {volume} {86}},\ \bibinfo
  {pages} {064313} (\bibinfo {year} {2012})}\BibitemShut {NoStop}%
\bibitem [{\citenamefont {Dobaczewski}\ \emph {et~al.}(2015)\citenamefont
  {Dobaczewski}, \citenamefont {Afanasjev}, \citenamefont {Bender},
  \citenamefont {Robledo},\ and\ \citenamefont {Shi}}]{DABRS.15}%
  \BibitemOpen
  \bibfield  {author} {\bibinfo {author} {\bibfnamefont {J.}~\bibnamefont
  {Dobaczewski}}, \bibinfo {author} {\bibfnamefont {A.~V.}\ \bibnamefont
  {Afanasjev}}, \bibinfo {author} {\bibfnamefont {M.}~\bibnamefont {Bender}},
  \bibinfo {author} {\bibfnamefont {L.~M.}\ \bibnamefont {Robledo}}, \ and\
  \bibinfo {author} {\bibfnamefont {Y.}~\bibnamefont {Shi}},\ }\href {\doibase
  http://dx.doi.org/10.1016/j.nuclphysa.2015.07.015} {\bibfield  {journal}
  {\bibinfo  {journal} {Nucl. Phys. A}\ }\textbf {\bibinfo {volume} {944}},\
  \bibinfo {pages} {388 } (\bibinfo {year} {2015})}\BibitemShut {NoStop}%
\bibitem [{\citenamefont {Robledo}\ \emph {et~al.}(2019)\citenamefont
  {Robledo}, \citenamefont {Rodr{\'i}guez},\ and\ \citenamefont
  {Rodr{\'i}guez-Guzm{\'a}n}}]{RRR.19}%
  \BibitemOpen
  \bibfield  {author} {\bibinfo {author} {\bibfnamefont {L.~M.}\ \bibnamefont
  {Robledo}}, \bibinfo {author} {\bibfnamefont {T.~R.}\ \bibnamefont
  {Rodr{\'i}guez}}, \ and\ \bibinfo {author} {\bibfnamefont {R.~R.}\
  \bibnamefont {Rodr{\'i}guez-Guzm{\'a}n}},\ }\href@noop {} {\bibfield
  {journal} {\bibinfo  {journal} {J. Phys. G}\ }\textbf {\bibinfo {volume}
  {46}},\ \bibinfo {pages} {013001} (\bibinfo {year} {2019})}\BibitemShut
  {NoStop}%
\bibitem [{\citenamefont {Kerman}(1961)}]{Kerman1961_APNY12-300}%
  \BibitemOpen
  \bibfield  {author} {\bibinfo {author} {\bibfnamefont {A.~K.}\ \bibnamefont
  {Kerman}},\ }\href@noop {} {\bibfield  {journal} {\bibinfo  {journal} {Ann.
  Phys. (N.Y.)}\ }\textbf {\bibinfo {volume} {12}},\ \bibinfo {pages} {300 }
  (\bibinfo {year} {1961})}\BibitemShut {NoStop}%
\bibitem [{\citenamefont {Gogny}(1975)}]{D1}%
  \BibitemOpen
  \bibfield  {author} {\bibinfo {author} {\bibfnamefont {D.}~\bibnamefont
  {Gogny}},\ }\href@noop {} {\bibfield  {journal} {\bibinfo  {journal} {in: G.~
  Ripka, M.~ Porneuf (Eds.), Nuclear Self-Consistent Fields, Proc. Internat.
  Conf. held at the Center for Theoretical Physics, Trieste, Italy, 1975,
  North-Holland, Amsterdam}\ ,\ \bibinfo {pages} {333}} (\bibinfo {year}
  {1975})}\BibitemShut {NoStop}%
\bibitem [{\citenamefont {Berger}\ \emph {et~al.}(1984)\citenamefont {Berger},
  \citenamefont {Girod},\ and\ \citenamefont {Gogny}}]{D1S-a}%
  \BibitemOpen
  \bibfield  {author} {\bibinfo {author} {\bibfnamefont {J.~F.}\ \bibnamefont
  {Berger}}, \bibinfo {author} {\bibfnamefont {M.}~\bibnamefont {Girod}}, \
  and\ \bibinfo {author} {\bibfnamefont {D.}~\bibnamefont {Gogny}},\
  }\href@noop {} {\bibfield  {journal} {\bibinfo  {journal} {Nucl.\ Phys.}\
  }\textbf {\bibinfo {volume} {A428}},\ \bibinfo {pages} {23c} (\bibinfo {year}
  {1984})}\BibitemShut {NoStop}%
\bibitem [{\citenamefont {Berger}\ \emph {et~al.}(1991)\citenamefont {Berger},
  \citenamefont {Girod},\ and\ \citenamefont {Gogny}}]{D1S}%
  \BibitemOpen
  \bibfield  {author} {\bibinfo {author} {\bibfnamefont {J.~F.}\ \bibnamefont
  {Berger}}, \bibinfo {author} {\bibfnamefont {M.}~\bibnamefont {Girod}}, \
  and\ \bibinfo {author} {\bibfnamefont {D.}~\bibnamefont {Gogny}},\
  }\href@noop {} {\bibfield  {journal} {\bibinfo  {journal} {Comp.\ Phys.\
  Comm.}\ }\textbf {\bibinfo {volume} {63}},\ \bibinfo {pages} {365} (\bibinfo
  {year} {1991})}\BibitemShut {NoStop}%
\bibitem [{\citenamefont {Gonzalez-Llarena}\ \emph {et~al.}(1996)\citenamefont
  {Gonzalez-Llarena}, \citenamefont {Egido}, \citenamefont {Lalazissis},\ and\
  \citenamefont {Ring}}]{GELR.96}%
  \BibitemOpen
  \bibfield  {author} {\bibinfo {author} {\bibfnamefont {T.}~\bibnamefont
  {Gonzalez-Llarena}}, \bibinfo {author} {\bibfnamefont {J.~L.}\ \bibnamefont
  {Egido}}, \bibinfo {author} {\bibfnamefont {G.~A.}\ \bibnamefont
  {Lalazissis}}, \ and\ \bibinfo {author} {\bibfnamefont {P.}~\bibnamefont
  {Ring}},\ }\href@noop {} {\bibfield  {journal} {\bibinfo  {journal} {Phys.
  Lett. B}\ }\textbf {\bibinfo {volume} {379}},\ \bibinfo {pages} {13}
  (\bibinfo {year} {1996})}\BibitemShut {NoStop}%
\bibitem [{RDF(2016{\natexlab{b}})}]{RDFNS.16}%
  \BibitemOpen
  \href@noop {} {\bibfield  {journal} {\bibinfo  {journal} {``Relativistic
  Density Functional for Nuclear Structure'', (World Scientific Publishing Co),
  Edited by Jie Meng, Int. Rev. Nucl. Phys.}\ }\textbf {\bibinfo {volume} {10}}
  (\bibinfo {year} {2016}{\natexlab{b}})}\BibitemShut {NoStop}%
\bibitem [{\citenamefont {Afanasjev}(2014)}]{A.14}%
  \BibitemOpen
  \bibfield  {author} {\bibinfo {author} {\bibfnamefont {A.~V.}\ \bibnamefont
  {Afanasjev}},\ }\href@noop {} {\bibfield  {journal} {\bibinfo  {journal}
  {Phys.\ Scr.}\ }\textbf {\bibinfo {volume} {89}},\ \bibinfo {pages} {054001}
  (\bibinfo {year} {2014})}\BibitemShut {NoStop}%
\bibitem [{\citenamefont {Lalazissis}\ \emph {et~al.}(2001)\citenamefont
  {Lalazissis}, \citenamefont {Vretenar},\ and\ \citenamefont {Ring}}]{LVR.01}%
  \BibitemOpen
  \bibfield  {author} {\bibinfo {author} {\bibfnamefont {G.~A.}\ \bibnamefont
  {Lalazissis}}, \bibinfo {author} {\bibfnamefont {D.}~\bibnamefont
  {Vretenar}}, \ and\ \bibinfo {author} {\bibfnamefont {P.}~\bibnamefont
  {Ring}},\ }\href@noop {} {\bibfield  {journal} {\bibinfo  {journal} {Nucl.\
  Phys. A}\ }\textbf {\bibinfo {volume} {679}},\ \bibinfo {pages} {481}
  (\bibinfo {year} {2001})}\BibitemShut {NoStop}%
\bibitem [{\citenamefont {Serra}\ and\ \citenamefont {Ring}(2002)}]{SR.02}%
  \BibitemOpen
  \bibfield  {author} {\bibinfo {author} {\bibfnamefont {M.}~\bibnamefont
  {Serra}}\ and\ \bibinfo {author} {\bibfnamefont {P.}~\bibnamefont {Ring}},\
  }\href {\doibase 10.1103/PhysRevC.65.064324} {\bibfield  {journal} {\bibinfo
  {journal} {Phys. Rev. C}\ }\textbf {\bibinfo {volume} {65}},\ \bibinfo
  {pages} {064324} (\bibinfo {year} {2002})}\BibitemShut {NoStop}%
\bibitem [{\citenamefont {Tian}\ \emph
  {et~al.}(2009{\natexlab{b}})\citenamefont {Tian}, \citenamefont {Ma},\ and\
  \citenamefont {Ring}}]{TMR.09a}%
  \BibitemOpen
  \bibfield  {author} {\bibinfo {author} {\bibfnamefont {Y.}~\bibnamefont
  {Tian}}, \bibinfo {author} {\bibfnamefont {Z.~Y.}\ \bibnamefont {Ma}}, \ and\
  \bibinfo {author} {\bibfnamefont {P.}~\bibnamefont {Ring}},\ }\href@noop {}
  {\bibfield  {journal} {\bibinfo  {journal} {Phys.\ Rev. C}\ }\textbf
  {\bibinfo {volume} {80}},\ \bibinfo {pages} {024313} (\bibinfo {year}
  {2009}{\natexlab{b}})}\BibitemShut {NoStop}%
\bibitem [{\citenamefont {Nik\v{s}i\'{c}}\ \emph {et~al.}(2010)\citenamefont
  {Nik\v{s}i\'{c}}, \citenamefont {Ring}, \citenamefont {Vretenar},
  \citenamefont {Tian},\ and\ \citenamefont {yu~Ma}}]{NRVTM.10}%
  \BibitemOpen
  \bibfield  {author} {\bibinfo {author} {\bibfnamefont {T.}~\bibnamefont
  {Nik\v{s}i\'{c}}}, \bibinfo {author} {\bibfnamefont {P.}~\bibnamefont
  {Ring}}, \bibinfo {author} {\bibfnamefont {D.}~\bibnamefont {Vretenar}},
  \bibinfo {author} {\bibfnamefont {Y.}~\bibnamefont {Tian}}, \ and\ \bibinfo
  {author} {\bibfnamefont {Z.}~\bibnamefont {yu~Ma}},\ }\href@noop {}
  {\bibfield  {journal} {\bibinfo  {journal} {Phys.\ Rev. C}\ }\textbf
  {\bibinfo {volume} {81}},\ \bibinfo {pages} {054318} (\bibinfo {year}
  {2010})}\BibitemShut {NoStop}%
\bibitem [{\citenamefont {Agbemava}\ \emph {et~al.}(2017)\citenamefont
  {Agbemava}, \citenamefont {Afanasjev}, \citenamefont {Ray},\ and\
  \citenamefont {Ring}}]{AARR.17}%
  \BibitemOpen
  \bibfield  {author} {\bibinfo {author} {\bibfnamefont {S.~E.}\ \bibnamefont
  {Agbemava}}, \bibinfo {author} {\bibfnamefont {A.~V.}\ \bibnamefont
  {Afanasjev}}, \bibinfo {author} {\bibfnamefont {D.}~\bibnamefont {Ray}}, \
  and\ \bibinfo {author} {\bibfnamefont {P.}~\bibnamefont {Ring}},\ }\href@noop
  {} {\bibfield  {journal} {\bibinfo  {journal} {Phys. Rev. C}\ }\textbf
  {\bibinfo {volume} {95}},\ \bibinfo {pages} {054324} (\bibinfo {year}
  {2017})}\BibitemShut {NoStop}%
\bibitem [{\citenamefont {Zhao}\ \emph {et~al.}(2015)\citenamefont {Zhao},
  \citenamefont {Lu}, \citenamefont {Vretenar}, \citenamefont {Zhao},\ and\
  \citenamefont {Zhou}}]{ZLVZZ.15}%
  \BibitemOpen
  \bibfield  {author} {\bibinfo {author} {\bibfnamefont {J.}~\bibnamefont
  {Zhao}}, \bibinfo {author} {\bibfnamefont {B.-N.}\ \bibnamefont {Lu}},
  \bibinfo {author} {\bibfnamefont {D.}~\bibnamefont {Vretenar}}, \bibinfo
  {author} {\bibfnamefont {E.-G.}\ \bibnamefont {Zhao}}, \ and\ \bibinfo
  {author} {\bibfnamefont {S.-G.}\ \bibnamefont {Zhou}},\ }\href@noop {}
  {\bibfield  {journal} {\bibinfo  {journal} {Phys. Rev. C}\ }\textbf {\bibinfo
  {volume} {91}},\ \bibinfo {pages} {014321} (\bibinfo {year}
  {2015})}\BibitemShut {NoStop}%
\bibitem [{\citenamefont {Taninah}\ \emph {et~al.}(2020)\citenamefont
  {Taninah}, \citenamefont {Afanasjev},\ and\ \citenamefont
  {Agbemava}}]{TAA.20}%
  \BibitemOpen
  \bibfield  {author} {\bibinfo {author} {\bibfnamefont {A.}~\bibnamefont
  {Taninah}}, \bibinfo {author} {\bibfnamefont {A.~V.}\ \bibnamefont
  {Afanasjev}}, \ and\ \bibinfo {author} {\bibfnamefont {S.~E.}\ \bibnamefont
  {Agbemava}},\ }\href@noop {} {\bibfield  {journal} {\bibinfo  {journal}
  {Phys. Rev. C}\ }\textbf {\bibinfo {volume} {102}},\ \bibinfo {pages}
  {054330} (\bibinfo {year} {2020})}\BibitemShut {NoStop}%
\bibitem [{\citenamefont {Wang}(2017)}]{Wang.17}%
  \BibitemOpen
  \bibfield  {author} {\bibinfo {author} {\bibfnamefont {Y.~K.}\ \bibnamefont
  {Wang}},\ }\href@noop {} {\bibfield  {journal} {\bibinfo  {journal} {Phys.
  Rev. C}\ }\textbf {\bibinfo {volume} {96}},\ \bibinfo {pages} {054324}
  (\bibinfo {year} {2017})}\BibitemShut {NoStop}%
\bibitem [{\citenamefont {Xiong}(2020)}]{Xiong.20}%
  \BibitemOpen
  \bibfield  {author} {\bibinfo {author} {\bibfnamefont {B.~W.}\ \bibnamefont
  {Xiong}},\ }\href {\doibase 10.1103/PhysRevC.101.054305} {\bibfield
  {journal} {\bibinfo  {journal} {Phys. Rev. C}\ }\textbf {\bibinfo {volume}
  {101}},\ \bibinfo {pages} {054305} (\bibinfo {year} {2020})}\BibitemShut
  {NoStop}%
\bibitem [{\citenamefont {Tian}\ \emph
  {et~al.}(2009{\natexlab{c}})\citenamefont {Tian}, \citenamefont {yu~Ma},\
  and\ \citenamefont {Ring}}]{TMR.09b}%
  \BibitemOpen
  \bibfield  {author} {\bibinfo {author} {\bibfnamefont {Y.}~\bibnamefont
  {Tian}}, \bibinfo {author} {\bibfnamefont {Z.}~\bibnamefont {yu~Ma}}, \ and\
  \bibinfo {author} {\bibfnamefont {P.}~\bibnamefont {Ring}},\ }\href@noop {}
  {\bibfield  {journal} {\bibinfo  {journal} {Phys.\ Rev. C}\ }\textbf
  {\bibinfo {volume} {79}},\ \bibinfo {pages} {064301} (\bibinfo {year}
  {2009}{\natexlab{c}})}\BibitemShut {NoStop}%
\bibitem [{\citenamefont {Prassa}\ \emph {et~al.}(2012)\citenamefont {Prassa},
  \citenamefont {Nik\ifmmode \check{s}\else \v{s}\fi{}i\ifmmode~\acute{c}\else
  \'{c}\fi{}}, \citenamefont {Lalazissis},\ and\ \citenamefont
  {Vretenar}}]{PNLV.12}%
  \BibitemOpen
  \bibfield  {author} {\bibinfo {author} {\bibfnamefont {V.}~\bibnamefont
  {Prassa}}, \bibinfo {author} {\bibfnamefont {T.}~\bibnamefont {Nik\ifmmode
  \check{s}\else \v{s}\fi{}i\ifmmode~\acute{c}\else \'{c}\fi{}}}, \bibinfo
  {author} {\bibfnamefont {G.~A.}\ \bibnamefont {Lalazissis}}, \ and\ \bibinfo
  {author} {\bibfnamefont {D.}~\bibnamefont {Vretenar}},\ }\href@noop {}
  {\bibfield  {journal} {\bibinfo  {journal} {Phys. Rev. C}\ }\textbf {\bibinfo
  {volume} {86}},\ \bibinfo {pages} {024317} (\bibinfo {year}
  {2012})}\BibitemShut {NoStop}%
\bibitem [{\citenamefont {Lu}\ \emph {et~al.}(2015)\citenamefont {Lu},
  \citenamefont {Li}, \citenamefont {Li}, \citenamefont {Yao},\ and\
  \citenamefont {Meng}}]{LLLYM.15}%
  \BibitemOpen
  \bibfield  {author} {\bibinfo {author} {\bibfnamefont {K.~Q.}\ \bibnamefont
  {Lu}}, \bibinfo {author} {\bibfnamefont {Z.~X.}\ \bibnamefont {Li}}, \bibinfo
  {author} {\bibfnamefont {Z.~P.}\ \bibnamefont {Li}}, \bibinfo {author}
  {\bibfnamefont {J.~M.}\ \bibnamefont {Yao}}, \ and\ \bibinfo {author}
  {\bibfnamefont {J.}~\bibnamefont {Meng}},\ }\href {\doibase
  10.1103/PhysRevC.91.027304} {\bibfield  {journal} {\bibinfo  {journal} {Phys.
  Rev. C}\ }\textbf {\bibinfo {volume} {91}},\ \bibinfo {pages} {027304}
  (\bibinfo {year} {2015})}\BibitemShut {NoStop}%
\bibitem [{\citenamefont {Shi}\ \emph {et~al.}(2019)\citenamefont {Shi},
  \citenamefont {Afanasjev}, \citenamefont {Li},\ and\ \citenamefont
  {Meng}}]{SALM.19}%
  \BibitemOpen
  \bibfield  {author} {\bibinfo {author} {\bibfnamefont {Z.}~\bibnamefont
  {Shi}}, \bibinfo {author} {\bibfnamefont {A.~V.}\ \bibnamefont {Afanasjev}},
  \bibinfo {author} {\bibfnamefont {Z.~P.}\ \bibnamefont {Li}}, \ and\ \bibinfo
  {author} {\bibfnamefont {J.}~\bibnamefont {Meng}},\ }\href@noop {} {\bibfield
   {journal} {\bibinfo  {journal} {Phys. Rev. C}\ }\textbf {\bibinfo {volume}
  {99}},\ \bibinfo {pages} {064316} (\bibinfo {year} {2019})}\BibitemShut
  {NoStop}%
\bibitem [{\citenamefont {Afanasjev}\ \emph {et~al.}(2003)\citenamefont
  {Afanasjev}, \citenamefont {Khoo}, \citenamefont {Frauendorf}, \citenamefont
  {Lalazissis},\ and\ \citenamefont {Ahmad}}]{A250}%
  \BibitemOpen
  \bibfield  {author} {\bibinfo {author} {\bibfnamefont {A.~V.}\ \bibnamefont
  {Afanasjev}}, \bibinfo {author} {\bibfnamefont {T.~L.}\ \bibnamefont {Khoo}},
  \bibinfo {author} {\bibfnamefont {S.}~\bibnamefont {Frauendorf}}, \bibinfo
  {author} {\bibfnamefont {G.~A.}\ \bibnamefont {Lalazissis}}, \ and\ \bibinfo
  {author} {\bibfnamefont {I.}~\bibnamefont {Ahmad}},\ }\href@noop {}
  {\bibfield  {journal} {\bibinfo  {journal} {Phys.\ Rev. C}\ }\textbf
  {\bibinfo {volume} {67}},\ \bibinfo {pages} {024309} (\bibinfo {year}
  {2003})}\BibitemShut {NoStop}%
\bibitem [{\citenamefont {Wang}\ \emph {et~al.}(2013)\citenamefont {Wang},
  \citenamefont {Sun}, \citenamefont {Dong},\ and\ \citenamefont
  {Long}}]{WSDL.13}%
  \BibitemOpen
  \bibfield  {author} {\bibinfo {author} {\bibfnamefont {L.~J.}\ \bibnamefont
  {Wang}}, \bibinfo {author} {\bibfnamefont {B.~Y.}\ \bibnamefont {Sun}},
  \bibinfo {author} {\bibfnamefont {J.~M.}\ \bibnamefont {Dong}}, \ and\
  \bibinfo {author} {\bibfnamefont {W.~H.}\ \bibnamefont {Long}},\ }\href@noop
  {} {\bibfield  {journal} {\bibinfo  {journal} {Phys.\ Rev. C}\ }\textbf
  {\bibinfo {volume} {87}},\ \bibinfo {pages} {054331} (\bibinfo {year}
  {2013})}\BibitemShut {NoStop}%
\bibitem [{\citenamefont {Huang}\ \emph {et~al.}(2016)\citenamefont {Huang},
  \citenamefont {Audi}, \citenamefont {Wang}, \citenamefont {Kondev},
  \citenamefont {Naimi},\ and\ \citenamefont {Xu}}]{AME2016-first}%
  \BibitemOpen
  \bibfield  {author} {\bibinfo {author} {\bibfnamefont {W.~J.}\ \bibnamefont
  {Huang}}, \bibinfo {author} {\bibfnamefont {G.}~\bibnamefont {Audi}},
  \bibinfo {author} {\bibfnamefont {M.}~\bibnamefont {Wang}}, \bibinfo {author}
  {\bibfnamefont {F.~G.}\ \bibnamefont {Kondev}}, \bibinfo {author}
  {\bibfnamefont {S.}~\bibnamefont {Naimi}}, \ and\ \bibinfo {author}
  {\bibfnamefont {X.}~\bibnamefont {Xu}},\ }\href@noop {} {\bibfield  {journal}
  {\bibinfo  {journal} {Chinese Physics}\ }\textbf {\bibinfo {volume} {C41}},\
  \bibinfo {pages} {030002} (\bibinfo {year} {2016})}\BibitemShut {NoStop}%
\bibitem [{\citenamefont {Pototzky}\ \emph {et~al.}(2010)\citenamefont
  {Pototzky}, \citenamefont {Erler}, \citenamefont {P.-G.Reinhard},\ and\
  \citenamefont {Nesterenko}}]{PERN.10}%
  \BibitemOpen
  \bibfield  {author} {\bibinfo {author} {\bibfnamefont {K.~J.}\ \bibnamefont
  {Pototzky}}, \bibinfo {author} {\bibfnamefont {J.}~\bibnamefont {Erler}},
  \bibinfo {author} {\bibnamefont {P.-G.Reinhard}}, \ and\ \bibinfo {author}
  {\bibfnamefont {V.~O.}\ \bibnamefont {Nesterenko}},\ }\href@noop {}
  {\bibfield  {journal} {\bibinfo  {journal} {Eur. Phys. J. A}\ }\textbf
  {\bibinfo {volume} {46}},\ \bibinfo {pages} {299} (\bibinfo {year}
  {2010})}\BibitemShut {NoStop}%
\bibitem [{\citenamefont {Bonneau}\ \emph {et~al.}(2007)\citenamefont
  {Bonneau}, \citenamefont {Quentin},\ and\ \citenamefont
  {M{\"o}ller}}]{BQM.07}%
  \BibitemOpen
  \bibfield  {author} {\bibinfo {author} {\bibfnamefont {L.}~\bibnamefont
  {Bonneau}}, \bibinfo {author} {\bibfnamefont {P.}~\bibnamefont {Quentin}}, \
  and\ \bibinfo {author} {\bibfnamefont {P.}~\bibnamefont {M{\"o}ller}},\
  }\href@noop {} {\bibfield  {journal} {\bibinfo  {journal} {Phys.\ Rev. C}\
  }\textbf {\bibinfo {volume} {76}},\ \bibinfo {pages} {024320} (\bibinfo
  {year} {2007})}\BibitemShut {NoStop}%
\bibitem [{\citenamefont {Afanasjev}\ and\ \citenamefont
  {Shawaqfeh}(2011)}]{AS.11}%
  \BibitemOpen
  \bibfield  {author} {\bibinfo {author} {\bibfnamefont {A.~V.}\ \bibnamefont
  {Afanasjev}}\ and\ \bibinfo {author} {\bibfnamefont {S.}~\bibnamefont
  {Shawaqfeh}},\ }\href@noop {} {\bibfield  {journal} {\bibinfo  {journal}
  {Phys.\ Lett. B}\ }\textbf {\bibinfo {volume} {706}},\ \bibinfo {pages} {177}
  (\bibinfo {year} {2011})}\BibitemShut {NoStop}%
\bibitem [{\citenamefont {Litvinova}\ and\ \citenamefont
  {Afanasjev}(2011)}]{LA.11}%
  \BibitemOpen
  \bibfield  {author} {\bibinfo {author} {\bibfnamefont {E.~V.}\ \bibnamefont
  {Litvinova}}\ and\ \bibinfo {author} {\bibfnamefont {A.~V.}\ \bibnamefont
  {Afanasjev}},\ }\href@noop {} {\bibfield  {journal} {\bibinfo  {journal}
  {Phys.\ Rev. C}\ }\textbf {\bibinfo {volume} {84}},\ \bibinfo {pages}
  {014305} (\bibinfo {year} {2011})}\BibitemShut {NoStop}%
\bibitem [{\citenamefont {Afanasjev}\ and\ \citenamefont
  {Litvinova}(2015)}]{AL.15}%
  \BibitemOpen
  \bibfield  {author} {\bibinfo {author} {\bibfnamefont {A.~V.}\ \bibnamefont
  {Afanasjev}}\ and\ \bibinfo {author} {\bibfnamefont {E.}~\bibnamefont
  {Litvinova}},\ }\href {\doibase 10.1103/PhysRevC.92.044317} {\bibfield
  {journal} {\bibinfo  {journal} {Phys. Rev. C}\ }\textbf {\bibinfo {volume}
  {92}},\ \bibinfo {pages} {044317} (\bibinfo {year} {2015})}\BibitemShut
  {NoStop}%
\bibitem [{\citenamefont {Decharg{\'e}}\ and\ \citenamefont
  {Gogny}(1980)}]{DG.80}%
  \BibitemOpen
  \bibfield  {author} {\bibinfo {author} {\bibfnamefont {J.}~\bibnamefont
  {Decharg{\'e}}}\ and\ \bibinfo {author} {\bibfnamefont {D.}~\bibnamefont
  {Gogny}},\ }\href@noop {} {\bibfield  {journal} {\bibinfo  {journal} {Phys.\
  Rev. C}\ }\textbf {\bibinfo {volume} {21}},\ \bibinfo {pages} {1568}
  (\bibinfo {year} {1980})}\BibitemShut {NoStop}%
\bibitem [{\citenamefont {Gareev}\ \emph {et~al.}(1973)\citenamefont {Gareev},
  \citenamefont {Ivanova}, \citenamefont {Soloviev},\ and\ \citenamefont
  {Fedotov}}]{GISF.73}%
  \BibitemOpen
  \bibfield  {author} {\bibinfo {author} {\bibfnamefont {F.~A.}\ \bibnamefont
  {Gareev}}, \bibinfo {author} {\bibfnamefont {S.~P.}\ \bibnamefont {Ivanova}},
  \bibinfo {author} {\bibfnamefont {V.~G.}\ \bibnamefont {Soloviev}}, \ and\
  \bibinfo {author} {\bibfnamefont {S.~I.}\ \bibnamefont {Fedotov}},\
  }\href@noop {} {\bibfield  {journal} {\bibinfo  {journal} {Phys. Elem. Part.
  and At. Nucl.}\ }\textbf {\bibinfo {volume} {4}},\ \bibinfo {pages} {357}
  (\bibinfo {year} {1973})}\BibitemShut {NoStop}%
\bibitem [{\citenamefont {Alikov}\ \emph {et~al.}(1988)\citenamefont {Alikov},
  \citenamefont {Badalov}, \citenamefont {Nesterenko}, \citenamefont
  {Sushkov},\ and\ \citenamefont {Wawryszczuk}}]{ABNSW.88}%
  \BibitemOpen
  \bibfield  {author} {\bibinfo {author} {\bibfnamefont {B.~A.}\ \bibnamefont
  {Alikov}}, \bibinfo {author} {\bibfnamefont {K.~N.}\ \bibnamefont {Badalov}},
  \bibinfo {author} {\bibfnamefont {V.~O.}\ \bibnamefont {Nesterenko}},
  \bibinfo {author} {\bibfnamefont {A.~V.}\ \bibnamefont {Sushkov}}, \ and\
  \bibinfo {author} {\bibfnamefont {J.}~\bibnamefont {Wawryszczuk}},\
  }\href@noop {} {\bibfield  {journal} {\bibinfo  {journal} {Z. Phys. A}\
  }\textbf {\bibinfo {volume} {331}},\ \bibinfo {pages} {265} (\bibinfo {year}
  {1988})}\BibitemShut {NoStop}%
\bibitem [{\citenamefont {Shirikova}\ \emph {et~al.}(2015)\citenamefont
  {Shirikova}, \citenamefont {Sushkov}, \citenamefont {Malov},\ and\
  \citenamefont {Jolos}}]{SSMJ.15}%
  \BibitemOpen
  \bibfield  {author} {\bibinfo {author} {\bibfnamefont {N.~Y.}\ \bibnamefont
  {Shirikova}}, \bibinfo {author} {\bibfnamefont {A.~V.}\ \bibnamefont
  {Sushkov}}, \bibinfo {author} {\bibfnamefont {L.~A.}\ \bibnamefont {Malov}},
  \ and\ \bibinfo {author} {\bibfnamefont {R.~V.}\ \bibnamefont {Jolos}},\
  }\href@noop {} {\bibfield  {journal} {\bibinfo  {journal} {Eur. Phys. J. A}\
  }\textbf {\bibinfo {volume} {51}},\ \bibinfo {pages} {21} (\bibinfo {year}
  {2015})}\BibitemShut {NoStop}%
\bibitem [{\citenamefont {Neerg{\aa}rd}\ and\ \citenamefont
  {Bentley}(2019)}]{NB.19}%
  \BibitemOpen
  \bibfield  {author} {\bibinfo {author} {\bibfnamefont {K.}~\bibnamefont
  {Neerg{\aa}rd}}\ and\ \bibinfo {author} {\bibfnamefont {I.}~\bibnamefont
  {Bentley}},\ }\href@noop {} {\bibfield  {journal} {\bibinfo  {journal} {Phys.
  Rev. C}\ }\textbf {\bibinfo {volume} {99}},\ \bibinfo {pages} {054315}
  (\bibinfo {year} {2019})}\BibitemShut {NoStop}%
\bibitem [{\citenamefont {Lalazissis}\ \emph {et~al.}(1997)\citenamefont
  {Lalazissis}, \citenamefont {K{\"o}nig},\ and\ \citenamefont {Ring}}]{NL3}%
  \BibitemOpen
  \bibfield  {author} {\bibinfo {author} {\bibfnamefont {G.~A.}\ \bibnamefont
  {Lalazissis}}, \bibinfo {author} {\bibfnamefont {J.}~\bibnamefont
  {K{\"o}nig}}, \ and\ \bibinfo {author} {\bibfnamefont {P.}~\bibnamefont
  {Ring}},\ }\href@noop {} {\bibfield  {journal} {\bibinfo  {journal} {Phys.\
  Rev. C}\ }\textbf {\bibinfo {volume} {55}},\ \bibinfo {pages} {540} (\bibinfo
  {year} {1997})}\BibitemShut {NoStop}%
\bibitem [{\citenamefont {Lalazissis}\ \emph {et~al.}(2009)\citenamefont
  {Lalazissis}, \citenamefont {Karatzikos}, \citenamefont {Fossion},
  \citenamefont {Arteaga}, \citenamefont {Afanasjev},\ and\ \citenamefont
  {Ring}}]{NL3*}%
  \BibitemOpen
  \bibfield  {author} {\bibinfo {author} {\bibfnamefont {G.~A.}\ \bibnamefont
  {Lalazissis}}, \bibinfo {author} {\bibfnamefont {S.}~\bibnamefont
  {Karatzikos}}, \bibinfo {author} {\bibfnamefont {R.}~\bibnamefont {Fossion}},
  \bibinfo {author} {\bibfnamefont {D.~P.}\ \bibnamefont {Arteaga}}, \bibinfo
  {author} {\bibfnamefont {A.~V.}\ \bibnamefont {Afanasjev}}, \ and\ \bibinfo
  {author} {\bibfnamefont {P.}~\bibnamefont {Ring}},\ }\href@noop {} {\bibfield
   {journal} {\bibinfo  {journal} {Phys.\ Lett.}\ }\textbf {\bibinfo {volume}
  {B671}},\ \bibinfo {pages} {36} (\bibinfo {year} {2009})}\BibitemShut
  {NoStop}%
\bibitem [{\citenamefont {Koepf}\ and\ \citenamefont {Ring}(1990)}]{KR.90}%
  \BibitemOpen
  \bibfield  {author} {\bibinfo {author} {\bibfnamefont {W.}~\bibnamefont
  {Koepf}}\ and\ \bibinfo {author} {\bibfnamefont {P.}~\bibnamefont {Ring}},\
  }\href {http://www.sciencedirect.com/science/article/pii/037594749090160N}
  {\bibfield  {journal} {\bibinfo  {journal} {Nucl.~ Phys.~ A}\ }\textbf
  {\bibinfo {volume} {511}},\ \bibinfo {pages} {279} (\bibinfo {year}
  {1990})}\BibitemShut {NoStop}%
\bibitem [{\citenamefont {Rutz}\ \emph {et~al.}(1999)\citenamefont {Rutz},
  \citenamefont {Bender}, \citenamefont {Reinhard},\ and\ \citenamefont
  {Maruhn}}]{RBRM.99}%
  \BibitemOpen
  \bibfield  {author} {\bibinfo {author} {\bibfnamefont {K.}~\bibnamefont
  {Rutz}}, \bibinfo {author} {\bibfnamefont {M.}~\bibnamefont {Bender}},
  \bibinfo {author} {\bibfnamefont {P.-G.}\ \bibnamefont {Reinhard}}, \ and\
  \bibinfo {author} {\bibfnamefont {J.~A.}\ \bibnamefont {Maruhn}},\
  }\href@noop {} {\bibfield  {journal} {\bibinfo  {journal} {Phys.\ Lett. B}\
  }\textbf {\bibinfo {volume} {468}},\ \bibinfo {pages} {1} (\bibinfo {year}
  {1999})}\BibitemShut {NoStop}%
\bibitem [{\citenamefont {Agbemava}\ \emph {et~al.}(2015)\citenamefont
  {Agbemava}, \citenamefont {Afanasjev}, \citenamefont {Nakatsukasa},\ and\
  \citenamefont {Ring}}]{AANR.15}%
  \BibitemOpen
  \bibfield  {author} {\bibinfo {author} {\bibfnamefont {S.~E.}\ \bibnamefont
  {Agbemava}}, \bibinfo {author} {\bibfnamefont {A.~V.}\ \bibnamefont
  {Afanasjev}}, \bibinfo {author} {\bibfnamefont {T.}~\bibnamefont
  {Nakatsukasa}}, \ and\ \bibinfo {author} {\bibfnamefont {P.}~\bibnamefont
  {Ring}},\ }\href {\doibase 10.1103/PhysRevC.92.054310} {\bibfield  {journal}
  {\bibinfo  {journal} {Phys. Rev. C}\ }\textbf {\bibinfo {volume} {92}},\
  \bibinfo {pages} {054310} (\bibinfo {year} {2015})}\BibitemShut {NoStop}%
\bibitem [{\citenamefont {M{\"o}ller}\ and\ \citenamefont {Nix}(1992)}]{MN.92}%
  \BibitemOpen
  \bibfield  {author} {\bibinfo {author} {\bibfnamefont {P.}~\bibnamefont
  {M{\"o}ller}}\ and\ \bibinfo {author} {\bibfnamefont {J.}~\bibnamefont
  {Nix}},\ }\href {\doibase http://dx.doi.org/10.1016/0375-9474(92)90244-E}
  {\bibfield  {journal} {\bibinfo  {journal} {Nucl. Phys. A}\ }\textbf
  {\bibinfo {volume} {536}},\ \bibinfo {pages} {20 } (\bibinfo {year}
  {1992})}\BibitemShut {NoStop}%
\bibitem [{\citenamefont {Afanasjev}\ \emph {et~al.}(2015)\citenamefont
  {Afanasjev}, \citenamefont {Agbemava}, \citenamefont {Ray},\ and\
  \citenamefont {Ring}}]{AARR.15}%
  \BibitemOpen
  \bibfield  {author} {\bibinfo {author} {\bibfnamefont {A.~V.}\ \bibnamefont
  {Afanasjev}}, \bibinfo {author} {\bibfnamefont {S.~E.}\ \bibnamefont
  {Agbemava}}, \bibinfo {author} {\bibfnamefont {D.}~\bibnamefont {Ray}}, \
  and\ \bibinfo {author} {\bibfnamefont {P.}~\bibnamefont {Ring}},\ }\href@noop
  {} {\bibfield  {journal} {\bibinfo  {journal} {Phys.\ Rev.\ C}\ }\textbf
  {\bibinfo {volume} {91}},\ \bibinfo {pages} {014324} (\bibinfo {year}
  {2015})}\BibitemShut {NoStop}%
\bibitem [{\citenamefont {Geng}\ \emph {et~al.}(2005)\citenamefont {Geng},
  \citenamefont {Toki},\ and\ \citenamefont {Meng}}]{GTM.05}%
  \BibitemOpen
  \bibfield  {author} {\bibinfo {author} {\bibfnamefont {L.}~\bibnamefont
  {Geng}}, \bibinfo {author} {\bibfnamefont {H.}~\bibnamefont {Toki}}, \ and\
  \bibinfo {author} {\bibfnamefont {J.}~\bibnamefont {Meng}},\ }\href@noop {}
  {\bibfield  {journal} {\bibinfo  {journal} {Prog.\ Theor.\ Phys.}\ }\textbf
  {\bibinfo {volume} {113}},\ \bibinfo {pages} {785} (\bibinfo {year}
  {2005})}\BibitemShut {NoStop}%
\end{thebibliography}%
\end{document}